\documentclass[11pt]{amsart}

\RequirePackage[OT1]{fontenc}

\usepackage[foot]{amsaddr}
\usepackage{natbib}
\usepackage{multirow}
\usepackage{enumitem}
\usepackage{caption,subcaption}

\usepackage{xr}
\makeatletter
\newcommand*{\addFileDependency}[1]{
  \typeout{(#1)}
  \@addtofilelist{#1}
  \IfFileExists{#1}{}{\typeout{No file #1.}}
}
\makeatother

\usepackage{tchdr}
\boldshortcuts

\usepackage{url}

\usepackage{graphicx}
\usepackage{wrapfig}
\usepackage{setspace}
\DeclareGraphicsExtensions{.eps, .ps}
\usepackage{amsmath, amsthm, amsfonts}

\numberwithin{equation}{section}
\theoremstyle{plain}
\newtheorem{theorem}{Theorem}[section]
 \newtheorem{remark}{Remark}

\newtheorem{lemma}[theorem]{Lemma}

\usepackage{setspace}
\def\logit{\text{logit}}
\def\pr{\text{pr}}

\def\I{\bf I}
\def\code#1{\texttt{#1}}
\def\new{\text{new}}

\begin{document}

\title[Addressing selection bias and measurement error in COVID-19 case counts]{Addressing selection bias and measurement error in COVID-19 case count data using auxiliary information} %

\author{Walter Dempsey}
\address{Department of Biostatistics, University of Michigan, Ann Arbor, MI 48109}

\begin{abstract}
  Coronavirus case-count data has influenced government policies and drives most epidemiological forecasts. Limited testing is cited as the key driver behind minimal information on the COVID-19 pandemic. While expanded testing is laudable, measurement error and selection bias are the two greatest problems limiting our understanding of the COVID-19 pandemic; neither can be fully addressed by increased testing capacity. In this paper, we demonstrate their impact on estimation of point prevalence and the effective reproduction number. We show that estimates based on the millions of molecular tests in the US has the same mean square error as a small simple random sample.  To address this, a procedure is presented that combines case-count data and random samples over time to estimate selection propensities based on key covariate information. We then combine these selection propensities with epidemiological forecast models to construct a \emph{doubly robust} estimation method that accounts for both measurement-error and selection bias.  This method is then applied to estimate Indiana's active infection prevalence using case-count, hospitalization, and death data with demographic information, a statewide random molecular sample collected from April 25--29th, and Delphi's COVID-19 Trends and Impact Survey.  We end with a series of recommendations based on the proposed methodology.
\end{abstract}

\maketitle

\newpage

\section{Introduction}
The World Health Organization has declared the coronavirus disease 2019 (COVID-19) a public health emergency.  As of July 29th, 2021, over 196 million cases have been confirmed worldwide with 34.8 million cases and over 612 thousand confirmed deaths across the United States. This pandemic
has become the focal point of everyday life; yet the data landscape for understanding COVID-19 remains limited.  Public databases~\citep{JHU_Lancet,NYT} provide incoming county-level information of confirmed cases and deaths.  Statisticians, epidemiologists, economists, and data scientists have used this granular data to forecast COVID-19 case-counts, deaths, and hospitalizations~\citep{Giordano2020,Song2020,Ray2020,2020.IHME,Wang2020.03,JTD36385}.

This paper has two main objectives.  The first objective is to express reservations at the use of observed case-counts as a proxy for disease prevalence and in estimation of standard epidemiological models for inference and forecasting.  The reason is straightforward: observed case-count data is plagued by selection bias and measurement error. Through a series of calculations, we will demonstrate that the information gained from increasing testing capacity is limited in the presence of selection bias and when testing inaccuracies persist.  In particular, the millions of tests in the US have a small effective sample size when compared to random sampling. These calculations demonstrate the importance of probabilistic sampling designs over time for estimation of point prevalence and effective reproduction number.

Selection bias in case-count data is primarily due to it being a \emph{diagnostic tool}, i.e., individuals who are symptomatic or have a known/suspected exposure are more likely to present for diagnostic testing.  Case-count data arising from non-random testing means it cannot provide valid prevalence or incidence estimates due to the significant proportion of asymptomatic and pauci-symptomatic cases. Random testing, on the other hand, is used for \emph{screening purposes}, i.e., is an appropriate tool for reconstructing prevalence/incidence estimates. Due to monetary and time constraints, however, random testing is performed infrequently.  As of June 2021, Indiana and Ohio are the only states to conduct statewide random sample testing\footnote{These are the only random samples to collect both seroprevalence and diagnostic testing results. The CDC and other states have conducted seroprevalence-only studies.}. Indiana's sample was collected from April 25--29, 2020~\citep{Yiannoutsos2021}.  Such infrequent random testing is likely to provide insufficient information to help researchers and policy makers better understand the disease trajectory which can change rapidly over time.  Therefore, while random testing may be preferable in theory, in practice governments, researchers, and policy makers continue to use coronavirus case-counts to understand the impact of COVID-19 on the population and make data-informed decisions.


The second objective is to demonstrate how random samples provide necessary \emph{auxiliary information} to address selection bias in coronavirus case-count data.  Random samples provide the necessary covariate information from a representative sample from the population to estimate selection propensities. These propensities can then used in an inverse-probability weighting scheme to construct estimators of disease prevalence that attempt to control for selection bias.  A doubly robust extension allows researchers to combine these estimates with epidemiological forecasts based on compartmental models that are common in the study of infectious diseases~\citep{Hao2020,Song2020,Ray2020,Johndrow2020}.

The proposed approach requires covariate information to be collected on individuals who receive a COVID-19 test.  Unfortunately, many states do not require or report auxiliary covariate information beyond basic demographic information (e.g., gender, age, race, and ethnicity).  We end with a brief list of suggestions of changes to current practice based on the proposed methodology. While we demonstrate empirical improvements over simple disease prevalence estimates, we also highlight how selection bias may persist and impact uncertainty quantification.

\begin{remark}[An evolving pandemic]
This paper focuses on COVID-19 case count, testing, and death data collected from April 2020 through February 2021.  Numbers presented on disease dynamics are therefore based on the original strain.  Selection bias and measurement error persist in data arising from the delta and omicron strains and will likely persist for future variants. While not discussed in this paper, the framework presented will remain an appropriate tool for addressing selection bias and measurement error in these settings.
\end{remark}

\subsection{Related work}

This article discusses the relationship between three statistical concepts: selection bias, measurement error, and population size.  Potential biases in observational studies of COVID-19 have been identified elsewhere in the literature~\citep{Kahn2021,Accorsi2020}.  While the impact of measurement error~\citep{Smeden2019} and selection bias~\citep{Keiding2016} on estimation are both well-studied topics in general, here we provide a new perspective by building on the work of \cite{Meng2018} who studied an error decomposition to understand the relationship between selection bias and population size.  Specifically, we quantify the interaction between measurement-error and selection bias on statistical error, showing how the sign and magnitude can change drastically.  We then discuss this relationship in the context of observational COVID-19 case-count data, showing the impact on the effective sample size can be quite large.

After demonstrating the limitations of case-count analysis when compared to random sampling, we then assess whether there is potential for combining the \emph{nonprobability samples} with \emph{probability samples} to improve point prevalence estimation. For any probability sampling design, the Horvitz-Thompson estimator~\citep{HT1952} incorporates design information via inverse-probability weights (IPW).  For nonprobability samples, the IPW estimator requires modelling the propensity scores.  Its use in the survey context is also referred to as quasi-randomization~\citep{Elliott2017}. \cite{Valliant2011} consider a weighted logistic regression procedure using the pooled probability and nonprobability samples.  \cite{Chen2019} consider a pseudo-likelihood approach that uses the random samples as a proxy for a term in the log-likelihood.  Here, we extend this approach to account for measurement-error as well as observing random samples at multiple times. We then provide an extension of the statistical error decomposition and discuss the trade-offs inherent in such a weighting approach. The proposed approach is distinct from validation studies~\citep{Fox2020} -- a traditional epidemiological method in which investigators compare measurement accuracy with a gold standard measure to mitigate bias.  Here, the \emph{gold standard} of a different test is replaced by a sample with a less biased selection mechanism.

One core component of coronavirus research is epidemiological compartmental modelling of case-count and death data. These models can be used to answer a variety of research questions including case-count forecasting, estimation of the effective reproduction number, and estimation of quarantine and other health policies on infectious disease dynamics.  The basic approach is a deterministic compartmental model called the susceptible-infectious-recovered (SIR) model.  A probabilistic extension was proposed by~\cite{Osthus2017} to model one-dimensional time series of infected proportions. \cite{Song2020} extended this approach to incorporate interventions and assess interventions on COVID-19 epidemic in China. \cite{Hao2020} extends this work further to account for various presymptomatic infectiousness, time-varying ascertainment rates, transmission rates and population movements.

Given a probability sampling design, individual predictions can be leveraged to improve estimation via model-assisted approaches~\citep{Breidt2017}.  For nonprobability samples, \cite{Chen2019} derive a doubly robust approach that uses outcome predictions given covariates on the nonprobability and probability samples.  Here, we combine the compartmental model of \cite{Song2020} but instead, as in \cite{Johndrow2020}, focus on COVID-19 confirmed death count data.  We generate epidemiological forecasts for active infection rates within each population strata.  We then demonstrate how to combine these forecasts with the IPW approach to construct doubly-robust estimates of active infection rates.  A derived statistical error decomposition guides this discussion.

Recent work by \cite{Zhao2021} pointed out that estimation of key epidemiological parameters such as the incubation time using standard epidemiological models can suffer from severe bias due to issues beyond selection bias and measurement error.  Right truncation and epidemic growth lead to patients ``being more likely to be infected towards the end of their exposure period''~\citep[pp. 3]{Zhao2021}.  Their approach constructs a study sample and statistical model to account for these issues.  In this paper, rather than focusing on sample construction, we ask whether one can collect auxiliary information to address selection bias in the observed case count data directly. Our approach is related to the concept of \emph{target validity}~\citep{Westreich2018}, in which the issues of internal and external validity are jointly addressed with respect to a specific population of interest.  This article is a concrete attempt to address both types of validity and extend the conversation on target validity within the context of analysis of observational COVID-19 studies.

 \section{COVID-19 testing and data}
 \label{section:data}

Here we provide the necessary background to understand COVID-19 diagnostic testing, its scientific use in managing the pandemic, and the data streams considered in this paper.

\subsection{Diagnostic testing}
\label{section:testinginfo}

 Upon infection with the original SARS-CoV-2 variant, an incubation period (time to symptom onset) starts and lasts approximately five days~\citep{Lauer2020}.  The viral load will be detectable by at least the end of the latent period (time to infectiousness), which for the original SARS-CoV-2 variant occurs before the end of the incubation period.  A \emph{molecular test} refers to diagnostic tests that aim to detect viral load above a certain threshold (e.g., RT-PCR or antigen tests); see~\cite{Mina2020} for a detailed discussion of cycle thresholds.  If someone has an active infection -- here defined as being infected with SARS-CoV-2 and having a viral load that has yet to fall below detectable levels by RT-PCR testing -- then after the incubation period, a molecular test with perfect sensitivity will yield a positive result while the patient has a viral load above the threshold of detection.  After that, the viral load will decrease below that threshold and a molecular test with perfect specificity will come back negative. While an individual infected with the original SARS-CoV-2 strain may yield a positive molecular test for several weeks, they will likely stop transmitting the disease within a few days of infection, meaning a positive molecular test does not imply transmissibility.

 A molecular test conducted on an actively infected individual may return a negative result.  Such false negatives are very strongly associated with when the test is conducted.  In the incubation phase, most molecular tests will return a false negative result.  Molecular tests are most sensitive when the viral loads are highest which for the original strain occurs during the first few days of transmissibility~\citep{Mina2020}.   Moreover, most molecular tests are performed via nasopharyngeal swab.  Specimen collection by swab is known to impact false negative/positive rates regardless of test timing.   Systematic reviews suggest that 87\% sensitivity and 97.6\% specificity are reasonable estimates for RT-PCR tests performed during the time window under consideration in this paper~\citep{Arevalo2020, Woloshin2020,Cohen2020}.  To the best of our knowledge, both Indiana Department of Health's molecular testing and the random state-wide RT-PCR tests were primarily collected via nasopharyngeal swab.

 The primary goal of molecular tests is diagnosis of active infections in the population.  Such diagnostic tests generate important information about the presence of SARS-CoV-2 in the population, and help scientists and policy-makers understand patterns of transmission and propagation. Rapid and frequent molecular testing is cited as a key component~\citep{OECD2021} in effective strategies to identify active infections and prevent systemic outbreaks.

\emph{Serological tests} look for an immunological response to the virus.  A week or so after an individual is infected with SARS-CoV-2, the individual will start producing antibodies.  At this point, a serological test with perfect sensitivity will come back positive.  This test provides evidence of past infection while the molecular test provides evidence of an \emph{active} infection.

To better estimate SARS-CoV-2 immunity in a population, seroprevalence studies that generate a probabilistic population sample and perform serological tests on the sample can be collected.  These studies are useful for disease surveillance. By June 2021, the CDC has conducted ten large-scale geographic serological surveys with three rounds.  While population-based sampling strategies provide a more representative nationwide sample, they are very time intensive and expensive.

 \subsection{Publicly available data on COVID-19}
 \label{subsection:testinginfo}

 The primary goal of this paper is to produce accurate estimates of the population-level active infection rates over time using \emph{publicly available} viral testing data.  Secondary goals include estimation of rates of change and the effective reproduction number which characterize disease trajectory.
 These quantities are fundamental to public health policy and provide critical information on the presence and transmission of SARS-CoV-2.

\begin{figure}[!th]
 \centering
 \begin{subfigure}{.45\textwidth}
  \centering
  \includegraphics[width=.9\linewidth]{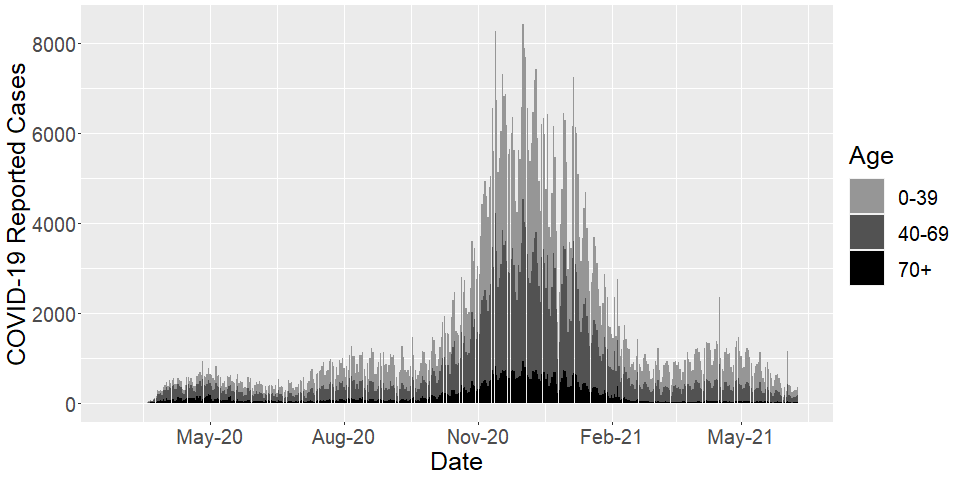}
  \caption{Case-counts per day}
  \label{fig:in-cases}
 \end{subfigure}
 \begin{subfigure}{.45\textwidth}
  \centering
  \includegraphics[width=.9\linewidth]{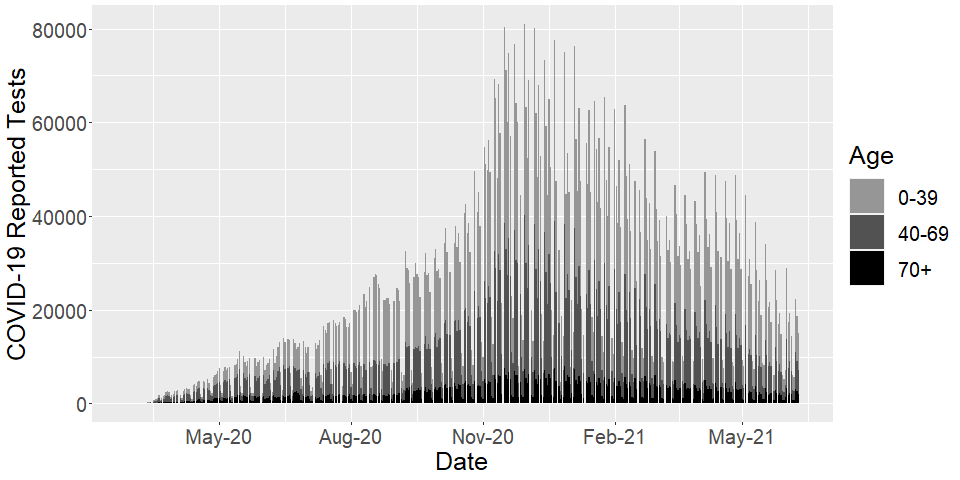}
 \caption{Tests per day}
 \label{fig:in-tests}
 \end{subfigure} \\ [1ex]
 \begin{subfigure}{\linewidth}
 \centering
 \includegraphics[width=.45\linewidth]{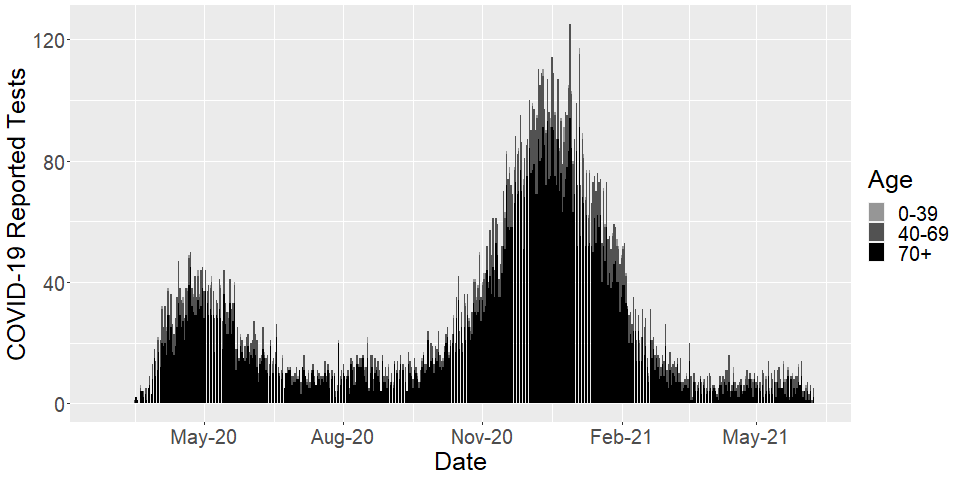}
  \caption{Deaths per day}
  \label{fig:in-deaths}
 \end{subfigure}
 \caption{Indiana daily COVID-19 case count, testing, and death data by age strata}
 \label{fig:in-data}
 \end{figure}

 Coronavirus case-count data refers to the number of positive molecular tests performed on each day.  Figure~\ref{fig:in-cases} plots the number of reported confirmed COVID-19 cases per day in the state of Indiana.  Figure~\ref{fig:in-tests} plots the total number of COVID-19 molecular tests performed per day. Figure~\ref{fig:in-deaths} plots the total number of COVID-19 related reported deaths per day. Public databases maintained by \href{https://bit.ly/2UqFSuA}{Johns Hopkins University} and the \href{https://bit.ly/2vUHfrK}{New York Times} provide accessible incoming county-level information of confirmed cases and deaths.

 Public databases most often, however, only contain aggregate information.  The Johns Hopkins dashboard, for example, provides demographic breakdown of case counts as well as the total confirmed cases and deaths by county.  Aggregate time series of case count and deaths can also be extracted.  Unfortunately, most dashboards do not provide demographic information on who requested a test nor on who tested positive for SARS-CoV-2 \emph{over time}. Working closely with the State of Indiana, we were able to access COVID-19 total tests, positive tests, and related deaths per day broken out marginally by age, gender, ethnicity, and race~\citep{IndianaData2021}.  These granular datasets are now publicly available and motivate the proposed approach.
 %

\begin{remark}[Reporting Delays]
Figure~\ref{fig:in-data} shows clear reduced testing and case count reporting on weekends compared to weekdays. Moreover, COVID-19 tests are reported on the day they were administered, while case counts are reported based on the date the positive test was reported to and confirmed by the Indiana Department of Health system.  To minimize the impact of reporting, testing and case-count data are aggregated at the weekly level for analysis.
 \end{remark}

\begin{remark}[Public versus government datasets]
In this paper, we focus on \emph{publicly-available data}, i.e., data that anyone can download directly from official government data portals such as the Indiana Data Hub (IDH). While the scientific community has contributed through independent COVID-19 observational and clinical studies, a significant component of public policy guidance has relied on testing and case count data, e.g., CDC and state guidelines based on test positivity rates and relative changes in the case counts over time.   Addressing selection bias and measurement-error in these public datasets is imperative for better informed public policy debates.

Note that Indiana's COVID-19 response team has access to official \emph{government data} which may not be publicly-available and is likely stored at the individual-level. Due to data privacy and legal concerns, some collected covariates may not be reported publicly and others are aggregated and reported marginally.  While data analytic decisions discussed in Section~\ref{section:applications} were made due to access of publicly-available data, the overall data analysis framework is designed to be a tool for health departments to assess the pandemic and guide responses, e.g.,  account for individual-level covariate information, rather than publicly-reported strata-level information, that may include important covariates such as symptom status and recent COVID-19 contact.
 \end{remark}


 \subsubsection{Testing restrictions and public health policy in Indiana}
 \label{section:publicpolicyindiana}

 Due to limited testing capacity, many US states instituted testing restrictions early on in the pandemic.
 Here, we reconstruct the testing restriction history for the state of Indiana.
 On March 6th, Indiana State Health Department of Health confirmed the first case of COVID-19 in Indiana.  From early March 2020 until April 28th, 2020, only symptomatic essential workers and their households, symptomatic high risk individuals, and individuals who had returned recently from overseas travel were eligible.  An individual was considered high risk if they were over the age of 65, diabetic, obese, pregnant, a member of a minority population at greater risk of severe illness, or had high blood pressure.
 On April 28th, 2020, the criteria expanded to include any symptomatic Indiana resident, people in close contact with those who had tested positive, and residents of congregant communities~\citep{wishtv2020}.
 As of May 12th, 2020, testing expanded to include any high risk individual regardless of symptom status~\citep{indystar2020}
 On June 15th, 2020, Indiana State Department of Health (ISDH) lifted all testing restrictions~\citep{indystar2020v2}.
 Testing restrictions impact the propensity of an individual to receive a test and therefore, if COVID-19 positive, contribute to the case count.  In this paper, testing restrictions are addressed by fitting time-varying testing propensities that depend on relevant covariate information.

 On March 23rd, 2020, Indiana's governor issued a \emph{stay at home} order effective March 26th through April 5th.  The order was extended until April 30th. On May 1st, 2020, a five-stage plan for gradual reopening was announced by Governor Holcomb~\citep{fivestageplan}.  Such policies target reduction in active infection rates. In this paper, Indiana's public health policy is incorporated in our construction of epidemiological forecasts in Section~\ref{section:modelbased}.

 \subsection{Probabilistic samples in Indiana}

 Due to testing restrictions and other potential selection biases, publicly reported COVID-19 case count data may not be sufficient to understand active infection rates or the disease trajectory.  Here, we discuss two random samples that provide auxiliary information and may help address selection bias.

 \subsubsection{Random statewide testing}

 Between April 25--29, 2020, Indiana conducted statewide random molecular testing of persons ages $\geq 12$ years to assess prevalence of active infection to SARS-CoV-2~\citep{Yiannoutsos2021}. A stratified random sampling design was conducted using Indiana’s 10 public health preparedness districts as sampling strata. 15,495 participants were contacted resulting in a final sample size of 3,658. Demographic data was collected (e.g., summary statistics on age, sex, and race) as well as data on whether they experienced any COVID-19 compatible symptoms during the past 2 weeks or had shared a household with someone who had a positive test result for SARS-CoV-2. During May 2--3, 2020, an additional non-random sample of 898 individuals was also collected.
 Table~\ref{tab:indiana} summarizes the data.

 \begin{table}[!th]
 \begin{tabular}{c | c | r r r | r | r r }
 & & \multicolumn{3}{c}{Total Tests (\%)} & IN & \multicolumn{2}{c}{Positive Test Rate (\%)}\\
 \cline{3-5} \cline{7-8}
 & & NonRandom & Random & CTIS & Census & NonRandom & Random \\ \hline
 \multirow{2}{*}{Sex} & Female & 58.2 & 55.0 & 54.0 & 50.7 & 21.7 (11.2) & 1.4 \\
 & Male & 41.8 & 45.0  & 46.0 & 49.3 & 24.2 (12.4) & 2.1 \\ \hline
 \multirow{3}{*}{Age} & $<40$ & 39.4 & 28.0  & 36.2 & 52.7
 & 29.7 (15.0) & 1.7 \\
 & $40-59$ & 41.1 & 36.0  & 34.3 & 25.2 & 24.9 (12.5) & 2.1 \\
 & $\geq 60$ & 19.5 & 36.0  & 29.5 & 22.1 & 6.7 (3.4) & 0.9 \\ \hline
 \multirow{2}{*}{Race} & White & 23.1 & 92.0  & - & 86.9 & 19.5 (9.6) & 1.5 \\
 & Nonwhite & 76.9 & 8.0  & - & 13.1 & 25.0 (12.3) & 3.4 \\ \hline
 \multirow{2}{*}{Fever} & Yes & 17.0 & 1.8  & 1.0 & - & 66.4 (32.1) & 4.5 \\
 & No & 83.0 & 98.2  & 99.0 & - & 15.6 (7.5) & 1.3 \\ \hline
 Household & Yes & 10.8 & 1.4  & 1.8 & - & 46.1 (22.4) & 29.4 \\
 $+$ Case & No & 89.2 & 98.6  & 98.2 & - & 21.6 (10.4) & 1.3 \\ \hline
 Prior $+$ & Yes & 6.1 & 1.4  & - & - & 39.2 (20.2) & 24.4 \\
 Test & No & 93.9 & 98.6  & - & - & 21.6 (11.1) & 1.3 \\ \hline
 \end{tabular}
 \caption{Estimated total tests ($\%$) and point prevalence of active infection with SARS-CoV-2 by demographics in Indiana~\citep{Yiannoutsos2021,doi:10.1073/pnas.2111454118}. NonRandom positive test rates in parentheses are adjusted rates to match the statewide rate of $11.7\%$ on August 30th.}
 \label{tab:indiana}
 \end{table}

 \subsubsection{Delphi's COVID-19 Trends and Impact Survey}
 \label{subsection:fbsymptom}
\begin{figure}[!th]
 \centering
 \includegraphics[width = 0.9\textwidth]{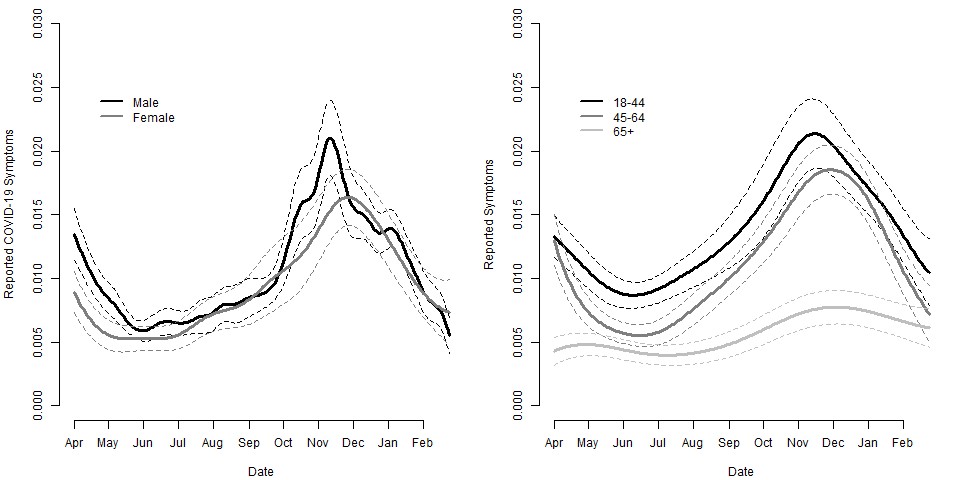}
 \caption{Rate of reported COVID-19 symptoms per strata.  Daily rates were estimated using weighted method suggested by~\citep{doi:10.1073/pnas.2111454118} on each day separately and then smoothed over time using local-linear nonparametric regression.}
 \label{fig:fbsymptoms}
 \vspace{-0.3cm}
 \end{figure}

 Since April 2020, in collaboration with Facebook, the Delphi group at Carnegie Mellon University has conducted the COVID-19 Trends and Impact Survey (CTIS) to monitor the spread and impact of the COVID-19 pandemic in the United States.  The survey is advertised through Facebook, who automatically select a random sample of its users to see the advertisement~\citep{doi:10.1073/pnas.2111454118}.  Data collected includes basic demographic information and if the respondent has symptoms such as fever, coughing, shortness of breath, or loss of smell which are associated with COVID-19.  The survey defines an individual as displaying \emph{COVID-like symptoms} if they exhibit a fever along with a cough, or shortness of breath, or difficulty breathing.  Figure~\ref{fig:fbsymptoms} displays smoothed estimates of the fraction of individuals who report COVID-like symptoms within the past 24-hours by age and gender.  The Delphi's COVID-19 Trends and Impact Survey (CTIS) is used as the main source of auxiliary information on time-varying characteristics of the population of Indiana, e.g., displaying COVID-like symptoms or in contact with COVID-19 positive individuals.

 \section{Analysis of case-count data}
 \label{section:casecount}

 Let $N$ denote the population size.  At a given time, let $Y_j$ denote COVID-19 status for the $j$th individual in the population, $j=1,\ldots, N$. Here, like in survey methodology~\citep{Cochran77}, we treat COVID-19 status as a fixed but unknown quantity of interest. For simplicity, we start by ignoring the dynamic nature of the outbreak and recoverability of individuals. We assume either individual $j$ is COVID-19 positive and $Y_j=1$ or is COVID-19 negative and $Y_j=0$. We also let $I_j \in \{0,1\}$ be an indicator that the individual was tested ($I_j = 1$) or not ($I_j=0$).

 To start, we assume the overall number of active COVID-19 cases and/or active infection rate (AIR) are of primary interest. That is, we are interested in either the population total $Y = \sum_{j=1}^N Y_j$ or the population average $\bar Y = Y/N$. Suppose that $n$ tests are performed and we observe the values $y_1, \ldots, y_n \in \{0,1\}$.  Then a natural candidate for AIR is the proportion of positive tests $\bar y = \frac{1}{n} \sum_{i=1}^n y_i$ -- commonly referred to as the test positivity rate -- and a natural candidate for overall active cases is $N \times \bar y$.
 Under simple random sampling (SRS) or any other epsem\footnote{equal probability of selection method} design, the above are unbiased estimators of the population-level quantities of interest.  Under SRS, the variance of the estimator can be expressed as $\frac{1}{N-1} \times \frac{1-f}{f} \times \sigma_Y^2$ where $f = n/N$ is the sampling fraction and $\sigma_Y^2 = \frac{1}{N} \sum_{j=1}^N (Y_j - \bar Y)^2 = \bar Y (1- \bar Y)$.

 These random selection mechanisms are independent of the outcome of interest. When this is not the case, selection effects may cause bias. To better understand this issue, \cite{Meng2018} recently provided the following intuitive and powerful statistical decomposition of the error between $\bar y$ and the true proportion $\bar Y$
 $$
 \bar y_n - \bar Y =  \rho_{I, Y} \times \sqrt{\frac{1-f}{f}} \times \sigma_Y.
 $$
 The first term represents \emph{data quality}, the second \emph{data quantity}, and the third \emph{problem difficulty}. The term $\rho_{I,Y}$ is the empirical correlation between the population values~$\{ Y_j \}_{j=1}^N$ and the selection values $\{ I_j \}_{j=1}^N$.  Under simple random sampling, $E_{\I} [ \rho_{I,Y} ] = 0$,
 so there is no bias.


 \subsection{Imperfect testing}
 \label{section:imperfecttesting}

 Tests are imperfect.  COVID-19 testing is no exception. Here we investigate the interplay between imperfect testing and selection bias.  Researchers often assume measurement error leads to parameter attenuation.  When paired with selection bias, however, the two sources become entangled, and resulting errors can be magnified, muted, or even switch signs.

\begin{figure}[!th]
 \centering
 \includegraphics[width = 0.7\textwidth]{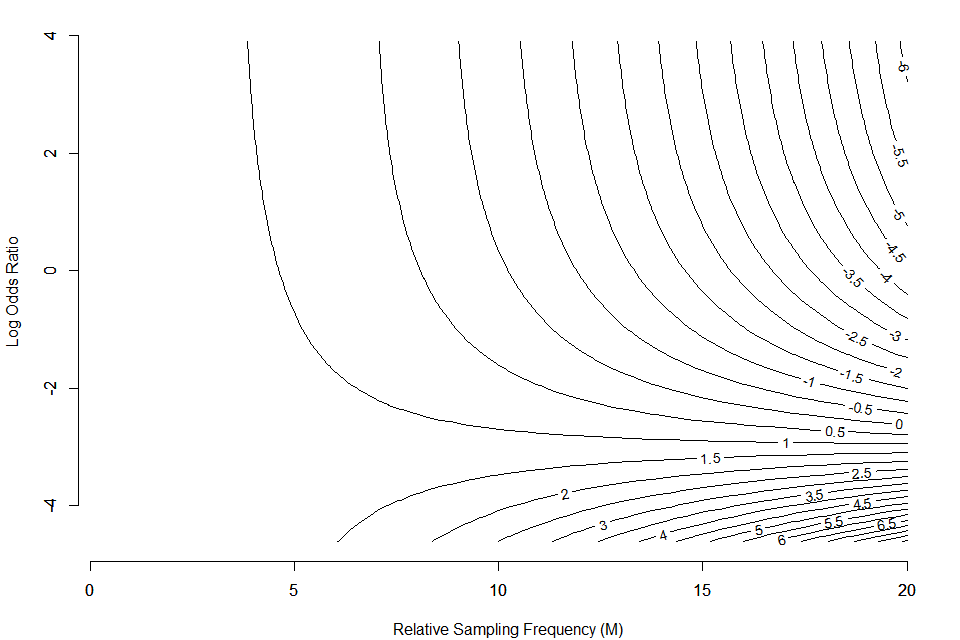}
 \caption{Imperfect testing adjustment~($D_M$) contour plot as a function of relative frequency $f_1/f_0$ (x-axis) and odds ratio (y-axis) for $FP=0.024$ and $FN=0.13$.}
 \label{fig:heatmap}
 \vspace{-0.3cm}
 \end{figure}

 Let $P_j$ be an indicator of measurement error, equal to $1$ when we incorrectly measure the binary outcome and $0$ otherwise. We suppose this is a stochastic variable where $\pr(P_j = 1 \mid Y_j = 1) =: FN$ is the false-negative rate and $\pr(P_j = 1 \mid Y_j = 0) =: FP$ is the false-positive rate.  If individual $j$ is selected (i.e., $I_j = 1$) then the observed outcome can be written as $Y_j^{\star} = Y_j(1-P_j) + (1-Y_j) P_j$.  The attentive data analyst will recognize the estimator $\bar y_n$ is now biased for simple random samples.  In Appendix~\ref{app:memestimator}, assuming sensitivity and specificity are known a priori, a novel iterative procedure is used to construct the estimator $\tilde y_n = (\bar y_n - FP)/(1-(FP+FN))$, which is unbiased under simple random sampling (SRS); see Appendix~\ref{app:modelbased} for a discussion of the connection to model-based estimators. In the language of the epidemiology literature, the estimator $\tilde y_n$ is the standard estimator for correcting for measurement error using false positive and negative rates that have been estimated from a validation sample, where $\bar y_n$ is the mis-measured average \& $\tilde y_n$ is the corrected average. To understand the impact of selection bias and imperfect testing, we derive the following statistical decomposition of the error between $\tilde y_n$ and $\bar Y$:
 \begin{equation}
 \label{eq:statdecomp}
 \rho_{I,Y} \times \sqrt{\frac{1-f}{f}} \times \sigma_{Y}
 \times \underbrace{\left[ 1 - \Delta \times \frac{\bar Y}{1-\bar Y} \times \frac{FP(1-\bar Y) + FN \cdot \bar Y}{f_0 (1-\bar Y) + f_1 \bar Y} \right] \times \frac{1}{1-(FP+FN)}}_{D_M},
 \end{equation}
 where $f=n/N$ is the sampling fraction, $f_1$ and $f_0$ are sampling fractions for COVID-19 positive and negative individuals respectively, and $\Delta = f_1 - f_0$ is the sampling rate differential.  See Appendix~\ref{app:memestimator} for the derivation. This extends work by \cite{Meng2018} to account for imperfect testing. The first three terms continue to represent \emph{data quality}, \emph{data quantity}, and \emph{problem difficulty} respectively.  The new term~$D_M$ represents the \emph{imperfect testing adjustment} which is a complex function of the sampling rate differential, the odds ratio, and the ratio of measurement error interaction with prevalence and sampling rates interaction with prevalence.
 For ease of comprehension, a notation glossary is provided in Section~\ref{app:notation} in the supplementary materials.
 \vspace{-0.25cm}
 \begin{remark}
 \label{rmk:testbinary}
 Note that test positivity is based on an underlying continuous cycle threshold and therefore not strictly binary.  We focus on the interplay of measurement error and selection bias in the context of binary outcomes due to the dichotomous nature of the COVID-19 testing data.  Also note that measurement error can lead to bias even in the absence of selection bias.
 \end{remark}
 \vspace{-0.25cm}
 Figure~\ref{fig:heatmap} shows that $D_M$, as a function of the relative frequency ($f_1/f_0$) and log odds ratio, can be both positive and negative as well as a range of magnitudes~\citep{Beesley2020,Beesley2019,Smeden2019}. Assuming no measurement error, $D_M = 1$ so the relation between estimation and selection bias is simple, e.g., if COVID-19 positive individuals were more likely to receive test then this implies upward bias in prevalence estimates. Under random testing (i.e., $f_0 = f_1$), $D_M = (1-FP-FN)^{-1}$ so measurement error simply magnifies this error. When tests are imperfect and selection bias exists, this simple relationship no longer holds.

 Comparing the mean-squared error (MSE) under a selection mechanism $\I$ with imperfect testing and SRS with perfect testing, we see that
 $$
 \frac{E_{\I} \left[ (\bar y_n - \bar Y)^2 \right]}{\sqrt{V_{SRS} (\bar Y)}} = (N-1) E_{\I} \left[ \rho_{I,Y}^2 D_M^2 \right].
 $$
 A key question is ``What is the (effective) sample size from a SRS with perfect testing that would yield equivalent MSE to the current testing strategy?'' In Appendix~\ref{app:effss}, we show the effective sample size $n_{eff}$ can be bounded by $\frac{f}{1-f} \times \frac{1}{E_{\I} \left[ \rho_{I,Y}^2 D_M^2 \right]}$. Between April 25th to 29th 2020, Indiana performed $95,879$ tests.  Indiana's population is roughly $6.732$ million, so $f = 0.003$.  The active infection rate was estimated to be~$1.81\%$~\citep{Yiannoutsos2021} in this time interval. Recent studies have suggested 87\% sensitivity~\citep{Arevalo2020} and 97.6\% specificity~\citep{Cohen2020} are reasonable measurement error rates for RT-PCR tests. Supposing COVID-19 positive individuals are $1.5$ times more likely to get tested, then the effective sample size is $168$. Recent proposals~\citep{Siddarth2020} have argued for increased testing capacity, which may likely reduce the relative sampling rate.  Even if the relative sampling rate drops to $1.2$ and $f$ increases to $0.01$ then the effective sample size will increase to $1025$.  Thus the effective sample size even in optimistic scenarios is equivalent to a moderate random sample from the population.  Moreover, increased testing capacity may alter false positive and negative rates due to changes in sample collection quality, e.g., a testing center switches from nasal swab to oropharyngeal swabs or saliva specimen to speed up data collection.  Consider the case where $f$ increasing from $0.003$ to $0.01$ and is associated with the false negative rate rising from $13\%$ to $20$\%.  Then the effective sample size is $863$, representing a $5.1$ factor increase rather than the expected $6.1$ factor increase. See Section~\ref{section:effss} in the Supplementary Materials for additional effective sample size calculations.  Note that while~\citep{Meng2018} argues the relative error increases as a function of population size, calculations in our setting indicate this is not true when relative frequency~$f_1/f_0$ is held fixed.


 \subsection{Regrettable rates: complex biases resulting from self-selection}
 \label{section:rates}

 The prior analysis demonstrates the potentially limited information regarding COVID-19 prevalence in observational case-count data.  Analysts may claim that daily observed case-counts simply undercount daily total cases by a constant multiple over time (i.e., undercounting).  If true then the ratio of case-counts at consecutive times may be a good estimate of the true change in prevalence, helping scientists understand the disease trajectory.  We next demonstrate how selection bias and imperfect testing impact such estimates.

 Let $\bar Y_{t-1}$ and $\bar Y_{t}$ denote the prevalence on two consecutive days and consider the estimator $r = \tilde y_t / \tilde y_{t-1}$.  Using a second-order Taylor series approximation, the error between ${\tilde y_t}/{\tilde y_{t-1}}$ and ${\bar Y_{t}}/{\bar Y_{t-1}}$ can be expressed approximately as
 $$
 \begin{aligned}
 \frac{\bar Y_t}{\bar Y_{t-1}} &\times \bigg[ \rho_{I_t,Y_t} D_{M_t} \sqrt{\frac{1-f_t}{f_t}} CV (Y_t)  -\rho_{I_{t-1},Y_{t-1}} D_{M_{t-1}} \sqrt{\frac{1-f_{t-1}}{f_{t-1}}} CV (Y_{t-1}) \bigg] \\
 &\times \left[ 1 - \rho_{I_{t-1},Y_{t-1}} D_{M_{t-1}} \sqrt{\frac{1-f_{t-1}}{f_{t-1}}} CV (Y_{t-1}) \right]
 \end{aligned}
 $$
 where $\rho_{I_j, Y_j}$ is the data quality, $f_j$ is the sampling fraction, $D_{M_j}$ is the measurement error adjustment, and $CV(Y_j) = \sigma_{Y_j}/ \bar Y_j$ is the coefficient of variation on day $j$.  See Appendix~\ref{app:ratio} for the derivation. The error magnitude depends on the true rate $\bar Y_{t} / \bar Y_{t-1}$ so a large decrease will have a small error relative to a large increase. The second term represents potential \emph{cancellation} which can  occur when data quality, sampling fraction, measurement error, and prevalence are constant across time.

\begin{figure}[!th]
 \centering
 \begin{subfigure}{.5\textwidth}
  \centering
  \includegraphics[width=.9\linewidth]{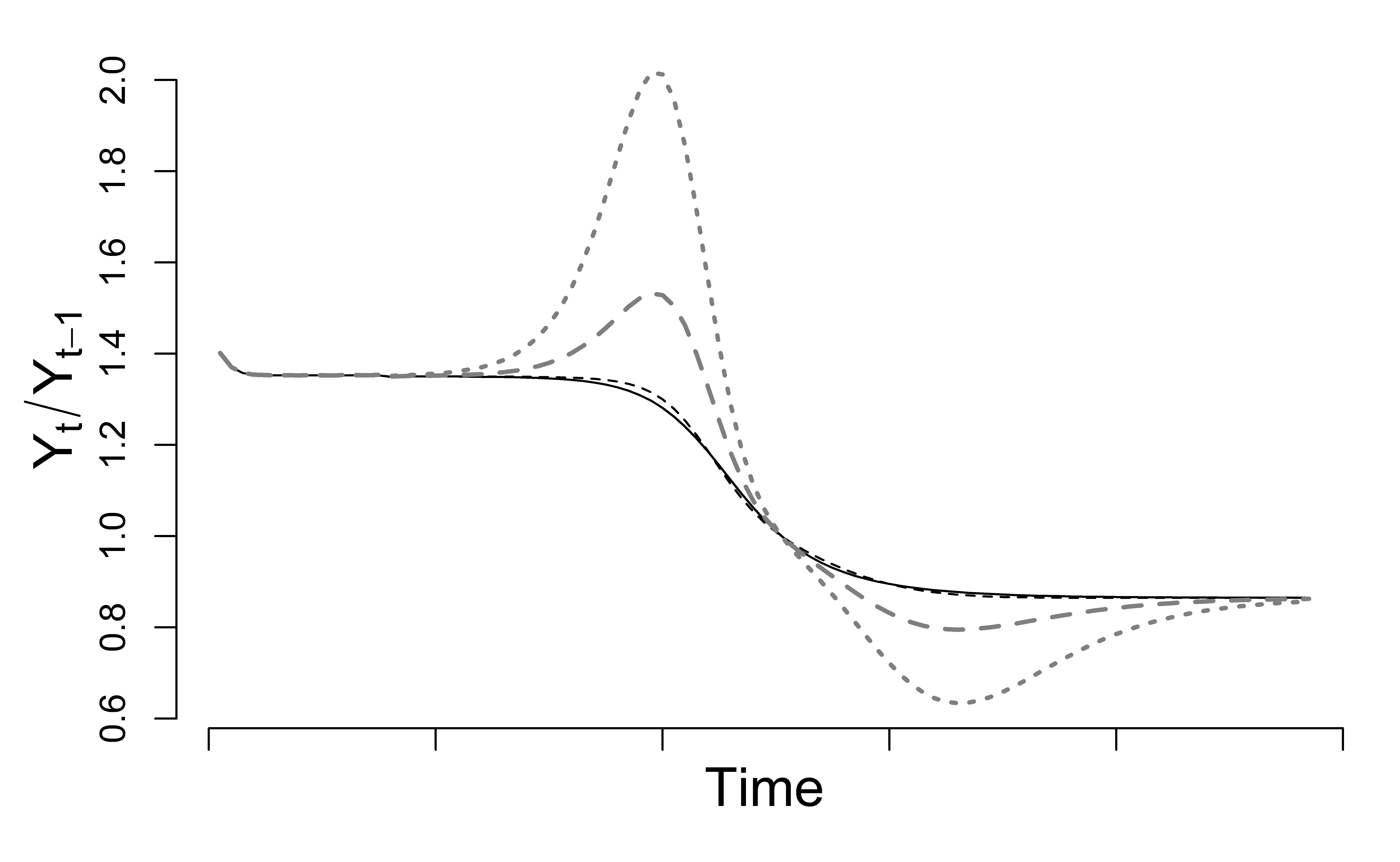}
  \caption{Ratio estimator}
  \label{fig:ratio-bias}
 \end{subfigure}%
 \begin{subfigure}{.5\textwidth}
  \centering
  \includegraphics[width=.9\linewidth]{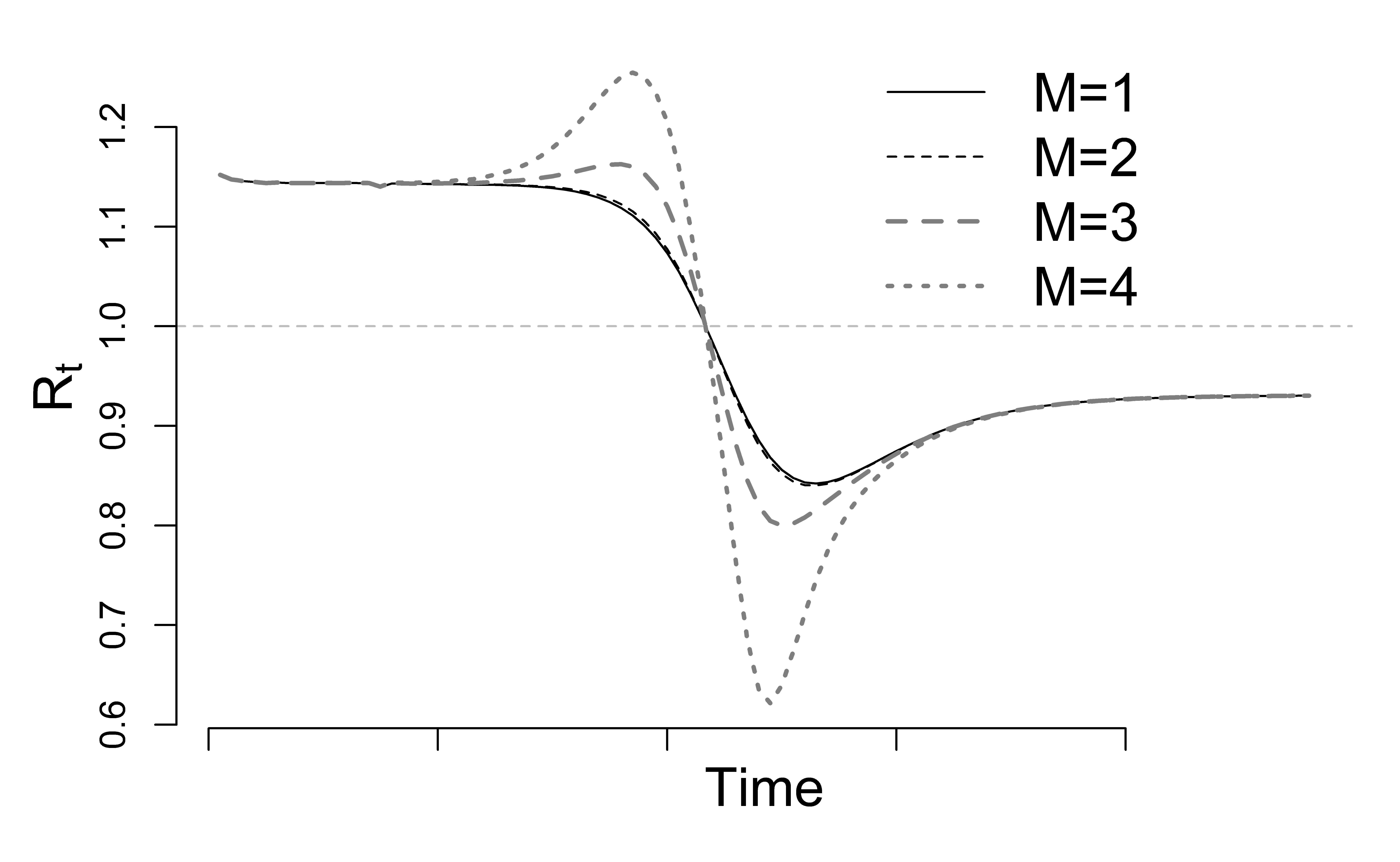}
  \caption{Effective reproductive rate estimator}
  \label{fig:r0-bias}
 \end{subfigure}
 \caption{Potential bias in ratio and effective reproductive rate estimators under an SEIR model with $\beta = 1.2$, $\gamma = 0.15$, and $\sigma = 0.3$.  Here, $f = 0.02$, $FP = 0.024$, $FN = 0.13$, and a range of relative sampling fractions $M = f_1/f_0$ are considered.}
 \label{fig:rates}
 \end{figure}

 Figure~\ref{fig:ratio-bias} displays the trajectory of the true ratio and the potential biased estimators under a susceptible-exposed-infected-recovered (SEIR) model~\citep{Pastor2001,Newman2002,Parshani2010} for the epidemic dynamics, with state evolution given by
 \begin{align}
 \frac{\partial s_t}{\partial t} &= - \beta s_t i_t; \quad \frac{\partial e_t}{\partial t} = \beta s_t i_t - \sigma e_t; \quad \label{eq:seir} \\
 \frac{\partial i_t}{\partial t} &= \sigma e_t - \gamma i_t; \quad
 \frac{\partial r_t}{\partial t} = - \gamma i_t. \nonumber
\end{align}
 where $s_t, e_t, i_t$ and $r_t$ are the fraction of susceptible, exposed, infected, and removed (recovered or deceased) individuals in the population at time $t$ respectively.  The SEIR model has been used extensively as a model for SARS-CoV-2 dynamics~\citep{Song2020}.  In terms of bias, the rate is overestimated prior to the peak in the fraction infected and underestimated afterwards; the bias increases dramatically when the relative fraction exceeds $2$.  Such biases may impact policy making.  Overestimation pre-peak may give policy makers more leverage in proposing aggressive actions to reduce prevalence.  Underestimation post-peak puts pressure on policy makers to prematurely relax social distancing measures.  Estimates at the peak time appear to have minimal bias.

 \subsection{Estimation of effective reproduction number}
 \label{section:r0-estimation}
 Many epidemiologists argue that tracking the effective reproduction number is the only way to manage through the crisis~\citep{Gabriel2020}.
Here, we study the instantaneous reproduction number~\citep{Cori20113,Fraser2007}, denoted $R_t$, which is the average number of secondary cases that each infected individual would infect if the conditions remained as they were at time~$t$.  This is distinct from the case reproduction number,~$R_t^c$, which is the average number of secondary cases that a case infected at time step $t$ will eventually infect~\citep{Wallinga2004}. The case reproduction number accounts for potential changes to contact rates and transmissibility, which include impact of control measures.  The instantaneous reproduction number~$R_t$ is the only reproduction number easily estimated in real time, and therefore has been a key focal point in the COVID-19 pandemic.

Under a Poisson likelihood, a simple relation between the trajectory of new cases and the instantaneous reproduction number can be derived \citep{Bettencourt2008}.  In particular, under an SIR model the number of case counts on day $t$, denoted $K_t$, is Poisson distributed with rate $K_{t-1} \exp \left( \gamma (R_t - 1) \right)$ where $K_{t-1} = Y_{t-1}-Y_{t-2}$ is the number of new cases on day $t-1$ and $\gamma$ is the serial interval, which is approximately $7$ days for COVID-19~\citep{Sanche2020}.

\cite{Heng2020} derive an approximate formula for the basic reproduction number under SEIR dynamics, which corresponds to the solution for the SIR model assuming an infectious period of $1/\gamma + 1/\sigma$ for $s_0 \approx 1$ and $i_0 \ll 1$. Using this connection, a moment-based estimator is given by
$$
R_t \approx 1 + \left( \frac{1}{\gamma} + \frac{1}{\sigma} \right) \log \left( \frac{K_t}{K_{t-1}} \right).
$$
This approximation ignores a term that is quadratic in the epidemic growth rate and inversely linear in $\sigma \times \gamma$; for realistic values of these quantities, however, the term is relatively negligible and therefore ignored. Of course, we do not observe $K_t$ and $K_{t-1}$.  Under SRS of new cases among those susceptible on day $t$, the natural estimator is $S_t \tilde y_t$.  Unfortunately the number of susceptible individuals on day $t$ is unknown. Here, we study the estimator $\hat R_t = 1 + \left( \frac{1}{\gamma} + \frac{1}{\sigma}\right) \log \left( \tilde y_t / \tilde y_{t-1} \right)$. We can again express the statistical error of $\hat R_t - R_t$ in useful terms as follows
 $$
 \begin{aligned}
 \left( \frac{1}{\gamma} + \frac{1}{\sigma} \right)\log &\bigg( 1 + \bigg[ \rho_{I_t,K_t} D_{M_t} \sqrt{\frac{1-f_t}{f_t}} CV (K_t)  -\rho_{I_{t-1},K_{t-1}} D_{M_{t-1}} \sqrt{\frac{1-f_{t-1}}{f_{t-1}}} CV (K_{t-1}) \bigg] \\
 &\times \left[ 1 - \rho_{I_{t-1},K_{t-1}} D_{M_{t-1}} \sqrt{\frac{1-f_{t-1}}{f_{t-1}}} CV (K_{t-1}) \right] \bigg) - \left( \frac{1}{\gamma} + \frac{1}{\sigma} \right) \log \left( \frac{S_t}{S_{t-1}} \right).
 \end{aligned}
 $$
 This implies a similar trade-off as before but on the logarithmic scale.  The error is no longer scaled by $\bar Y_t/\bar Y_{t-1}$ but by the serial interval and does not depend on prevalence but on the fraction of new cases out of those susceptible. This leads to differences in when the bias is most pronounced. Figure~\ref{fig:r0-bias} displays the bias as a function of the relative sampling fraction ignoring the final term (i.e., $S_t / S_{t-1} \approx 1$). Section~\ref{app:cori_rt} in the supplementary materials presents an alternative estimator of the instantaneous effective reproductive number and shows similar bias~\cite{Cori20113}.

 \subsection{Rate comparisons}

 \begin{figure}[!th]
 \centering
 \begin{subfigure}{.5\textwidth}
  \centering
  \includegraphics[width=.9\linewidth]{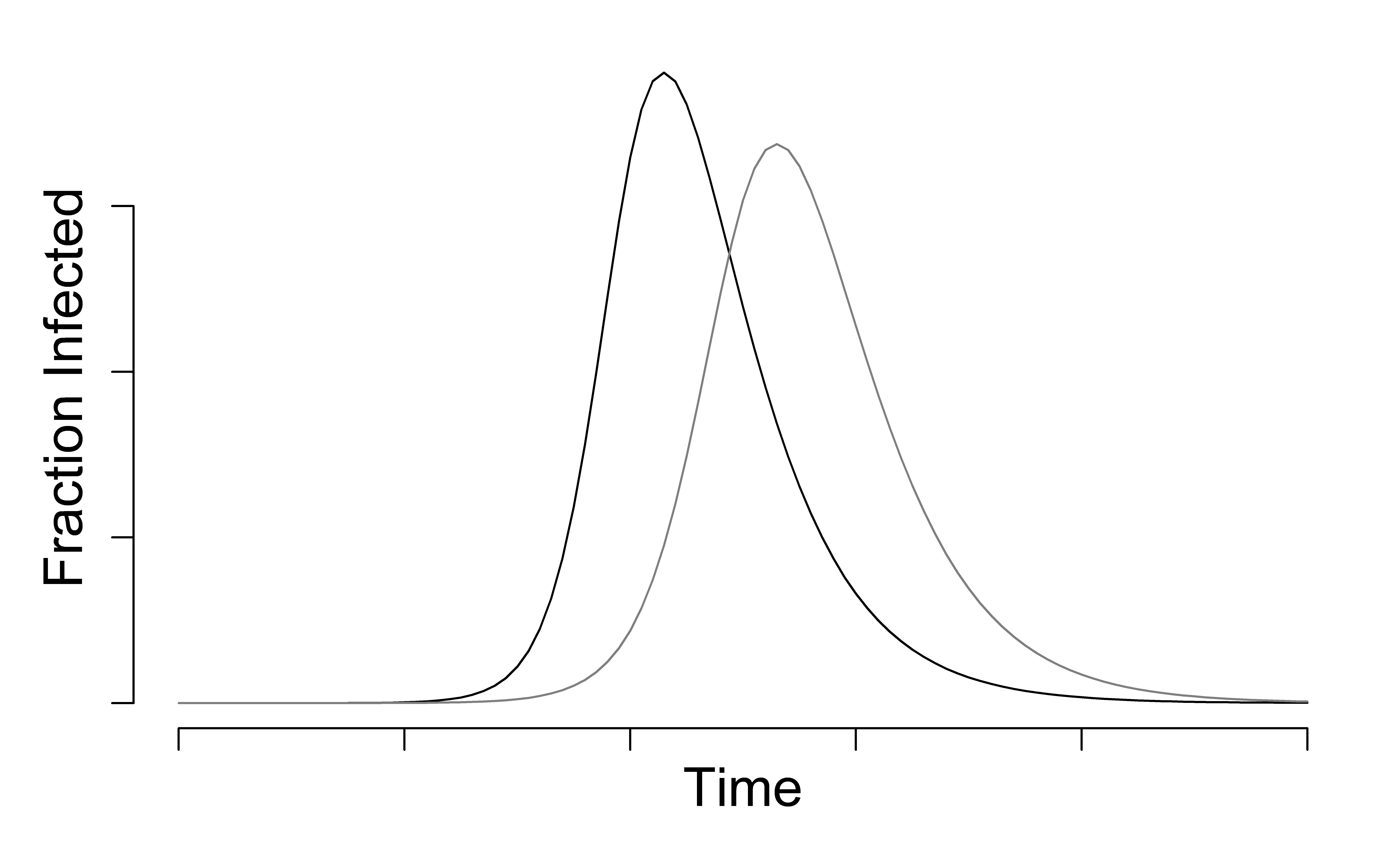}
  \caption{Fraction of new cases in population}
  \label{fig:fracpop}
 \end{subfigure}%
 \begin{subfigure}{.5\textwidth}
  \centering
  \includegraphics[width=.9\linewidth]{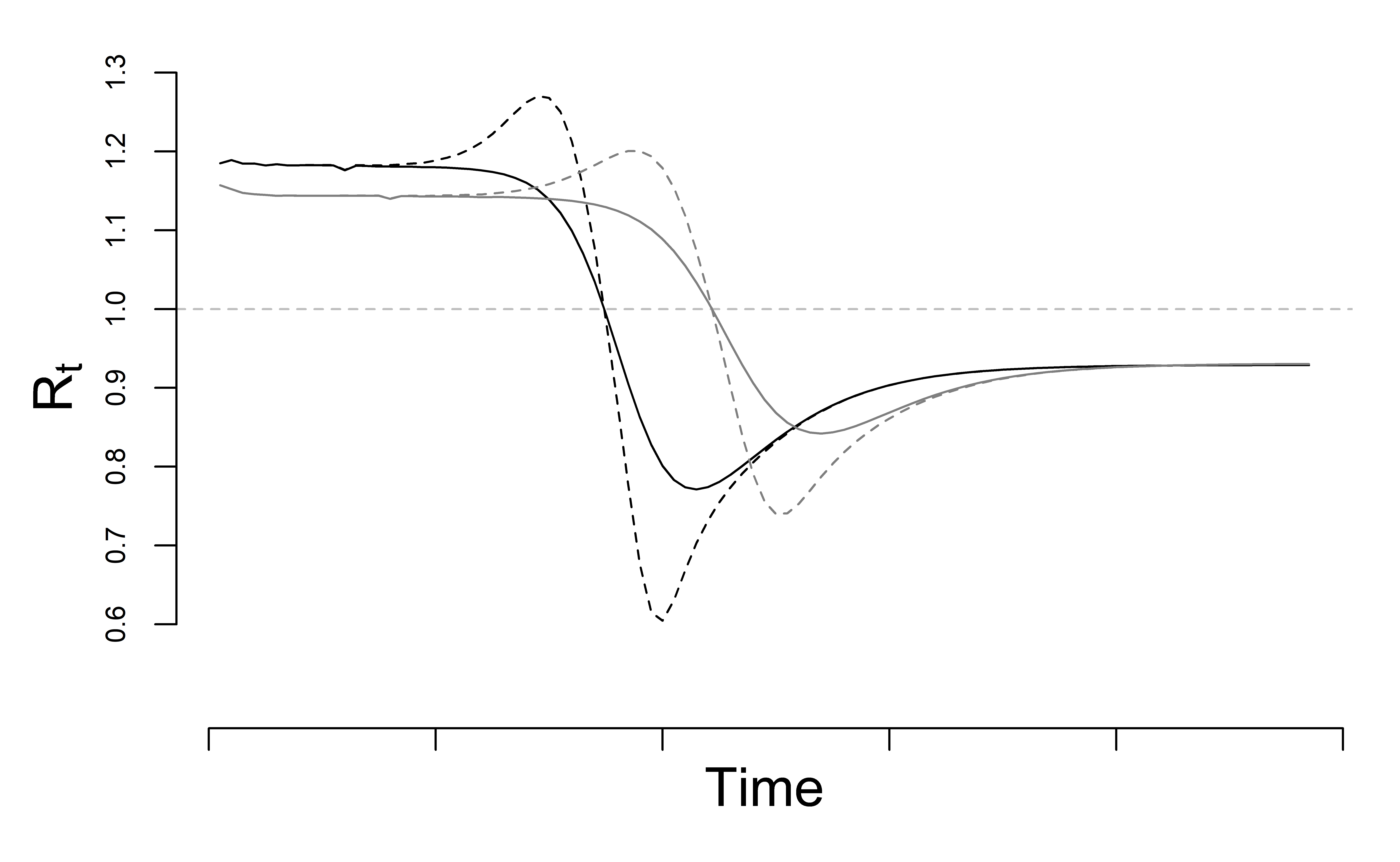}
  \caption{Effective reproduction rate estimators}
  \label{fig:eff}
 \end{subfigure}
 \caption{Left: fraction infected in two SEIR models with $\beta = 1.2$ and $0.9$ respectively, $\sigma = 0.3$, and $\gamma = 0.15$ with same initial conditions. Right: comparison of $\hat R_t$ across time with $FN = 0.30$, $FP = 0.024$, and $M = 4$.}
 \label{fig:comparison}
 \end{figure}

 So far we have focused on understanding the limitations of using case-count data to understand population quantities of interest for a \emph{single} population.  Many are interested in cross-population comparisons to contrast the impact of countries' mitigation policies.  Here, for simplicity, we focus on comparing the estimated effective reproductive rate.  We assume the two time-series are aligned so that $t=0$ is the time of first case in each population respectively.

 Consider two countries (A and B) in which the peak occurs 2 weeks prior for country A than country B.  Figure~\ref{fig:comparison} presents such a comparison where each country's disease trajectory follows an SEIR model (A=black and B=grey). Figure~\ref{fig:eff} shows how biases interact in complex ways.  At first, the difference is correctly estimated; then the gap is over-estimated as country A sees a rapid rise in cases; then the magnitude of over-estimation increases as country A sees declining case-count while country B sees rapidly increasing case-count; then country A's rate is correctly estimated while country B's rate is under-estimated as it sees declining case-count; finally, the gap disappears.  While this may not always be the case, the analysis demonstrates how estimates can tell a more complex story than the truth (i.e., country A's peak is 2 weeks prior to country B's peak).

 \section{Potential improvements to prevalence estimation}
 \label{section:improvedcasecount}

 The prior section presented negative consequences of selection-bias and measurement error when estimating infection prevalence, rates of change, and the effective reproduction number from observed case-count data.  In this section, we consider two directions to improve upon these estimators.
 The proposed methods and statistical error decompositions guide our recommendations in Section~\ref{section:discussion}.

 \subsection{Selection propensity estimation}

 With non-probability samples, bias can be reduced by modelling the self-selection propensity and using inverse probability weighting (IPW)~\citep{Elliott2017} to adjust for selection bias.
 In the current context, however, one does not observe those who are not tested.  To estimate selection propensities, auxiliary information is needed.

 To see this mathematically, let $X_j$ denote a vector of covariates for the $j$th individual in the population, $j=1,\ldots,N$.  For simplicity, the dynamic nature of the outbreak and recoverability of individuals is ignored for now. Let~$I_j^{NR}$ denote the selection indicator for individual~$j$ in the population into the non-probability sample.  Assume it is a Bernoulli random variable that depends only on these covariates, i.e., $P(I^{NR}_j = 1 \mid X_j = x) = \pi (x; \theta)$. Here, we focus on logistic regression models, i.e., $\pi(x; \theta) = \text{expit} \left( x^\top \theta \right)$. Maximum likelihood estimation follows by maximizing
 \begin{equation}
 \label{eq:propensity}
 \sum_{j=1}^N I_j^{NR} \log \left( \frac{\pi (X_j; \theta)}{1-\pi(X_j; \theta)} \right) + \sum_{j=1}^N \log \left( 1 - \pi (X_j; \theta) \right).
 \end{equation}
 The first sum only involves individuals observed in the nonprobability sample.
 The second sum is over the entire population.  Maximum likelihood estimation therefore requires knowledge of covariate information for every individual in the population.  Typically, this is not possible.  Here we present a method that uses auxiliary information obtained from probability samples.

 \subsubsection{Auxiliary information through probability samples}
 \label{subsec:auxprob}

 Here we assume access to a probability sample measuring the same set of covariates.  Let $I_j^{R}$ be an indicator that the individual was included in the probability sample and $W_j^R$ be the probability of inclusion for the $j$th individual. \cite{Chen2019} use a probability sample to construct a design-unbiased estimator of the second term
 \begin{equation}
 \label{eq:auxinfoprob}
 \sum_{j=1}^N I_j^{NR} \log \left( \frac{\pi (X_j; \theta)}{1-\pi(X_j; \theta)} \right)  + \sum_{i=1}^N I_j^R W_j^R \log ( 1 - \pi (X_j; \theta)).
 \end{equation}
 Expectation of~\ref{eq:auxinfoprob} with respect to the sampling design yields~\eqref{eq:propensity}. Solving~\eqref{eq:auxinfoprob} is done by iteratively re-weighted least squares (IRLS); see Section~\ref{app:irls} and Section~\ref{app:notation} in the supplementary materials for details and a notation glossary respectively.
 \vspace{-0.5cm}
 \begin{remark}
 Weights built from selection propensities~$\{\pi(x_i; \theta)\}_{i=1}^n$ are common practice in the epidemiology literature.  Survey sampling and transportability weights require knowledge of the selection mechanism~\citep{10.1093/aje/kwx164,ColeStuart2010}.  Here auxiliary information is required to estimate the selection propensities.
 \end{remark}
 \vspace{-0.5cm}

 \subsubsection{An IPW estimator and statistical error decomposition}
 \label{section:IPWerrordecomp}

 Given a selection propensity, define the inverse probability weight~$w^{NR}(x) = \pi (x; \hat \theta)^{-1}$. Then the IPW estimator adjusted for measurement error is given by
 \begin{equation}
 \label{eq:ipwest}
 \bar y_n^\star
 = \frac{1}{1-FP-FN} \cdot \frac{\sum_{i=1}^n w^{NR} (x_i) (y_i - FP)}{\sum_{i=1}^n w^{NR} (x_i)}
 \stackrel{(2)}{=} \frac{1}{1-FP-FN} \sum_{k=1}^K \frac{d_k w_k}{w} (\bar y_k - FP),
 \end{equation}
 where equality (2) is under the assumption that $x_i$ is a stratification variable with $k$ indexing the strata, $w_k$ is the weight and $d_k$ is the number of samples in strata $k$, and $w = \sum_{k=1}^K d_k w_k$.

 Let $I_j^{NR} (X_j) = I_j^{NR}  \cdot w^{NR}(X_j)$ for $j=1,\ldots,N$.  Then the error when comparing weighted estimator~$\bar y_n^\star$ to the true prevalence $\bar Y$ can be expressed as:
 \begin{equation}
 \label{eq:statdecomp2}
 \rho_{I^{NR} (X), Y} \times \sqrt{\frac{1-f+ CV^2_W}{f}} \times \sigma_{Y} \times \underbrace{\left[ 1 - \tilde \Delta \times \frac{\bar Y}{1-\bar Y} \times \frac{FP(1-\bar Y) + FN \cdot \bar Y}{\tilde f_0 (1-\bar Y) + \tilde f_1 \bar Y} \right] \times \frac{1}{1-(FP+FN)}}_{\tilde D_M}
 \end{equation}
 where $CV_W$ is the coefficient of variation (i.e., standard deviation/mean) of $w^{NR} (X_J)$ given $I_J = 1$, $\rho_{I^{NR} (X), Y}$ is the empirical correlation which here depends on covariate distribution, $\tilde f_k = E[ w^{NR} (X_J) I_J \mid Y_J = k]$ for $k=0,1$, and $\tilde \Delta = \tilde f_1 - \tilde f_0$.  See Appendix~\ref{app:ipwderivation} for the derivation.

 Comparing~\eqref{eq:statdecomp2} to~\eqref{eq:statdecomp} shows that weighting impacts the estimation error in three ways.  First, there is a negative impact on the data quantity component; taking the ratio of these quantities yields
 $\sqrt{1 + \frac{CV_W^2}{1-f}} \geq 1$.  Hence, if the data quality does not increase (i.e., $| \rho_{I^{NR} (X), Y} | = | \rho_{I,Y}|$ ) then weighting increases the error magnitude.  Second, the relative error increase depends on the fraction of population sampled $f$, implying that for large samples there is a larger potential increase in the error if the weights do not improve data quality. Third, the impact of measurement-error on data quality is changed when considering a weighted estimand. In particular, weighting may result in $\text{sgn}(\Delta) \neq \text{sgn} (\tilde \Delta)$ which implies the impact of measurement-error may be in a different direction.

 As demonstrated below in Lemma~\ref{lemma:ipw}, if the propensity model is correctly specified then the $E [ \rho_{I^{NR} (X), Y} ] = 0$ and therefore $\rho_{I^{NR} (X), Y} = O(N^{-1})$; however, if the weights are not correctly specified then the data quality index is unlikely to inversely scale with population size.
 Similar to~\cite{Meng2018}, if the data quality is not at the level of $N^{-1}$, then confidence intervals constructed from an IPW estimator are likely to put too much confidence in the sheer data size.

 \subsubsection{Time-varying propensities}

 Here we extend the IPW approach to the temporal setting to account for the dynamic nature of the outbreak by considering the joint likelihood
 \begin{equation}
 \label{eq:tvpropensity}
 \sum_{t=1}^T \left[ \sum_{j=1}^N I^{NR}_{j,t} \log \left( \frac{\pi_t (X_{j,t}; \theta)}{1-\pi_t(X_{j,t}; \theta)} \right) + \sum_{j=1}^N \log \left( 1 - \pi_t (X_{j,t}; \theta) \right) \right]
 \end{equation}
 where $t=1,\ldots,T$ are the days when case-count data is reported, and $I^{NR}_{j,t}$ denotes self-selection into testing on day $t$, which is highly correlated with prior testing and results.  For example, an individual who tests positive may be unlikely to seek testing in the subsequent few days/weeks.  Moreover, an individual in a high prevalence area may be more likely to seek out testing.  Here, we assume that the covariate vector $X_{j,t}$ contains all features of the past relevant for selection.

 If sufficiently large random samples are collected at each time $t =1,\ldots,T$, then the pseudo-likelihood can be re-written as in~\eqref{eq:auxinfoprob} and propensities estimated separately per time point.  Unfortunately, large probabilistic samples are not available at every time within a given region.  To address this, here we consider a non-parametric kernel-based approach where the selection propensity at time~$t$, denoted $\hat \theta_t$, maximizes the smoothed pseudo-likelihood
 $$
 \sum_{t^\prime=1}^T K_h(|t^\prime - t|) \left[ \sum_{j=1}^N I_{j,t^\prime}^{NR} \log \left( \frac{\pi_t (X_{j,t^\prime}; \theta)}{1-\pi_t(X_{j,t^\prime}; \theta)} \right) + \sum_{j=1}^N W^{R}_{j,t^\prime} I^R_{j,t^\prime} \log \left( 1 - \pi_t (X_{j,t^\prime}; \theta) \right) \right]
 $$
 where $K_h$ is a kernel function with tunable parameter $h$. Given $\pi (x;\hat \theta_t)$, the prevalence estimator $\bar y_{n,t}^\star$ is given by~\ref{eq:ipwest} using case-count data observed on day~$t$.

 \subsubsection{Asymptotics}
 \label{section:asymptotics}

 Here we suppose there is a sequence of finite populations of size $N_{\nu}$ indexed by $\nu$ and that there are $L_\nu$ non-probability samples and probability samples of size $n$ drawn at equally spaced times $\{ t^\prime_l \}_{l=1}^{L_{\nu}}$ over the study window~$[0,T]$. Assuming correct selection model specification, then under the probability sample design and nonprobability sample propensities, Lemma~\ref{lemma:ipw} shows that the IPW estimator at a time~$t \in \{ t^\prime_l \}$ is consistent as $\nu$ goes to infinity (i.e., $N_{\nu}, L_{\nu} \to \infty$) and calculates the estimator's variance.  In Lemma~\ref{lemma:ipw}, sensitivity/specificity are unknown and estimated using a pseudo-likelihood on two random samples whose sizes (denoted~$n_{FP}$ and $n_{FN}$) reflect uncertainty in these quantities.  See Appendix~\ref{app:asympderivations} for additional details.

 \begin{lemma}[Variance of IPW estimator] \normalfont
 \label{lemma:ipw}
The estimates~$\hat \mu_t := \bar y_{n,t}^\star$, $\hat \pi_{j,t}$, $\hat{FP}$, and $\hat{FN}$ at a time~$t \in \{t_l^\prime\}$ are solutions to the following set of estimating equations
 $$
 \Phi_n (\eta_t) =
 \left (
 \begin{array}{c}
 \frac{1}{N} \sum_{j=1}^N I^{NR}_{j,t} \frac{Y_{j,t} - FP - (1-FP-FN) \cdot \mu_t}{\pi_{j,t}} \\
 \frac{1}{N \times \sum_{l = 1}^L K_h \left(|t - t_l^\prime| \right)} \sum_{l = 1}^L K_h \left(|t - t_l^\prime| \right) \left[ \sum_{j=1}^N I^{NR}_{j,t_l^\prime} X_{j,t_l^\prime} - \sum_{j = 1}^N I^{R}_{j,t_l^\prime} W^{R}_{j,t_l^\prime}  \pi_{j,t_l^\prime} X_{j,t_l^\prime}  \right] \\
 \frac{1}{n_{FP}} \sum_{i=1}^{n_{FP}} \frac{Z_j}{FP} - \frac{1-Z_j}{1-FP} \\
 \frac{1}{n_{FN}} \sum_{i=1}^{n_{FN}} \frac{\tilde Z_j}{FN} - \frac{1-\tilde Z_j}{1-FN} \\
 \end{array}
 \right ) = {\bf 0}
 $$
 where $\eta_t = (\mu_t, \pi_t, FP, FN)$. Under certain regularity assumptions (see Appendix~\ref{app:asympderivations}), we have $\bar y_{n,t}^\star - \bar Y_{t} = O_p (\bar n^{-1/2})$  and $\text{var} (\hat y_{n,t}) = V_{t}^{(IPW)} + o (\bar n^{-1})$ where $\bar n = n \sum_{l=1}^L K_{h} (|t - t_l^\prime|)$ and $V_t^{(IPW)}$ is the first diagonal element of $E [\phi_n(\eta_0)]^{-1} \text{Var}(\phi_n(\eta_0))E [\phi_n(\eta_0)]^{-1}$ where $\phi_n = \frac{\partial \Phi (\eta)}{\partial \eta}$ with $E[\cdot]$ and $\text{Var} (\cdot)$ are under the joint randomization of propensity score and sampling designs.
 \end{lemma}

 \subsection{Model-based estimation}
 \label{section:modelbased}

 Up to this point, the primary focus has been selection bias in coronavirus case-count data from a survey sampling perspective.  Here, we consider compartmental model approaches from infectious disease epidemiology.  Our primary objective is a model-based forecast of strata-level active infection rate~$\bar y_k$. To do this, a probabilistic extension of a standard epidemiological state-space model -- the susceptible, exposed, infected, and removed (recovered and death) model, or SEIR model -- is presented. A probabilistic SIR model was originally proposed by~\cite{Osthus2017} with only one-dimensional time series of infected proportions; this formulation was extended by~\cite{Song2020} to model coronavirus case-counts.

 Let $s_t$, $e_t$, $i_t$, and $r_t$ denote the proportion of survivors, exposed, infected, and removed cases (i.e., including both recovered cases and deaths) at time $t$.  The population-level SEIR dynamics are given by the set of differential equations in~\eqref{eq:seir}.
Here, we consider covariate information that takes the form of a stratification variable with $K$ strata.  Based on these population-level dynamics, we can compute the number of \emph{new infections} at time~$t$, i.e., $I^\new_{t} := - N \cdot (e_{t+1} - e_{t} + s_{t+1} - s_t)$ to denote the number of new infections on day $t$ in the $k$th strata.  The joint distribution of strata-specific new infections~$ \{ I^{\new}_{t,k} \}_{k=1}^K$ follows a multinomial distribution with
 $$
 ( I^\new_{1,t}, \ldots, I^\new_{K,t} ) \sim \text{Multinomial} \left( I^\new_t, (p_{1,t}, \ldots, p_{K, t}) \right).
 $$
 where $( {\bf p}_t )_{t=1}^T = ( (p_{1,t}, \ldots, p_{K, t}) )_{t=1}^T$ are a sequence of parameters on the simplex, i.e., $\sum_{k=1}^K p_{k,t} = 1$.

 Selection bias is addressed by analyzing COVID-19 death data rather than case counts.  Let $D_{k,r}$ denote the number of individuals who pass away on day $r$ from strata $k$.  Then
 $$
 D_{k,r} \mid p, \theta, \nu \sim \text{Poisson} \left( \sum_{t=1}^r \text{IFR}_{k} \cdot I^\new_{k,t} \theta_{(r-t)} \right)
 $$
 where $\text{IFR}_{k}$ is the infection fatality rate for the $k$th strata and $\theta_{j}$ is a discrete-time distribution for time from infection to death.  This extends prior analysis of death data~\cite{Johndrow2020} by allowing the infection fatality rate to depend on a stratification variable, which is important as IFR depends heavily on age~\citep{Levin2020}. See Section~\ref{section:application_of_modelbased} for discussion of parameter choices.  

 %

 \subsection{Doubly robust estimation.}
 Rather than relying solely on IPW or epidemiological forecasts, here we combine forecasting and inverse-probability weighting by extending recent work by~\cite{Chen2019} to account time-varying propensities and measurement-error.
 Let~$\hat \mu(x)$ denote the posterior mean of the active infection rate individuals with covariate value~$x$ based on the SEIR model described in Section~\ref{section:modelbased}. In this context, the doubly-robust estimator is given by
 $$
 \bar y_{n}^{(DR)} = \frac{1}{N} \sum_{j=1}^N \hat \mu (x_j) + \frac{1}{\sum_{j=1}^N I_j^{NR} w (x_j)} \sum_{j=1}^N I_j^{NR} w(x_j) \left( \frac{Y_j - FP}{1 - FP - FN} - \hat \mu(x_j) \right).
 $$
 where $\hat \mu(x_j)$ is not corrected for measurement-error as it estimates true active infections. This estimator is called ``doubly-robust'' because it is consistent if either the model-based forecasts or the time-varying propensities are correctly specified. A statistical error decomposition can be derived
 \begin{align}
 \label{eq:statdecomp3}
 \begin{split}
 \rho_{I^{NR} (X), Y-\mu(X)} &\times \sqrt{\frac{1-f+ CV^2_W}{f}} \times \sigma_{Y-\mu(X)}  \\
 &\times \underbrace{\left[ 1 - \frac{\rho_{I^{NR}(X),Y} \sigma_{Y}}{\rho_{I^{NR} (X),Y-\mu(X)} \sigma_{Y-\mu(X)}} \times \tilde \Delta \times \frac{\bar Y}{1-\bar Y} \times \frac{FP(1-\bar Y) + FN \cdot \bar Y}{\tilde f_0 (1-\bar Y) + \tilde f_1 \bar Y} \right]}_{\tilde D_M}.
 \end{split}
 \end{align}
 See Appendix~\ref{app:DRderivation} for the derivation.
 If the model is adequate, then one may expect a reduction in the problem difficulty and (potentially) in the data quality components.
 Interestingly, the impact of measurement-error on data quality now depends on a relative comparison of the data quality and problem difficulty of the weighted estimator and the doubly-robust estimator.

 Unfortunately, the first term of the doubly robust estimator cannot be computed as covariate information is not collected on the entire population.  Here, we suppose the same asymptotic regime as in Section~\ref{section:asymptotics} and use the probability sample at the time~$t \in \{t_l^\prime\}$ to estimate this term:
 $$
 \frac{1}{\sum_{j=1}^N I^{R}_{j,t} W^R_{j,t}} \sum_{j=1}^N  I_{j,t}^{R} W_{j,t}^{R} \hat \mu_{t} (X_{j,t}) + \frac{1}{\sum_{j=1}^N I_{j,t}^{NR} w (X_{j,t})} \sum_{j=1}^N I_{j,t}^{NR} w(X_{j,t}) \left( \frac{Y_{j,t} - FP}{1 - FP - FN} - \hat \mu_t(X_{j,t}) \right).
 $$
 Assuming correct selection propensity model specification, then under the probability sample design and nonprobability sample design, Lemma~\ref{lemma:dr} shows that the doubly robust estimator is consistent and calculates the estimator's variance. See Appendix~\ref{app:asympderivations} for additional details.

 \begin{lemma}[Variance of doubly-robust estimator] \normalfont
 \label{lemma:dr}
 The estimates~$\mu_t := \bar y_{n,t}^\star$ and $\hat \pi_{j,t}$
 are solutions to the following set of estimating equations
 $$
 \Phi_n (\eta_t) =
 \left (
 \begin{array}{c}
 \frac{1}{N} \left[ \sum_{j=1}^N \frac{I^{NR}_{j,t}}{\pi_{j,t}} \left( \left( \frac{Y_{j,t} - FP}{1-FP-FN} - \hat \mu_{j,t}  \right)  - \mu_t \right) + \frac{\sum_{l=1}^L K_{h} (|t - t_l^\prime|) \sum_{j=1}^N I^{R}_{j,t_l^\prime} W^{R}_{j,t_l^\prime} \hat \mu_{j,t_l^\prime}}{\sum_{l=1}^L K_{h} (|t - t_l^\prime|) \sum_{j=1}^N I^{R}_{j,t_l^\prime} W^{R}_{j,t_l^\prime}} \right] \\
 \frac{1}{N \times \sum_{l=1}^L K_{h} (|t - t_l^\prime|)} \sum_{l=1}^L K_{h} (|t - t_l^\prime|) \sum_{j=1}^N I^{NR}_{j,t^\prime} X_{j,t^\prime} - \frac{1}{N} \sum_{j = 1}^N I^{R}_{j,t^\prime} W^{R}_{j,t^\prime}  \pi_{j,t^\prime} X_{j,t^\prime}  \\
 \frac{1}{n_{FP}} \sum_{j=1}^{n_{FP}} \frac{Z_j}{FP} - \frac{1-Z_j}{1-FP} \\
 \frac{1}{n_{FN}} \sum_{j=1}^{n_{FN}} \frac{\tilde Z_j}{FN} - \frac{1-\tilde Z_j}{1-FN} \\
 \end{array}
 \right ) = {\bf 0}
 $$
 where $\eta_t = (\mu_t, \pi_t, FP, FN)$. Under regularity assumptions (see Appendix~\ref{app:asympderivations}), we have $\bar y_{n,t}^\star - \bar Y_{t} = O_p (\bar n^{-1/2})$ and $\text{var} (\hat y_{n,t}) = V_{t}^{(DR)} + o (\bar n^{-1})$ where $\bar n = n \sum_{l=1}^L K_h(|t-t_l|)$ and $V_t^{(DR)}$ is the first diagonal element of $E [\phi_n(\eta_0)]^{-1} \text{Var}(\Phi_n(\eta_0))E [\phi_n(\eta_0)]^{-1}$ where $\phi_n = \frac{\partial \Phi (\eta)}{\partial \eta}$ with $E[\cdot]$ and $\text{Var} (\cdot)$ are under the joint randomization of propensity score and sampling designs.
 \end{lemma}

 \section{COVID-19 active infection prevalence in Indiana}
 \label{section:applications}


 We next consider estimation of the active infection rate in Indiana using unweighted, IPW, model-based, and doubly robust estimates. To start, we recap the data sources used in estimation:
 \begin{itemize}[leftmargin=*]
 \item {\bf Testing data}:  The number of daily tests performed and number of daily positive tests are reported.  These counts are broken out jointly by age, gender, and racial demographic information.  Figure~\ref{fig:in-cases} and~\ref{fig:in-tests} plots daily COVID-19 positive cases and tests by age strata.

 \item {\bf Random/Nonrandom statewide sample}: Table~\ref{tab:indiana} summarizes data from a random sample from April 25--29th as well as a nonrandom sample obtained between May 2--3 in racial/ethnic minority communities~\citep{Yiannoutsos2021}. The nonrandom sample is not a random subsample of the overall case-counts; however, this data provides important supplementary covariate information.
 To account for the nonrandom sample being from high risk areas, we adjust estimates to match the statewide positivity rate of 11.7\% on August 30th. These adjusted rates are presented in parentheses in Table~\ref{tab:indiana} along with the relevant covariate information.

 \item {\bf Delphi's COVID-19 Trends and Impact Survey}: From the daily symptom surveys completed as part of Delphi's CTIS (see Section~\ref{subsection:fbsymptom} for details), we extract survey responses for individuals who identify as living in Indiana.  We collect age, sex, and demographic information as well as COVID-19 related symptoms, which Delphi defines as having (1) a fever and cough, (2) shortness of breath, \emph{or} (3) difficulty breathing. Table~\ref{tab:comparison} in Section~\ref{app:in_add_details} of the supplementary materials shows minimal bias with respect to symptom distributions when comparing to Indiana's random sample to the CTIS data collected from April 25th to April 29th.



 \item {\bf Death data}: Daily COVID-19 related deaths are observed by age, gender, and racial demographics.  Figure~\ref{fig:in-deaths} plots COVID-19 related deaths over time by age group. While most tests and positive cases are within the younger age strata, most deaths are within the 70+ age strata.
 \end{itemize}

\begin{remark}
\label{rmk:limitations}
 We acknowledge several limitations with respect to representativeness of these samples.  First, the random sample had significant missingness ($3,658/15,495 \approx 24\%$ response rate). See~\cite{Yiannoutsos2021} for how the scientific team handled missing data in prevalence estimation. Second, the hit rate on randomly selected Facebook users is likely low and Facebook's Indiana user population may differ from the Indiana population of interest.  To address these concerns, the Delphi team provided respondent weights calculated by Facebook.  See~\cite{Barkay2020} for a description of the weights and corresponding methodology.  Further adjustments are beyond the scope of this paper but are considered important future work. We urge health policy experts and government officials to assess these issues when applying this framework in their own work.
 \end{remark}

 \subsection{Inverse-probability weighting approach}
 \label{section:ipwapproach}


 We start by using Table~\ref{tab:indiana} to compute IPW weights for end of April, early May.  The random sample is $n=3,658$. IPW weights are computed per strata using Indiana Census and random survey data, and are allowed to depend on Gender, Age, Race, Fever, Positive Case in Household, and Prior Positive Test. The weight for the strata defined as Male, 40--59, White, has a Fever, no household cases, and no prior tests, for example, is proportional to:
 $$
 \frac{\left( 0.493 \times 0.252 \times 0.869 \times 0.018 \times 0.986 \times 0.986 \right)}{\left(0.418 \times 0.411 \times 0.231 \times 0.17 \times 0.892 \times 0.939 \right)} \approx 0.334
 $$
 Using the constructed weights and the adjusted strata-level prevalence estimates, the IPW estimate under no measurement-error is $7.7\%$, a decrease of four percentage points from the observed prevalence of $11.7\%$. We also make use of CTIS sample over that window of time to construct a second IPW estimate conditional on demographics and fever but ignoring the household and prior testing information, which under no measurement-error is $6.5$\%.

 The Indiana study did not report sensitivity and specificity; therefore, we take the suggested measures from~\cite{Arevalo2020} which report 87\% sensitivity is a reasonable estimates and~\cite{Cohen2020} which report 97.6\% specificity.  This corresponds to a false negative rate of $13$\% and false positive rate of $2.4\%$, resulting in estimates of $6.2\%$ and $4.9\%$ using the random and CTIS samples respectively.

 \subsubsection{Disease prevalence by April 2020}

 Indiana's population as of 2019 was $6.732$ million.  A total of $19,649$ tests were administered in Indiana between April 25th to 29th. Subtracting off the $95,879$ tests that had already performed yields a sampling fraction of $f = 2.96 \times 10^{-3}$. Here we consider estimation of data quality $E_{\I} \left[ \rho_{I^{NR} (X),Y} \right]$. In \cite{Meng2018}, estimation relied on observing the true outcome (i.e., election vote totals). Here, we rely on~\cite{Yiannoutsos2021} who use the random sample to estimate the true prevalence of active COVID-19 disease at 1.81\% after accounting for non-response and measurement error.  Our goal is to build sensitivity analyses to understand the amount of information in observational case-count data.

 The unweighted estimate for the true prevalence of active COVID-19 disease between April 25th and April 29th is 11.7\%.  Assuming a false negative rate of $13$\% and false positive rate of $2.4$\%, the unweighted estimate is $11.0$\%, leading to an error of $9.2$\%.  Using~\eqref{eq:statdecomp}, an estimate of the relative sampling rate is
 \begin{equation*}
 \rho D_M = \sqrt{\frac{f}{1-f}} \frac{\text{0.092}}{\sigma_Y} = 3.75 \times 10^{-2} \Rightarrow \Delta = 1.39 \times 10^{-2} \Rightarrow M = 6.1.
 \end{equation*}
 The IPW estimates for the prevalence of active COVID-19 disease between April 25th and April 29th adjusted for measurement-error are $6.2$\% and $4.9$\%, leading to errors $4.4$\% and $3.1$\% respectively.  Using~\eqref{eq:statdecomp2}, an estimate of the relative sampling rate is
 \begin{equation*}
 \tilde \rho D_M = \sqrt{\frac{f}{1-f+CV_W^2}} \frac{\text{0.044}}{\sigma_Y} = 8.14 \times 10^{-3} \Rightarrow \tilde \Delta = 8.22 \times 10^{-3}  \Rightarrow \tilde M = 3.9,
 \end{equation*}
 using the Indiana non-random sample.  For the IPW estimate using CTIS data, $\tilde M = 3.3$. Therefore, bias is high using unweighted data and remains moderate using the weighted estimates.

 A sensitivity analysis can be performed by considering the range of false negative/positive rates.  Here, sensitivity ranges between $81$\% and $91$\% and specificity between $95$\% and $98.8$\%~\cite{Katz2020}. This leads to a range for $M$ of $(5.60, 6.44)$ for the unweighted analysis, and $(3.63,4.07)$ and $(3.10, 3.45)$ for the weighted analyses.  Note that in calculating the sampling fraction~$f$, we assumed individuals who tested recently are unlikely to test again in this time window. If we instead do not subtract off these tests, the sampling fraction is $f = 2.92 \times 10^{-3}$ and the above calculations change by a negligible amount.

 \subsubsection{Time-varying IPW estimator}
 \label{section:tvipw}

 The IPW analysis is next extended to the time-varying setting.  To do so, we make use of the CTIS, hospitalization, and testing data.  Recall the COVID-19 testing and positive case counts are known by age, gender, and racial demographics. Symptom status (e.g., fever, coughing, shortness of breath) and related covariate information (COVID-19 positive contact), however, are currently unavailable, which is likely due to data privacy considerations.  As COVID-19 symptom status is likely to alter the testing propensity and likelihood of active infection, for illustrative purposes CTIS and hospitalization data are used in two ways to impute symptom status per strata. We outline the imputation methods below; see Section~\ref{section:discussion} for additional discussion.

 In the first approach, we estimate two logistic regressions using weighted pseudo-likelihoods with the CTIS data.  The first estimates the probability of contact with a COVID-19 positive individual given demographic and COVID-19 test status (positive or negative COVID-19 test). Figure~\ref{fig:contactlik1} and~\ref{fig:contactlik2} in Section~\ref{app:in_add_details} in the supplementary materials presents the estimated likelihood of contact with a COVID-19 positive individual given a negative and positive COVID-19 test in the past 24 hours respectively.  The second estimates the probability of fever given demographic, COVID-19 test status, \emph{and} whether the individual has had contact with a COVID-19 positive individual.  Figure~\ref{fig:symptomlik1} and~\ref{fig:symptomlik2} in Section~\ref{app:in_add_details} in the supplementary materials presents the estimated likelihood of reporting a fever with a COVID-19 positive individual given a negative and positive COVID-19 test in the past 24 hours respectively.  We see that likelihood of fever varies greatly based on whether the individual also reported contact with a COVID-19 positive individual. Using these two models, mean imputation is used to calculate the number of positive and negative tests given demographic strata, fever status, and COVID-19 contact status. Inverse probability weights are computed based on the resulting dataset (termed IPW1).

 In the second approach, we leverage hospitalization information. We first estimate the probability of fever given demographic and COVID-19 test status.  When the COVID-19 test status is positive, the probability of fever is allowed to depend on whether the individual was hospitalized.  Figure~\ref{fig:symptomlik1_model2} and~\ref{fig:symptomlik2_model2} in Section~\ref{app:in_add_details} in the supplementary materials presents the estimated likelihood of reporting a fever with a COVID-19 positive individual given a negative and positive COVID-19 test in the past 24 hours respectively.  We see that hospitalization significantly increases the likelihood of fever as expected.  We use these models and the case-to-hospitalization rates by demographic strata to perform mean imputation of the number of positive and negative tests given demographic strata, fever status, and COVID-19 contact status. The use of hospitalization data to estimate the probability of fever may be subject to survivor bias as well as biases driven by differential care.  Inverse probability weights are then computed based on the resulting dataset (termed IPW2).

 \begin{figure}[!th]
 \centering
 \begin{subfigure}{.5\textwidth}
  \centering
  \includegraphics[width=.9\linewidth]{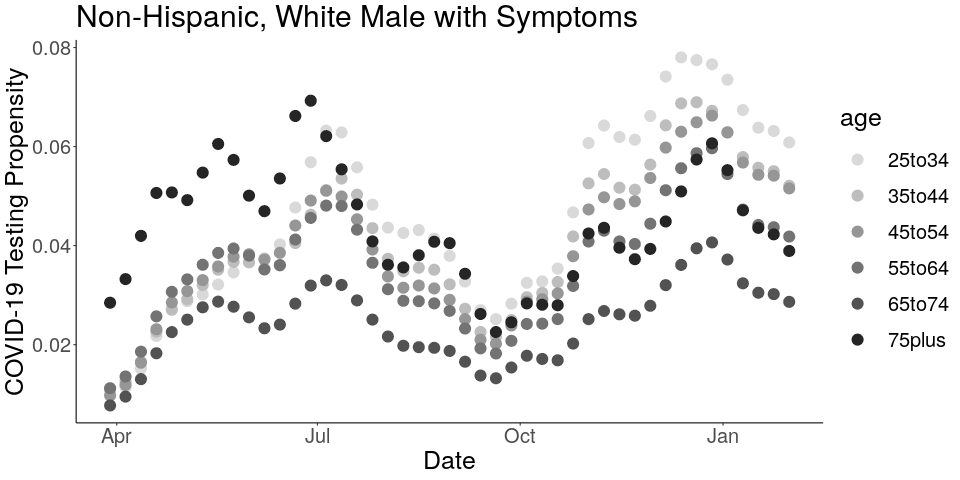}
  \caption{Testing Likelihood Given Symptoms}
  \label{fig:testinglik1_mainpaper}
 \end{subfigure}%
 \begin{subfigure}{.5\textwidth}
  \centering
 \includegraphics[width=.9\linewidth]{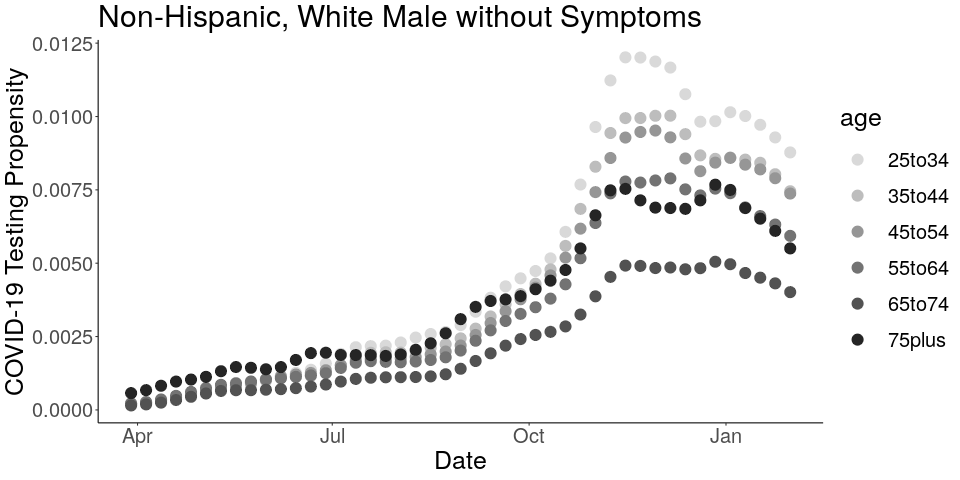}
  \caption{Testing Propensity Given No Symptoms}
  \label{fig:testinglik2_mainpaper}
 \end{subfigure}
 \caption{Testing Propensity across age strata and COVID-19 symptom status}
 \label{fig:testinglik_mainpaper}
 \end{figure}

 Figure~\ref{fig:testinglik_mainpaper} presents the testing propensity for non-Hispanic, white males across age strata and fever status using the second approach.  Note that the ratio of testing propensity within strata across fever status is time-varying, starting 50 times more likely in early April and dropping to ten times more likely by end of 2020.  Similar plots for other strata and a discussion of their relative testing propensities can be found in Section~\ref{section:propplots} of the supplementary materials.

 \subsection{Application of SEIR model}
 \label{section:application_of_modelbased}

 Here we consider the multinomial SEIR model using observed death data as presented in Figure~\ref{fig:in-deaths}. The model, presented in Section~\ref{section:modelbased}, requires infection fatality rates to be specified per-strata.  In this paper, similar to~\cite{Johndrow2020}, we specify fixed per-strata infection fatality rates.  Based on published age-specific IFRs~\citep{Levin2020}, IFR closely follows a log-linear relationship with age. \cite{Ironse2103272118} uses Indiana's random survey and death data to estimate a marginal IFR of $0.84\%$.  Combining across published age-specific IFRs and anchoring our analysis to the estimated marginal IFR of $0.84\%$ for Indiana, Table~\ref{tab:ifrperage} presents the age-specific IFRs used in the analysis. \begin{table}[!th]
 \begin{tabular}{c | c c c c c c c c}
 Age Strata & 0-40 & 50-59 & 60-69 & 70-79 & 80+ \\ \hline
 IFR & $0.014\%$ & $0.120\%$ & $1.206\%$ & $3.815\%$ & $12.920\%$
 \end{tabular}
 \caption{IFR by age-strata}
 \label{tab:ifrperage}
 \vspace{-1cm}
 \end{table}
 The discrete-time distribution~$\{ \theta_{j} \}$ is a discretized version of a truncated normal distribution with mean $25$, standard deviation~$5$, minimum value $0$, and maximum value $44$.  This closely mirrors the distribution from~\cite{Johndrow2020}. Strata for ages less than $40$ are collapsed due to the limited number of COVID-19 related deaths in this age range. For similar reasons, the multinomial parameter~${\bf p}_t$ is assumed to be constant in time with a symmetric Dirichlet prior.
 \vspace{-0.3cm}
 \begin{remark}[Sensitivity to IFR]
 To account for potential misspecification of the marginal IFR, we perform the same analysis with a 10\% increase and decrease in the marginal IFR under the log-linear model. Figure~\ref{fig:tv_air_sens} in the supplementary materials presents the model-based and doubly robust estimates in these scenarios.
 \end{remark}
\vspace{-0.3cm}

 The SEIR model in equation~\eqref{eq:seir} is extended to allow for time-varying transmission rates~$\beta_t$ to account for public policy changes discussed in Section~\ref{section:publicpolicyindiana} over the time period considered.  Here, we insert three change points $(t_1,t_2,t_3)$ on March 23rd, June 15th, and October 1st of 2020 respectively.  Given sensitivity to these choices, instead of considering a simple multiplicative structure, we allow for delayed implementation and slow change in the transmission rate by using weights:
 $$
 \beta_t = \sum_{k=1}^{3} \left(\beta_{k-1} (1-w^{(k)}_t ) + \beta_{k} w_t\right),\quad w^{(k)}_t = (1 + \exp \left( -\xi (t - t_k - \nu) \right))^{-1} {\bf 1}[t \geq t_k]
 $$
 where~$(\beta_0,\beta_1, \beta_2, \beta_3)$ is the vector of transmission rate parameters, $\nu \geq 0$ is the potential delay, $\xi$ is the rate of change, and ${\bf 1}[\cdot]$ is an indicator function.  Each weight~$w_t^{(k)} = 0$ for $t < t_k$ and $w_t^{(k)} \to 1$ as $t \to \infty$. See Section~\ref{app:prior_modelbased} in the supplementary materials for details on prior specification.

 \begin{figure}[!th]
 \centering
 \begin{subfigure}{.5\textwidth}
  \centering
  \includegraphics[width=.9\linewidth]{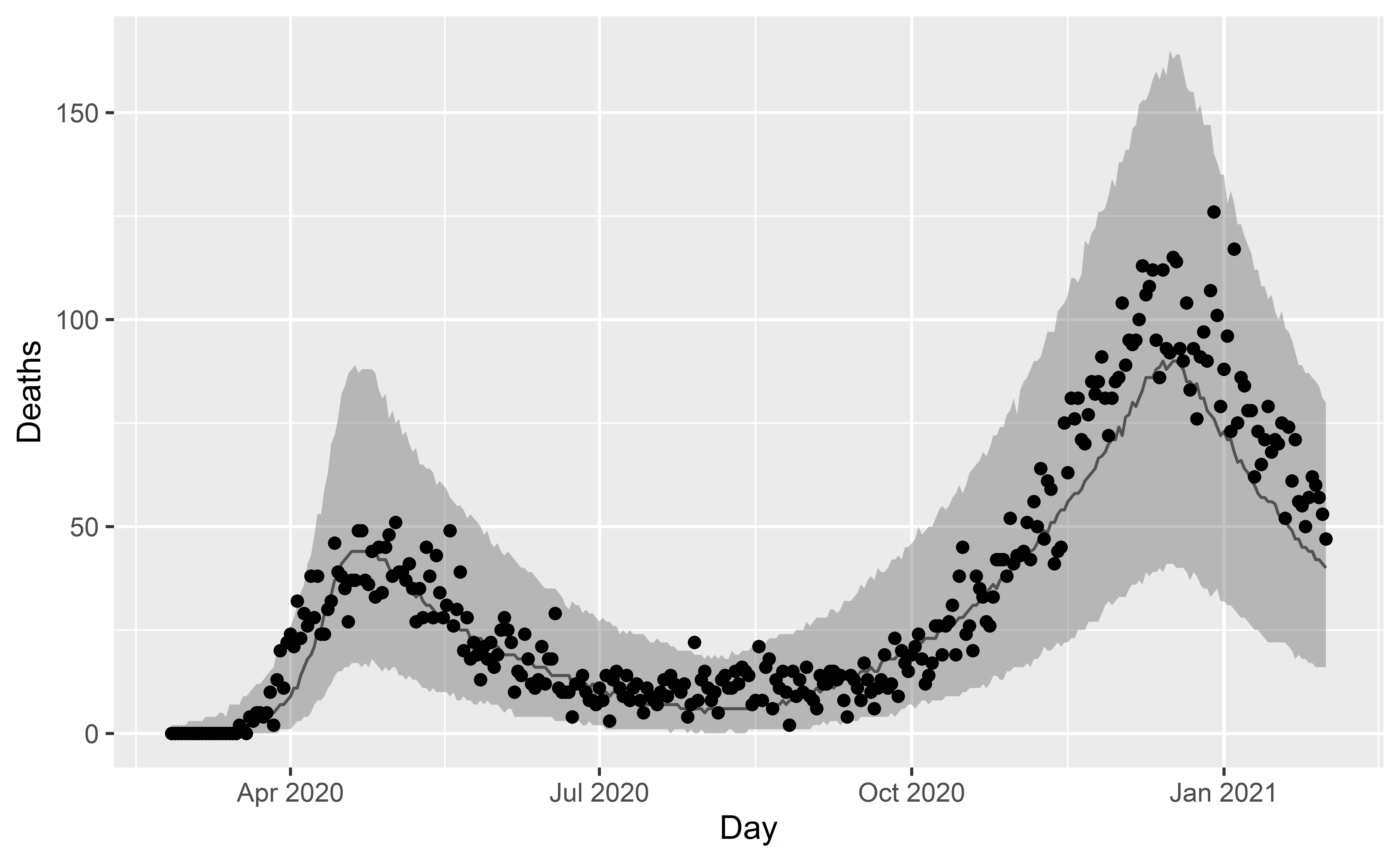}
 \caption{PPC on aggregate daily death counts.}
 \label{fig:deathsppc}
 \end{subfigure}%
 \begin{subfigure}{.5\textwidth}
  \centering
  \includegraphics[width=.9\linewidth]{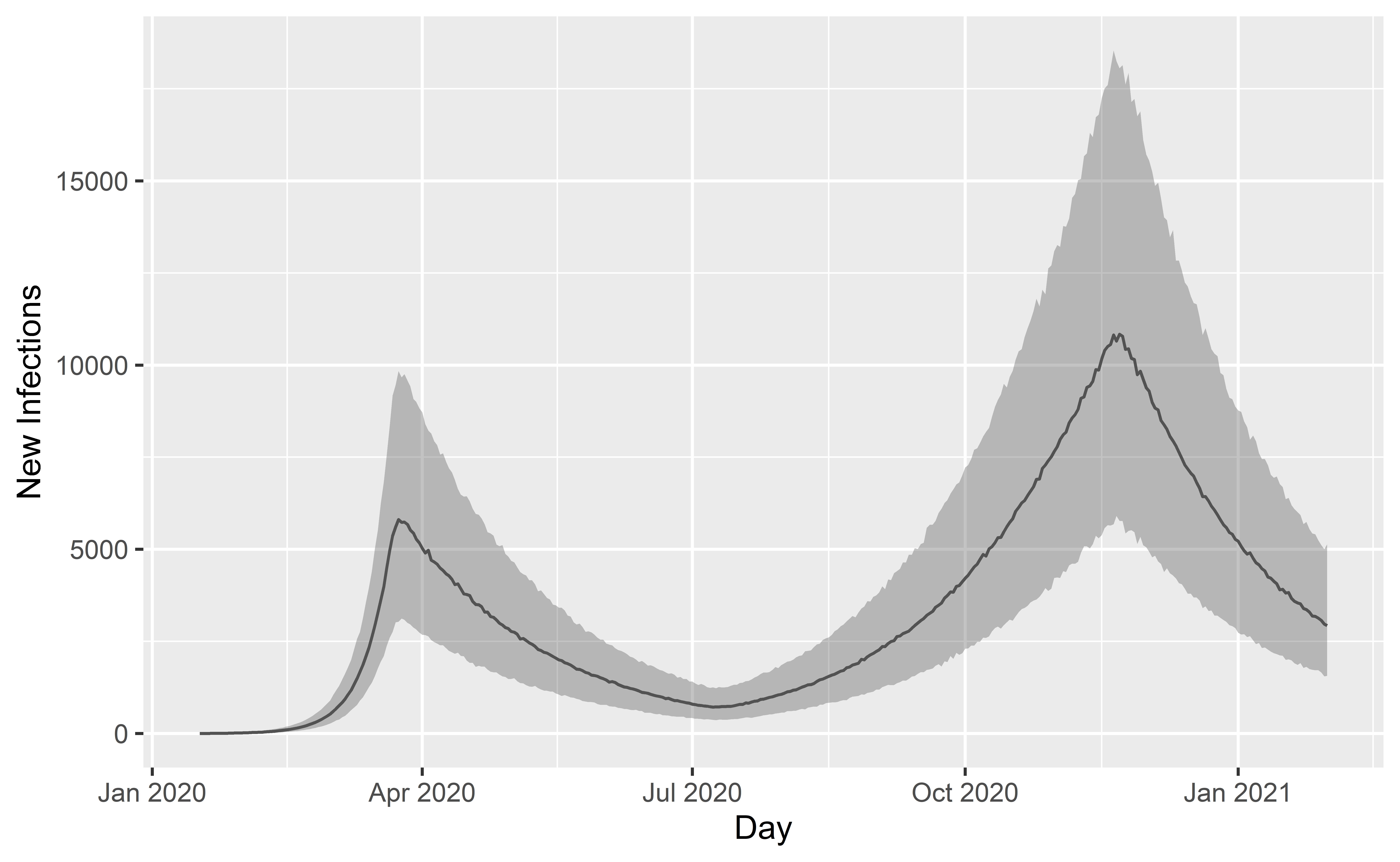}
 \caption{Posterior on daily new infections.}
 \label{fig:casesppc}
 \end{subfigure}
 \caption{Posterior predictive check (PPC) on death counts and posterior distribution of infections.}
 \label{fig:ppcs}
 \end{figure}

 To check model fit, a posterior predictive check on aggregate death counts was performed. Figure~\ref{fig:deathsppc} shows the results which suggests chosen SEIR model fits the observed data well. Figure~\ref{fig:casesppc} presents the posterior distribution of daily new infections. As the active infection rate is defined as all infected individuals whose viral load has yet to fall below detectable levels by RT-PCR testing, to construct the active infection estimates we exponentially discount the number of newly infected individuals to estimate the number of these newly infected individuals who still have an active infection on future days. That is, the number of active infections on day~$t$ is given by~$\sum_{s=0}^t I_t^{\new} e^{-\lambda (t-s)}$. Strata-specific active infections are calculated similarly. In our analysis, the exponential discounting parameter is set to $20$ days to match with prior evidence that ``30\% to 40\% of people will still test positive at three weeks''~\citep{wlocneg}. To compare with case count data, Figure~\ref{fig:undercounting} in the supplementary materials computes the cumulative undercount factor of observed case counts as compared to new infections.  Posterior distributions for the age-specific new infections are used as part of the doubly-robust estimator of the active infection rate.

 \subsection{Time-varying prevalence estimates}

\begin{figure}[!th]
 \centering
 \includegraphics[width=.7\linewidth]{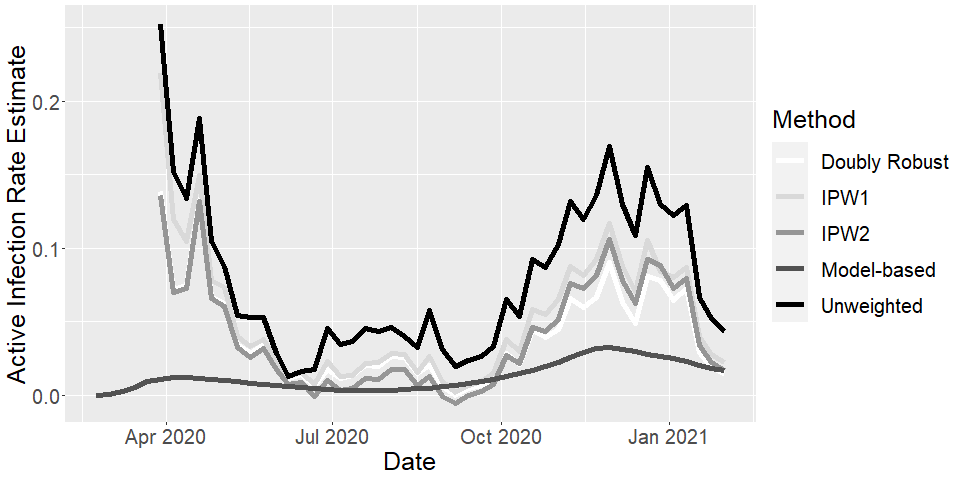}
 \caption{Time-varying active infection rate estimates based on unweighted, model-based, IPW and doubly robust methods.}
 \label{fig:tvestimates}
 \end{figure}

 Here, we construct active infection rate estimates using the unweighted, inverse-probability weighted, model-based, and doubly robust estimators.  Figure~\ref{fig:tvestimates} presents the results. The inverse-probability weighting methods does reduce bias compared to the unweighted estimates. Bias likely remains which we conjecture is due to the limited availability of symptom and other important covariate information to estimate strata-specific infection rates. This suggests better data collection in the future may strongly improve performance.

 Model based estimates appear more reasonable.  The estimate on the last week of April is $1.17\%$ which is an underestimate when compared to the estimated prevalence using the stratified random sample of $1.81\%$. Caution is warranted when interpreting these estimates.  Infection fatality rates are likely time-varying and may vary by other factors such as quality of healthcare and access to vaccines which were made available starting November 2020.  Therefore, improvements could be made by using more accurate IFR estimates, but there is currently minimal publicly available data to do so.  Moreover, the (thankfully) low number of deaths per strata make uncertainty in these estimates quite high.  Finally, the doubly robust estimate appears similar to the IPW estimate, with the largest differences occurring in November and December 2020.

Figure~\ref{fig:air_cis} in the supplementary materials presents the confidence intervals per time point for the IPW2 estimator.  The confidence interval length decreases substantially over time, reflecting the increased testing capacity over this window of time.  Due to the number of surveys per week, there is minimal uncertainty in the parameter estimates.  As the number of tests per week increases to over one hundred thousand, there is therefore minimal uncertainty in the active infection rate estimates. This points to the importance of the statistical decomposition~\eqref{eq:statdecomp2} and the discussion in Section~\ref{section:IPWerrordecomp}. Finally, given the reliance on a probabilistic sample with low response rate (see Remark~\ref{rmk:limitations}), a sensitivity analysis showing the impact of a potential unmeasured confounder is presented in Section~\ref{sec:sensitivity_confounders} of the supplementary materials.

 \section{Discussion}
 \label{section:discussion}

 There is nothing routine about COVID-19, including the corresponding statistical questions.  The goal of this paper was to point out questionable statistical routines.  Precision in reported case-count data gives the illusion of information when what what is needed is quantification of uncertainty. Extensions of recent statistical error decompositions~\citep{Meng2018} demonstrate how selection bias leads data analysts to feel certain about incorrect conclusions.  As case-count data is routinely used for public health policy making, we presented an inverse-probability weighting method and a doubly-robust estimation method that leverage auxiliary information collected through random samples to overcome these issues. We end with a brief discussion of important related topics.

 \subsection*{Data quantity versus quality}

 Governments and policy makers often implicitly argue that increased testing capacity will alleviate selection bias.  Without complete compliance, however, our understanding of future outbreaks may be plagued by self-selection bias, compounded by changing sensitivity and specificity rates of RT-PCR testing. Random testing removes these effect modifiers, giving governments more information to fight the disease.  This paper emphasizes that data quantity is secondary to both data quality and methodological considerations to account for selection bias.

 While we emphasize the importance of random sampling, we do not view it as a panacea. With access to \emph{only} a non-probabilistic sample, we cannot address self-selection bias.  Probability samples provide necessary auxiliary information to do so.  Data analysts can then focus on statistical issues related to the random samples such as how to handle non-response bias.  Covariates that the scientific team think are correlated with non-response (e.g., political affiliation, rural/urban location) should be collected as part of the random sampling protocol and corrected for in the data analysis. On the other hand, collecting covariate information that correlates with testing propensity (e.g., symptom status) in the non-probabilistic sample is insufficient.



 \subsection*{Model-based solutions.}

 A common argument is that the SEIR model can be extended to account for selection bias and measurement error directly; therefore, there is no need for auxiliary random sampling.  Without strong assumptions on the selection mechanism, however, the estimates are often not identifiable.  When an issue ``cannot be resolved nonparametrically then it is usually dangerous to resolve it parametrically''\citep{CoxHink74}. Absent some type of random sampling, the best route forward for all data analyses is careful associated sensitivity analyses and humility in data-driven conclusions.  \cite{Ironse2103272118}, for example, use the Indiana seroprevalence random survey and to anchor their analysis; however, infrequent random surveys imply time-varying aspects such as the IFR may not be resolved by use of a single anchor.


 \subsection*{Real-world implementation of the proposed method}

 The proposed methodology relied on access to two critical datasets: (1) COVID-19 testing, case, hospitalization, and death data by demographic strata and (2) CTIS data.  Public access to (1) required submission of data requests to the state of Indiana.  The author's request was one of only five data requests to be approved by Indiana for public release.  Most states do not release such granular data, making selection bias adjustments difficult.  Coordinated, systematic data collection and reporting is critical.
 Lack of covariate information on individuals seeking COVID-19 tests is unacceptable.  Random samples over time to supplement this data with auxiliary information is desperately needed.  We acknowledge that this proposal will lead to data privacy concerns that will need to be addressed.



 A valid criticism of the proposed approach is that recording covariates on every tested individual is time-consuming and costly.  We argue that the list of relevant factors is of reasonable length.  For example, symptom status and COVID-19 contact are clearly relevant factors in testing selection.  Future work may consider how to include/exclude factors over time that are not significant to limit citizen reporting burden.  Moreover, equation~\eqref{eq:auxinfoprob} presupposes covariate information is collected for all individuals in the nonprobability sample, which is often not feasible for state/local governments where rapid sample collection is prioritized.  One can easily extend~\eqref{eq:auxinfoprob} by randomly sampling a subset of the nonprobability sample on which to collect the covariate information.
 This would balance goals of rapid testing and auxiliary data collection.

\newpage
\bibliographystyle{plainnat}
\bibliography{covid-refs}

\newpage
\appendix

\section{Reproducible code}

All relevant code can be found at \url{https://github.com/wdempsey/covid-umich}.

\section{Notation Glossary}
\label{app:notation}

\paragraph{Notation for Section~\ref{section:casecount} (non-temporal setting)}
\begin{itemize}
\item[$Y_j$] \quad Binary COVID-19 status of individual $j$ in the population.
\item[$N$] \quad Population size
\item[$\bar Y$] \quad Population average, $\bar Y = N^{-1} \sum_{j=1}^N Y_j$
\item[$I_j$] \quad Selection indicator of individual $j$ in the population into the sample
\item[$y_j$] \quad Binary COVID-19 status of sample individual $j=1,\ldots,n$.
\item[$\bar y_n$] \quad Sample average, $\bar y_n = n^{-1} \sum_{j=1}^N I_j Y_j = n^{-1} \sum_{j=1}^n y_j$
\item[$\rho_{I,Y}$] \quad \emph{Data quality}, i.e., correlation between selection indicator and population outcome
\item[$f$] \quad \emph{Data quantity}, i.e., sampling fraction $f = n/N$
\item[$\sigma_Y$] \quad \emph{Problem Difficulty}, i.e., population variance $\sigma_Y^2 = (N)^{-1} \sum_{j=1}^N (Y_j - \bar Y)^2$.
\item[$FP$] \quad False Positive Rate
\item[$FN$] \quad False Negative Rate
\item[$\tilde y_n$] \quad Sample average adjusted for false negative and positive rates, i.e., $\tilde y_n = (1-FP-FN)^{-1} (\bar y_n - FP)$
\item[$f_0, f_1$] \quad Sampling fraction among COVID-19 negative and positive individuals respectively, i.e., $f_1 := \sum_{j=1}^N I_j Y_j / \sum_{j=1}^N Y_j$
and $f_0 := \sum_{j=1}^N I_j (1-Y_j) / \sum_{j=1}^N (1-Y_j)$.
\item[$\Delta, M$] \quad Sampling rate differential on additive -- $\Delta = f_1 - f_0$ -- and multiplicative -- $M = f_1/f_0$ -- scales.
\end{itemize}

\paragraph{\bf Notation for Section~\ref{section:improvedcasecount} (temporal setting)}
\begin{itemize}
\item[$Y_{j,t}$] \quad Binary COVID-19 status of individual $j$ in the population at time~$t$.
\item[$I^{NR}_{j,t}$] \quad Selection indicator of individual $j$ in the population into the non-probability sample.
\item[$X_{j,t}$] \quad Feature vector for individual $j$ in the population at time~$t$ that impact self-selection into the non-probability sample.
\item[$I_{j,t}^{R}$] \quad Selection indicator of individual $j$ in the population into the probability sample.
\item[$W_{j,t}^{R}$] \quad Weight of individual $j$ in the population in the probability sample (i.e., inverse-probability of selection)
\item[$\pi (X_{j,t}; \theta)$] \quad Self-selection propensity of individual~$j$ into the non-probability sample given feature vector~$X_{j,t}$ and parameter~$\theta$.
\item[$w (X_{j,t})$] \quad Weight of individual~$j$ in the population in the non-probability sample, i.e., $w(x) = 1/\pi(x; \theta)$.
\item[$\hat \mu(X_{j,t})$] \quad Model-based estimate of the active infection rate at time~$t$ for individuals with feature vector~$X_{j,t}$
\end{itemize}

\section{Technical details}

Recall that $P_j$ is  an indicator of measurement error, equal to $1$ when we incorrectly measure the outcome and $0$ when we observe the true outcome. We suppose this is a stochastic variable where $\pr(P_j = 1 \mid Y_j = 1) =: FN$ is the false-negative rate and $\pr(P_j = 1 \mid Y_j = 0) =: FP$ is the false-positive rate.  If individual $j$ is selected (i.e., $I_j = 1$) then the observed outcome can be written as $Y_j^{\star} = Y_j(1-P_j) + (1-Y_j) P_j$.

\subsection{Derivations for imperfect testing framework.}
\label{app:imperfect}
We start by considering the empirical mean estimator under imperfect testing,
$$
\bar y_n^\star = \frac{\sum_{j=1}^N Y_j^\star I_j}{\sum_{j=1}^N I_j} = \frac{\sum_{i=1}^N  I_j Y_j^\star }{\sum_{j=1}^N  I_j } = \frac{\sum_{i=1}^N  I_j \left[ Y_j (1-P_j) + (1-Y_j) P_j \right]}{\sum_{j=1}^N  I_j }
$$
For any set of numbers $\{ A_1, \ldots, A_N \}$ we can view it as the support of a random variable $A_J$ induced by the random index $J$ defined on $\{1,\ldots, N\}$.  When $J$ is uniformly distributed $E_J (A_J) = \sum_{j=1}^N A_j / N \equiv \bar A_N$. Then
$$
\begin{aligned}
\bar y_n^\star  - \bar Y_N &= \frac{E_J \left[ I_J \left[ Y_J (1-P_J) + (1-Y_J) P_J \right] \right]}{E_J [ I_J ] } - E_J[Y_J] \\
&= \frac{E_J \left[ I_J P_J (1-2Y_J) \right]}{E_J [ I_J ] } + \left( \frac{E_J [I_J Y_J]}{E_J [ I_J ] } - \frac{E_J[Y_J] E_J[I_J]}{E_J[I_J]} \right) \\
\end{aligned}
$$
The term in parentheses can be re-written as
$$
\begin{aligned}
\frac{E_J [I_J Y_J]- E_J[Y_J] E_J[I_J]}{E_J[I_J]} &=  \frac{E_J [I_J Y_J]- E_J[Y_J] E_J[I_J]}{\sqrt{V_J(I_J) V_J(Y_J)}} \frac{\sqrt{V_J(I_J)}}{E_J[I_J]} \times \sqrt{V_J(Y_J)} \\
&= \rho_{I,Y} \times \sqrt{\frac{(1-f)}{f}} \times \sigma_Y
\end{aligned}
$$
which agrees with Meng's (2019) decomposition. For the other term, first we define $Z_j := 1 - 2 Y_j $. Then $Z_j = 1$ if $Y_j = 0$ and $Z_j = -1$ if $Y_j = 1$. Then the term can be re-written as
$$
\begin{aligned}
\frac{E_J \left[ I_J P_J (1-2Y_J) \right]}{E_J [ I_J ] } &= \left( \frac{E_J \left[ I_J P_J Z_J \right]}{E_J [ I_J ] } -  \frac{E_J \left[ P_J Z_J \right] E_J[ I_J]}{E_J [ I_J ] } \right) +  \frac{E_J \left[ P_J Z_J \right] E_J[ I_J]}{E_J [ I_J ] } \\
\end{aligned}
$$
The term in parentheses can be re-expressed using the previous technique as:
$$
\rho_{I, PZ} \times \sqrt{\frac{1-f}{f}} \times \sigma_{PZ}
$$
where now the ``data defect'' and ``problem difficulty'' are with respect to $PZ$ rather than $Y$. The final term is equal to
$$
\begin{aligned}
E_J [P_J Z_J ] &= E_J [ E_J [ P_J Z_J \mid Y_J ] ] \\
&= \pr (P = 1 \mid Y = 0) (1-\bar Y) - \pr(P=1 \mid Y = 1) \bar Y \\
&= FP - (FP + FN) \cdot \bar Y
\end{aligned}
$$
Combining these yields:
$$
\bar y_n^\star - \bar Y = \sqrt{\frac{1-f}{f}} \left(\rho_{I,Y} \sigma_Y + \rho_{I, PZ} \sigma_{PZ} \right) + \left( FP - (FP+FN) \bar Y \right)
$$

\subsubsection{Derivation of an estimator unbiased under SRS}
\label{app:memestimator}
We see the final term is given by $FP (1-\bar Y) - FN \bar Y$ is the bias associated with using the unadjusted prevalence estimate $\bar y_n^\star$.  This motivates an adjusted estimate
$$
\tilde y_n^{(0)} = \bar y_n^\star - FP (1- \bar y_n^\star ) + FN \bar y_n^\star
= \bar y_n^\star (1 + FN + FP) - FP.
$$
Now considering the error for the adjusted estimate, $\tilde y_n^{(0)} - \bar Y$, we have
$$
\begin{aligned}
 &\bar y_n^\star (1 + FN + FP) - FP - \bar Y \\
 =&(\bar y_n^\star - \bar Y) +  (FN + FP) \bar y_n^\star - FP  \\
 =&\underbrace{\sqrt{\frac{1-f}{f}} \left[\rho_{I,Y} \sigma_Y + \rho_{I, PZ} \sigma_{PZ} \right]}_{\Psi} + (  FN + FP ) (\bar y_n^\star -  \bar Y) \\
 =& \Psi + (FN + FP) \Psi + (FN + FP) (FP - (FP+FN) \bar Y).
 \end{aligned}
$$
The final term is a (smaller) bias term and so we propose another adjusted estimator $\tilde y_n^{(1)} = \tilde y_n^{(0)} + (FN+FP)  ( (FN+FP) \bar y_n^\star - FP)$, with associated error $\tilde y_n^{(1)} - \bar Y$ given by
$$
\begin{aligned}
 &(\bar y_n^{(0)} - \bar Y) + (FN+FP)  ( (FN+FP) \bar y_n^\star - FP)\\
 =&\Psi + (FN + FP) \Psi + (FN + FP) (FP - (FP+FN) \bar Y) + (FN + FP) ((FP+FN) \bar y_n^\star - FP)  \\
  =& \Psi + (FN + FP) \Psi + (FN + FP)^2 \Psi + (FN+FP)^2 (FP - (FP+FN) \bar Y).
 \end{aligned}
$$
This motivates recursively defining estimators $\tilde y_n^{(t)} = \tilde y_n^{(t-1)} + (FN+FP)  ( (FN+FP) \bar y_n - FP)$ for $t=1,2,\ldots$ where $\tilde y_n^{(0)} = \bar y_n^\star$.  Then
$$
\tilde y_n^{(t)} = \bar y_n^\star \sum_{s=0}^{t+1} (FP+FN)^s - FP \sum_{s=0}^{t} (FP+FN)^s
$$
and the associated error at iteration $t$ given by
$$
\Psi \sum_{s=0}^t (FN+FP)^s = \Psi \frac{1 - (FN+FP)^t}{1 - (FN+FP)}.
$$
We can then get an estimator with no residual bias term by taking the limit as $t$ goes to infinity; that is, define
$$
\tilde y_n = \lim_{t \to \infty} \tilde y_n^{(t)} = \frac{\bar y_n^\star - FP}{1 -(FN+FP)}.
$$
Then the associated error $\tilde y_n - \bar Y$ can be expressed as $\frac{\Psi}{1-(FN+FP)}$.

\subsection{Model-based derivation of the estimator.}
\label{app:modelbased}
The estimator $\tilde y$ was derived as a limit of a process that removes the residual bias term at each step.  Here we consider a model-based explanation.  Let $\theta = \pr( Y = 1)$ and $\phi = \pr( \text{test is positive})$.  Given a known false negative (FN) and false positive (FP) rates, we have
$$
\begin{aligned}
\phi &= \theta \cdot (1-FN) + (1-\theta) \cdot FP = \theta (1-FN-FP) + FP \\
\Rightarrow \theta &= \frac{\phi - FP}{1-FN-FP}.
\end{aligned}
$$
Thus, the estimator $\tilde y$ is also the appropriate estimator under a model-based approach.  While the derivation here is more straightforward, the derivation in the prior section provides a simple formula for the associated error $\tilde y_n - \bar Y$ and gives a novel connection between the empirical estimator $\bar y_n^\star$ and the adjusted estimator $\tilde y_n$ without reference to the model-based approach.

\subsection{Further simplification.}
For the binary outcome $Y$, we have $\sigma_Y = \sqrt{\bar Y (1-\bar Y)}$. Moreover,
$$
\begin{aligned}
V_J(P_J Z_J) &= E_J[(P_J Z_J)^2] - E_J[P_J] E_J[Z_J] \\
&= E_J[P_J] - E_J[P_J] (1 - 2 \bar Y) = 2 \bar Y E_J [ P_J ] \\
&= 2 \bar Y \left( FP (1-\bar Y) + FN \bar Y \right) \\
\Rightarrow \sigma_{PZ} &= \sqrt{ 2 \bar Y \left( FP (1-\bar Y) + FN \cdot  \bar Y \right) }
\end{aligned}
$$
Then the formula for the error is given by:
\begin{equation}
\label{eq:finalstep}
\sqrt{\frac{1-f}{f}} \left[\rho_{I,Y} \sqrt{\bar Y (1-\bar Y)} + \rho_{I, PZ} \sqrt{ 2 \bar Y \left( FP (1-\bar Y) + FN \cdot \bar Y \right )} \right] \times \frac{1}{1- (FN+FP)}
\end{equation}
By definition, we have
$$
\begin{aligned}
\rho_{I,PZ} &= \frac{C(I, PZ)}{\sqrt{V(PZ) V(I)}} \\
&= \frac{C(I, PZ)}{\sqrt{V(Y) V(I)}} \sqrt{\frac{V(Y)}{V(PZ)}} \\
&= \rho_{I,Y} \frac{C(I,PZ)}{C(I,Y)} \sqrt{ \frac{(1-\bar Y)}{2 ( FP (1-\bar Y) + FN \cdot \bar Y)} }
\end{aligned}
$$

$$
\begin{aligned}
C(I, PZ) &= E[ I P Z ] - E[I] E[PZ] \\
&=  [FP f_0 - (FP f_0 + FN f_1) \bar Y] - f [ FP - (FP+FN) \bar Y ] \\
&=  - FP \Delta \bar Y + FP \bar Y^2 \Delta - FN \bar Y^2 \Delta \\
&=  - \Delta \bar Y (FP \cdot (1-\bar Y) + FN \cdot \bar Y) \\
\end{aligned}
$$
where $f = f_1 \bar Y + f_0 (1-\bar Y)$ so $f_0 - f = -\Delta \bar Y$ and $f_1 - f = \Delta (1-\bar Y)$.
$$
\begin{aligned}
C(I, Y) &= E[ I Y ] - f \bar Y \\
&=  f_1 \bar Y + f_0 (1-\bar Y) - f \bar Y \\
&=  f_0 (1-\bar Y) + \Delta (1-\bar Y) \bar Y \\
&= (1-\bar Y) (f_0 + \Delta \bar Y)
\end{aligned}
$$
Combining yields
$$
\begin{aligned}
\rho_{I,PZ} &= \rho_{I,Y} \times \frac{- \Delta \bar Y (FP \cdot (1-\bar Y) + FN \cdot \bar Y) }{(1-\bar Y) (f_0 + \Delta \bar Y)} \times \sqrt{ \frac{(1-\bar Y)}{2 ( FP (1-\bar Y) + FN \cdot \bar Y)} } \\
&= - \rho_{I, Y} \times \Delta \times \sqrt{\frac{\bar Y}{1-\bar Y}} \frac{\sqrt{FP(1-\bar Y) + FN \cdot \bar Y}}{f_0 (1-\bar Y) + f_1 \bar Y} \times \sqrt{\frac{\bar Y}{2}}
\end{aligned}
$$
We can then re-write $\rho_{I,Y} \sigma_Y + \rho_{I,PZ} \sigma_{PZ}$ as
$$
\rho_{I,Y} \sigma_Y \left( 1 - \Delta \times \frac{\bar Y}{1-\bar Y} \times \frac{FP(1-\bar Y) + FN \cdot \bar Y}{f_0 (1-\bar Y) + f_1 \bar Y} \right).
$$
Inserting into equation~\eqref{eq:finalstep} yields the desired result.

\subsection{Derivation of effective sample size}
\label{app:effss}

Let $S_Y^2 = (N-1)^{-1} \sum_{j=1}^N (Y_j - \bar Y)^2$ be the population variance as defined in survey sampling~\citep{Cochran77}.  Then $\sigma_Y^2 = (N-1)/N \cdot S_Y^2$.  Under SRS, the MSE is the variance as the estimate is unbiased and the variance is given by $(1-f)/n S_Y^2$.  Then setting the MSE under general selection and SRS equal we have
$$
\begin{aligned}
\underbrace{\frac{1-f}{f} \times E_{\I} \left[ \rho_{I, Y}^2 \times D_M^2 \right]}_{1/n_{eff}^\star} \times \sigma_Y^2 &= \frac{1-f}{n_{eff}} S_Y^2 \\
\frac{1}{n_{eff}^\star} \times \frac{N-1}{N} S_Y^2 &= \frac{1-f}{n_{eff}} S_Y^2 \\
\frac{1}{n_{eff}^\star} &=  \left( \frac{1}{n_{eff}} - \frac{1}{N} \right) \left( \frac{N}{N-1} \right) \\
\frac{1}{n_{eff}^\star} \left[ 1 - \frac{1}{N} + \frac{n_{eff}^\star}{N} \right]  &=  \frac{1}{n_{eff}} \\
n_{eff}^\star \left[ 1 - \frac{1}{N} + \frac{n_{eff}^\star}{N} \right]^{-1}  &=  n_{eff} \\
\frac{n_{eff}^\star}{ 1 + (n_{eff}^\star -1) N^{-1}} &=  n_{eff} \\
\end{aligned}
$$
Then if $n_{eff}^\star \geq 1$, we have that
$$
n_{eff} \leq n_{eff}^\star = \frac{f}{1-f} \times \frac{1}{E_{\I} \left[ \rho^2_{I, Y} \times D_M^2 \right]}
$$

\subsection{Ratio estimator}
\label{app:ratio}

Let ${\bf u} = (u_{t-1},u_t) \in \mathbb{R}^2$ and $g({\bf u}) = \frac{u_t}{u_{t-1}}$, i.e., a differentiable function $g:\mathbb{R}^2 \to \mathbb{R}$. Centering a Taylor series expansion of second-order around coordinates $(U_2, U_1) \in \mathbb{R}^2$ yields
$$
\begin{aligned}
g({\bf u}) =& g(U_{t-1}, U_t) - \frac{U_t}{U_{t-1}^2} (u_{t-1} - U_{t-1}) + \frac{1}{U_{t-1}} (u_t - U_t) \\
&+ \frac{1}{2} \left[ \frac{2 U_t}{U_{t-1}^3} (u_{t-1} - U_{t-1})^2 + 0 \times (u_t - U_t)^2 - 2 \times (u_{t-1} - U_{t-1}) (u_t - U_t) \frac{1}{U_t^2} \right]
\end{aligned}
$$
Plugging in $(\bar y_{t-1}, \bar y_t)$ for $(u_{t-1}, u_t)$ and $(\bar Y_{t-1}, \bar Y_t)$ for $(U_{t-1}, U_t)$ yields the $\frac{\bar y_t}{\bar y_{t-1}} - \frac{\bar Y_t}{\bar Y_{t-1}} $ is equal to
$$
\begin{aligned}
=&
- \frac{\bar Y_t}{\bar Y_{t-1}^2} (\bar y_{t-1} - \bar Y_{t-1}) + \frac{1}{\bar Y_{t-1}} (\bar y_t - \bar Y_t) \\
&+ \frac{\bar Y_t}{\bar Y_{t-1}^3} (\bar y_{t-1} - \bar Y_{t-1})^2 -  (\bar y_{t-1} - \bar Y_{t-1}) (\bar y_t - \bar Y_t) \frac{1}{\bar Y_{t-1}^2} \\
&= \frac{\bar Y_t}{\bar Y_{t-1}} \bigg[  \rho_{I_t,Y_t} \sqrt{\frac{1-f_t}{f_t}} CV (Y_t)  -\rho_{I_{t-1},Y_{t-1}} \sqrt{\frac{1-f_{t-1}}{f_{t-1}}} CV (Y_{t-1}) \\
&+ \rho^2_{I_{t-1},Y_{t-1}} \frac{1-f_{t-1}}{f_{t-1}} CV^2 (Y_{t-1}) -  \rho_{I_{t-1},Y_{t-1}} \sqrt{\frac{1-f_{t-1}}{f_{t-1}}} CV (Y_{t-1}) \times
\rho_{I_t,Y_t} \sqrt{\frac{1-f_t}{f_t}} CV (Y_t)   \bigg] \\
&= \frac{\bar Y_t}{\bar Y_{t-1}} \bigg[ \rho_{I_t,Y_t} \sqrt{\frac{1-f_t}{f_t}} CV (Y_t)  -\rho_{I_{t-1},Y_{t-1}} \sqrt{\frac{1-f_{t-1}}{f_{t-1}}} CV (Y_{t-1}) \bigg] \left[ 1 - \rho_{I_{t-1},Y_{t-1}} \sqrt{\frac{1-f_{t-1}}{f_{t-1}}} CV (Y_{t-1}) \right]
\end{aligned}
$$
where the second equality is obtained by plugging in the statistical decomposition of the error for both time points and the coefficient of variation being defined as $CV(Y) := \sigma_Y/\mu_Y$.  Under measurement error, the extra terms $D_{t}$ and $D_{t-1}$ can be inserted in the correct locations.

\subsection{Estimation of effective reproduction number}

Let
$$
\delta_t := \bigg[ \rho_{I_t,K_t} D_{M_t} \sqrt{\frac{1-f_t}{f_t}} CV (K_t)  -\rho_{I_{t-1},K_{t-1}} D_{M_{t-1}} \sqrt{\frac{1-f_t}{f_t}} CV (K_{t-1}) \bigg].
$$
Then the previous sections derivation shows that the estimate of the number of new cases on day t is given by
$$
\frac{S_t \cdot \bar y_t}{S_{t-1} \cdot \bar y_{t-1}} =
\frac{K_t}{K_{t-1}} \left( 1 + \delta_t \times \left[ 1 - \rho_{I_{t-1},K_{t-1}} D_{M_{t-1}} \sqrt{\frac{1-f_t}{f_t}} CV (K_{t-1}) \right] \right)
$$
Then setting $e_t = \delta_t \times [1 - \rho_{I_{t-1},K_{t-1}} D_{M_{t-1}} \sqrt{\frac{1-f_t}{f_t}} CV (K_{t-1}) ]$, we have
$$
\begin{aligned}
\log \left( \frac{S_t \bar y_t}{S_{t-1} \bar y_{t-1}} \right) - \log \left( \frac{K_t}{K_{t-1}} \right) &= \log (1 + e_t) \\
\log \left( \frac{\bar y_t}{\bar y_{t-1}} \right) - \log \left( \frac{K_t}{K_{t-1}} \right) &= 1 + e_t - \log \left( \frac{S_t}{S_{t-1}} \right) \\
1 + \frac{1}{\gamma} \log \left( \frac{\bar y_t}{\bar y_{t-1}} \right) - \left[ 1 + \frac{1}{\gamma} \log \left( \frac{K_t}{K_{t-1}} \right) \right] &= \frac{1}{\gamma} \left[ \log \left( 1 + e_t \right) - \log \left( \frac{S_{t}}{S_{t-1}} \right) \right] \\
\Rightarrow \hat R_t - R_t &= \frac{1}{\gamma} \left[ \log \left( 1 + e_t \right) - \log \left( \frac{S_{t}}{S_{t-1}} \right) \right]
\end{aligned}
$$

\subsection{Computing the effective sample size}
\label{section:effss}

For binary outcomes, we have
\begin{equation} \label{eq:binaryrho}
\rho_{I,Y} = \Delta \sqrt{\frac{\bar Y (1 - \bar Y)}{f (1-f)} }
\end{equation}
where $\Delta = P_J (I_J = 1 \mid Y_J = 1) - P(I_J = 1 \mid Y_J = 0) = f_1 - f_0$.  Suppose that $M = f_1/f_0$; then $f_0 = f / (\bar Y \cdot (M-1) + 1)$.  Using the upper bound $E_{\I} [ \rho_{I,Y}^2 ] \leq E_{\I} [\rho_{I,Y} ]^2$, we compute effective sample size under a range of prevalences $\bary$, and relative sample rates $M$ given $f = 0.003$ (i.e., current sampling fraction).

\begin{table}[ht]
\centering
\begin{tabular}{rrrrrrrr}
& \multicolumn{7}{c}{$M$} \\ \cline{2-8}
$\bar y$ & 1.05 & 1.15 & 1.25 & 1.35 & 1.45 & 1.55 & 1.65 \\
  \hline
0.01 & 40444 & 4503 & 1624 & 830 & 503 & 338 & 242 \\
0.03 & 13787 & 1541 & 558 & 286 & 174 & 117 & 85 \\
0.05 & 8463 & 950 & 345 & 178 & 109 & 73 & 53 \\
0.07 & 6187 & 697 & 254 & 132 & 81 & 55 & 40 \\
0.09 & 4928 & 557 & 204 & 106 & 65 & 44 & 32 \\
0.11 & 4131 & 469 & 173 & 90 & 56 & 38 & 28 \\
   \hline
\end{tabular}
\end{table}

We also present the same plot under $FP = 0.024$ and $FN = 0.13$ to show the impact of measurement error on effective sample size.

\begin{table}[ht]
\centering
\begin{tabular}{rrrrrrrr}
& \multicolumn{7}{c}{$M$} \\ \cline{2-8}
 & 1.05 & 1.15 & 1.25 & 1.35 & 1.45 & 1.55 & 1.65 \\
  \hline
0.01 & 29019 & 3247 & 1177 & 605 & 368 & 248 & 179 \\
0.03 & 9894 & 1112 & 405 & 209 & 128 & 87 & 63 \\
0.05 & 6075 & 686 & 251 & 130 & 80 & 54 & 39 \\
0.07 & 4442 & 504 & 185 & 96 & 59 & 41 & 30 \\
0.09 & 3539 & 403 & 149 & 78 & 48 & 33 & 24 \\
0.11 & 2967 & 339 & 126 & 66 & 41 & 28 & 21 \\
   \hline
\end{tabular}
\end{table}

For binary outcomes, we have
\begin{equation} \label{eq:weightedbinaryrho}
\rho_{\tilde I (X),Y} = \tilde \Delta \sqrt{\frac{\bar Y (1 - \bar Y)}{f (1-f) \text{E}(W_J \mid I_J = 1)^2 + f \text{Var}(W_J \mid I_J = 1)} }
\end{equation}

\section{IPW statistical error decomposition derivation}
\label{app:ipwderivation}
We start by considering the empirical weighted mean estimator under imperfect testing,
$$
\bar y_n^\star = \frac{\sum_{j=1}^N W_j Y_j^\star I_j}{\sum_{j=1}^N W_j I_j} = \frac{\sum_{i=1}^N  I_j W_j Y_j^\star }{\sum_{j=1}^N  I_j } = \frac{\sum_{i=1}^N  I_j \left[ Y_j (1-P_j) + (1-Y_j) P_j \right]}{\sum_{j=1}^N  I_j }
$$
Then
$$
\begin{aligned}
\bar y_n^\star  - \bar Y_N &= \frac{E_J \left[ I_J W_J \left[ Y_J (1-P_J) + (1-Y_J) P_J \right] \right]}{E_J [ I_J W_J ] } - E_J[Y_J] \\
&= \frac{E_J \left[ I_J W_J P_J (1-2Y_J) \right]}{E_J [ I_J W_J ] } + \left( \frac{E_J [I_J W_J Y_J]}{E_J [ I_J W_J ] } - \frac{E_J[Y_J] E_J[I_J W_J]}{E_J[I_J W_J]} \right) \\
\end{aligned}
$$
The term in parentheses can be re-written as
$$
\begin{aligned}
\frac{E_J [I_J W_J Y_J]- E_J[Y_J] E_J[I_J W_J]}{E_J[I_J W_J]} &=  \frac{E_J [I_J W_J Y_J]- E_J[Y_J] E_J[I_J W_J]}{\sqrt{V_J(I_J W_J) V_J(Y_J)}} \frac{\sqrt{V_J(I_J W_J)}}{E_J[I_J]} \times \sqrt{V_J(Y_J)} \\
&= \rho_{\tilde I (X),Y} \times \frac{\sqrt{V_J(I_J W_J)}}{E_J[I_J]} \times \sigma_Y
\end{aligned}
$$
Then
\begin{align*}
E_J (I_J W_J) &= P(I_J = 1) E_J [W_J \mid I_J = 1] = f \times E_J [W_J \mid I_J = 1] \\
V_J(I_J W_J) &= E[ I_J W_J^2] - E[I_J W_J]^2 \\
&= f \cdot \left( E [W_J^2 \mid I_J = 1] - f E[W_J \mid I_J = 1 ]^2 \right) \\
&= f \cdot \left( E [W_J^2 \mid I_J = 1] \pm E[W_J \mid I_j = 1]^2 - f E[W_J \mid I_J = 1 ]^2 \right) \\
&= f \cdot \left( V (W_J \mid I_J = 1) + E[W_J \mid I_J = 1]^2 (1-f) \right)
\end{align*}
Taking the ratio:
$$
\sqrt{\frac{f \cdot \left( V (W_J \mid I_J = 1) + E[W_J \mid I_J = 1]^2 (1-f) \right)}{f^2 E_J [W_J \mid I_J = 1]^2}} = \sqrt{\frac{1 - f + CV(W)^2}{f}}
$$
which agrees with Meng's (2019) decomposition of weighted outcome. For the other term, first we define $Z_j := 1 - 2 Y_j $. Then $Z_j = 1$ if $Y_j = 0$ and $Z_j = -1$ if $Y_j = 1$. Then the term can be re-written as
$$
\begin{aligned}
\frac{E_J \left[ I_J W_J P_J (1-2Y_J) \right]}{E_J [ I_J W_J ] } &= \left( \frac{E_J \left[ I_J W_J P_J Z_J \right]}{E_J [ I_J W_J ] } -  \frac{E_J \left[ P_J Z_J \right] E_J[ I_J W_J]}{E_J [ I_J W_J ] } \right) +  \frac{E_J \left[ P_J Z_J \right] E_J[ I_J W_J]}{E_J [ I_J W_J ] } \\
\end{aligned}
$$
The term in parentheses can be re-expressed using the previous technique as:
$$
\rho_{\tilde I, PZ} \times \sqrt{\frac{1-f+CV(W)^2}{f}} \times \sigma_{PZ}
$$
where now the ``data defect'' and ``problem difficulty'' are with respect to $PZ$ rather than $Y$. The final term is equal to
$$
\begin{aligned}
E_J [P_J Z_J ] &= E_J [ E_J [ P_J Z_J \mid Y_J ] ] \\
&= \pr (P = 1 \mid Y = 0) (1-\bar Y) - \pr(P=1 \mid Y = 1) \bar Y \\
&= FP - (FP + FN) \cdot \bar Y
\end{aligned}
$$
Combining these yields:
$$
\bar y_n^\star - \bar Y = \sqrt{\frac{1-f+CV(W)^2}{f}} \left(\rho_{\tilde I(X),Y} \sigma_Y + \rho_{\tilde I(X), PZ} \sigma_{PZ} \right) + \left( FP - (FP+FN) \bar Y \right)
$$
By previous arguments, we have
$$
\rho_{\tilde I (X), PZ} = \rho_{\tilde I(X), Y} \times \frac{C(\tilde I(X), PZ)}{C(\tilde I(X), Y)} \sqrt{ \frac{1-\bar Y}{ 2 (FP(1-\bar Y) + FN \bar Y)} }
$$
Then
\begin{align*}
C(IW, PZ) &= E[IWPZ] - E[IW] E[PZ] \\
&=  [FP \tilde f_0  - (FP \tilde f_0 + FN \tilde f_1) \bar Y]  - \tilde f \cdot ( FP - (FP+FN) \bar Y )\\
&= -FP \tilde \Delta \bar Y + FP \bar Y^2 \Delta - FN \bar Y^2 \Delta \\
&= -\tilde \Delta \bar Y ( FP \cdot (1-\bar Y) + FN \cdot \bar Y)
\end{align*}
where $\tilde f_i = E[ I_J W_J \mid Y_J = i]$ for $i \in \{0,1 \}$ and $f = \tilde f_1 \bar Y + \tilde f_0 (1-\bar Y)$. Moreover,
\begin{align*}
C(\tilde I,Y) &= E[ I W Y ] - \tilde f \bar Y \\
&= \tilde f_1 \bar Y + \tilde f_0 (1-\bar Y) - \tilde f \bar Y \\
&= \tilde f_0 (1-\bar Y) + \tilde \Delta (1-\bar Y) \bar Y \\
&= (1- \bar Y) (\tilde f_0 + \tilde \Delta \bar Y)
\end{align*}
Combining yields
$$
\begin{aligned}
\rho_{\tilde I (X),PZ} &= \rho_{\tilde I(X),Y} \times \frac{- \tilde \Delta \bar Y (FP \cdot (1-\bar Y) + FN \cdot \bar Y) }{(1-\bar Y) (\tilde f_0 + \tilde \Delta \bar Y)} \times \sqrt{ \frac{(1-\bar Y)}{2 ( FP (1-\bar Y) + FN \cdot \bar Y)} } \\
&= - \rho_{\tilde I(X), Y} \times \Delta \times \sqrt{\frac{\bar Y}{1-\bar Y}} \frac{\sqrt{FP(1-\bar Y) + FN \cdot \bar Y}}{\tilde f_0 (1-\bar Y) + \tilde f_1 \bar Y} \times \sqrt{\frac{\bar Y}{2}}
\end{aligned}
$$
We can then re-write $\rho_{\tilde I (X),Y} \sigma_Y + \rho_{\tilde I(X),PZ} \sigma_{PZ}$ as
$$
\rho_{\tilde I(X),Y} \sigma_Y \left( 1 - \tilde \Delta \times \frac{\bar Y}{1-\bar Y} \times \frac{FP(1-\bar Y) + FN \cdot \bar Y}{\tilde f_0 (1-\bar Y) + \tilde f_1 \bar Y} \right).
$$
Thus yielding the desired result.

\section{Doubly robust statistical error decomposition derivation}
\label{app:DRderivation}
We start by considering the empirical weighted mean estimator under imperfect testing,
\begin{align*}
&\frac{1}{N} \sum_{j=1}^N \mu(X_j) - \frac{\sum_{j=1}^N I_j W_j \mu(X_j)}{\sum_{j=1}^N I_j W_j} + \frac{\sum_{j=1}^N I_j W_j Y_j^\star}{\sum_{j=1}^N I_j W_j} \\
=& \frac{\sum_{i=1}^N  I_j W_j \left( \left[ Y_j (1-P_j) + (1-Y_j) P_j \right] - \mu (X_j) \right)}{\sum_{j=1}^N  I_j W_j } + \frac{1}{N} \sum_{j=1}^N \mu(X_j
)
\end{align*}
Then
$$
\begin{aligned}
&\frac{E_J \left[ I_J W_J \left( \left[ Y_J (1-P_J) + (1-Y_J) P_J \right] - \mu(X_J) \right) \right]}{E_J [ I_J W_J ] } - E_J[Y_J -\mu (X_J)] \\
= &\frac{E_J \left[ I_J W_J P_J (1-2Y_J) \right]}{E_J [ I_J W_J ] } + \left( \frac{E_J [I_J W_J (Y_J -\mu (X_J))]}{E_J [ I_J W_J ] } - \frac{E_J[Y_J-\mu(X_J)] E_J[I_J W_J]}{E_J[I_J W_J]} \right) \\
\end{aligned}
$$
The term in parentheses can be re-written as
$$
\begin{aligned}
&\frac{E_J [I_J W_J (Y_J-\mu(X_J))]- E_J[(Y_J-\mu(X_J))] E_J[I_J W_J]}{E_J[I_J W_J]} \\
=  &\frac{E_J [I_J W_J (Y_J-\mu(X_J))]- E_J[(Y_J-\mu(X_J))] E_J[I_J W_J]}{\sqrt{V_J(I_J W_J) V_J(Y_J-\mu(X_J))}} \frac{\sqrt{V_J(I_J W_J)}}{E_J[I_J]} \times \sqrt{V_J(Y_J-\mu(X_J))} \\
= &\rho_{\tilde I (X),Y-\mu(X)} \times \frac{\sqrt{V_J(I_J W_J)}}{E_J[I_J]} \times \sigma_{Y-\mu(X)}
\end{aligned}
$$
From prior derivations, we then have the desired result.

\section{Asymptotic derivations}
\label{app:asympderivations}


Here, we take the suggested measures from~\cite{Arevalo2020} which report 87\% sensitivity is a reasonable estimates and~\cite{Cohen2020} which report 97.6\% specificity.  This corresponds to a false negative rate of $13\%$ and false positive rate of $2.4\%$.  To allow for potential uncertainty, we assume two \emph{synthetic} simple random samples.  The first is a simple random sample of true negative cases denoted $S_{FP}$; the second is simple random sample of true positive cases denoted $S_{FN}$.  As $\mu_t$ and $\theta_{t}$ are estimated independently at each time $t$, we can focus on the estimating equations per time point~$t$ separately.

Suppose there is a sequence of finite populations of size $N_{\nu}$ indexed by $\nu$.  Each finite population has $L_\nu$ non-probability samples and  probability samples of fixed size $n$ drawn at equally spaced times $\{ t^\prime_l \}_{l=1}^L$ over the study window~$[0,T]$. Let~$\Delta_\nu$ denote the gap times between consecutive sample times such that as $\nu \to \infty$ we have that $\Delta_\nu \to 0$. Finally, let~$n_{FP}$ and $n_{FN}$ denote the sample sizes of $S_{FP}$ and $S_{FN}$ respectively.  For simplicity, dependency on $\nu$ is suppressed and the limiting process is represented by $N \to \infty$. The asymptotic argument below relies on regularity conditions C1-C6 from~\cite{Chen2019}, smoothness of the population-level distribution of the time-varying covariates~$\{ X_{j,t} \}$ as a function of $t > 0$, and smoothness of the propensity function~$\pi(x; \theta)$ as a function of $x \in \mathbb{R}^p$.

For each~$\nu$ and $t$, the estimates of $\eta = (\mu_t, \theta_t, FP, FN)$ are given by:
$$
\Phi (\eta) = \left (
\begin{array}{c}
\frac{1}{N} \sum_{j=1}^N I_{j,t} \frac{Y_{j,t} - FP - (1-FP-FN) \cdot \mu_t}{\pi_{j,t}} \\
\frac{1}{N \times \sum_{l=1}^L K_{h} \left(|t - t_l^\prime| \right)} \sum_{l=1}^L K_{h} \left(|t - t_l^\prime| \right) \left[ \sum_{j=1}^N I_{j,t^\prime} X_{j,t^\prime_l} - \sum_{j = 1}^N \tilde I_{j,t_l^\prime} \tilde W_{j,t_l^\prime}  \pi_{j,t_l^\prime} X_{j,t_l^\prime} \right] \\
\frac{1}{n_{FP}} \sum_{i=1}^{n_{FP}} \frac{Z_j}{FP} - \frac{1-Z_j}{1-FP} \\
\frac{1}{n_{FN}} \sum_{i=1}^{n_{FN}} \frac{\tilde Z_j}{FN} - \frac{1-\tilde Z_j}{1-FN} \\
\end{array}
\right ) = {\bf 0}
$$
where $\pi_{j,t^\prime} = \pi (X_{j,t^\prime}; \theta_t)$, $Z_j$ is an indicator of a false positive in the sample $S_{FP}$  and $\tilde Z_j$ is an indicator of a false negative in the sample $S_{FN}$.  We assume the $|S_{FP}|^{-1} \sum_{j \in S_{FP}} Z_j = 0.024$ and $|S_{FP}|^{-1} \sum_{j \in S_{FP}} Z_j = 0.30$.  Sample sizes, denoted $n_{FP}$ and $n_{FN}$ respectively, are chosen to achieve desired estimator variance, i.e., $\frac{0.024 \cdot (0.976)}{n_{FP}} = \sigma^2_{FP}$ and $\frac{0.13 \cdot 0.87}{n_{FN}} = \sigma^2_{FN}$. In this paper, we set $\sigma^2_{FP} = 0.02^2$ and $\sigma_{FN}^2 = 0.05^2$ which means $n_{FP} \approx 59$ and $n_{FN} \approx 45$ respectively.

To prove consistency, first consider the setting where a single random sample of size $n$ is observed at time $t$.  Let $\tilde \Phi (\eta)$ denote the corresponding estimating equations, which is the same as $\Phi(\eta)$ except for the second component which is now given by
$$
\frac{1}{N} \left[ \sum_{j=1}^N I_{j,t} X_{j,t} - \sum_{j = 1}^N \tilde I_{j,t} \tilde W_{j,t}  \pi_{j,t} X_{j,t} \right].
$$
Then under joint randomization of propensity score model and sampling designs, we have $E [ \tilde \Phi (\eta_0) ] = {\bf 0}$.  Consistency then follows by arguments in Section 3.2 of Tsiatis (2006). Under conditions C1-C6 from~\cite{Chen2019} and as $n$ increases, we have $\tilde \Phi (\hat \eta^\prime) = 0$ and $\tilde \Phi(\eta_0) = O_p ( n^{-1/2} )$ where $\hat \eta^\prime$ here refers to the solution to $\tilde \Phi (\eta)$.

To prove consistency under the sequence of sampling regimes, we need to show that $E[ \Phi (\eta_0) ] \to 0$ as $N \to \infty$.  First, we study the difference in the 2nd term:
\begin{align*}
&\frac{1}{N}  E \left[ \sum_{j = 1}^N \tilde I_{j,t} \tilde W_{j,t}  \pi_{j,t} X_{j,t} - \sum_{l=1}^L  \sum_{j = 1}^N K_{h,l} \tilde I_{j,t^\prime_l} \tilde W_{j,t^\prime_l}  \pi_{j,t^\prime_l} X_{j,t^\prime_l} \right] \\
= &\frac{1}{N} \sum_{j = 1}^N \left[ \pi_{j,t} X_{j,t} - \sum_{l=1}^L K_{h,l} \pi_{j,t^\prime_l} X_{j,t^\prime_l} \right], \\
= &\frac{1}{N} \sum_{j = 1}^N \left[ \sum_{l=1}^L K_{h,l} \pi_{j,t} X_{j,t} \pm \sum_{l=1}^L K_{h,l} \pi_{j,t_l^\prime} X_{j,t}  - \sum_{l=1}^L K_{h, l} \pi_{j,t^\prime_l} X_{j,t^\prime_l} \right], \\
= &\frac{1}{N} \sum_{j = 1}^N \left[ \left( \sum_{l=1}^L K_{h,l}  \left( \pi (X_{j,t}; \theta) - \pi (X_{j,t^\prime_l}; \theta) \right) \right) X_{j,t} + \sum_{l=1}^L K_{h,l} \pi_{j,t^\prime_l} (X_{j,t} - X_{j,t^\prime_l} ) \right],
\end{align*}
where~$K_{h,l} := K_h(|t - t_l^\prime|) / \sum_{l=1}^L K_{h} \left(|t - t_l^\prime| \right)$ is the normalized kernel for shorthand. The difference in the 1st term is similarly defined.

Let~$X_{j,t,l}$ denote the $l$th component of the vector $X_{j,t}$.  Then consider the population-level distribution $\{ X_{j,t,l} \}_{j=1}^N$.  We make the additional assumption that
$$
\lim_{\epsilon \to 0} \lim_{N \to \infty} \frac{1}{N} \sum_{j=1}^N X_{j,t+\epsilon, l}  = \lim_{N \to \infty} \frac{1}{N} \sum_{j=1}^N X_{j,t, l}.
$$
That is, the limiting distributions are continuous as a function of $t$. If $\pi(x; \theta)$ is a continuous function in~$x$ and under suitable conditions on the kernel density~$K_{h_\nu} (\cdot)$ with the bandwidth $h_\nu \to \infty$ such that $\sum_{l=1}^L K_{h_{\nu}} \left( | t - t_l^\prime | \right) \to \infty$ as $\nu \to \infty$, we have
$$
\left | \sum_{l=1}^L K_{h, l} \pi (X_{j, t_l^\prime}; \theta) - \pi (X_{j, t}; \theta) \right|  \to 0
$$
Combining this, $E \left[ \Phi(\eta_0)  - \tilde \Phi (\eta_0) \right] \to 0$ which implies consistency as desired.

Then by a first-order Taylor expansion we have $\hat \eta - \eta_0 = \left[ \phi (\eta) \right]^{-1} \Phi(\eta_0) + o_p (\bar n^{-1/2})$ where $\phi(\eta) = \partial \Phi (\eta)/\partial \eta$ and $\bar n = n \sum_{l=1}^L K_{h,l}$.  To calculate
$$
\frac{\partial \Phi (\eta)}{\partial \mu_t} =
\left (
\begin{array}{c}
-\frac{(1-FP-FN)}{N} \sum_{j=1}^N \frac{I_{j,t}}{\pi_{j,t}} \\
0 \\
0 \\
0
\end{array}
\right )
$$

$$
\frac{\partial \Phi (\eta)}{\partial \theta_t} =
\left (
\begin{array}{c}
-\frac{1}{N} \sum_{j=1}^N I_{j,t} (Y_{j,t} - FP - (1-FP-FN) \cdot \mu_t) \frac{1-\pi_{j,t}}{\pi_{j,t}} X_{j,t}^\top \\
- \frac{1}{N} \sum_{t^\prime = 1}^T K_{t,t^\prime} \sum_{j = 1}^N \tilde I_{j,t^\prime} \tilde W_{j,t^\prime}  \pi_{j,t^\prime} (1-\pi_{j,t^\prime}) X_{j,t^\prime} X_{j,t^\prime}^\top\\
0 \\
0
\end{array}
\right )
$$

$$
\frac{\partial \Phi (\eta)}{\partial FP} =
\left (
\begin{array}{c}
-\frac{1}{N} \sum_{j=1}^N I_{j,t} \frac{(1-\mu_t)}{\pi_{j,t}} \\
0\\
-\frac{1}{n_{FP}} \sum_{j=1}^{n_{FP}} \left( \frac{Z_j}{FP^2} + \frac{1-Z_j}{(1-FP)^2} \right)  \\
0
\end{array}
\right )
$$

$$
\frac{\partial \Phi (\eta)}{\partial FN} =
\left (
\begin{array}{c}
- \frac{1}{N} \sum_{j=1}^N I_{j,t} \frac{\mu_t}{\pi_{j,t}} \\
0\\
0 \\
-\frac{1}{n_{FN}}  \sum_{j =1}^{n_{FN}} \left( \frac{\tilde Z_j}{FN^2} + \frac{1-\tilde Z_j}{(1-FN)^2} \right)
\end{array}
\right )
$$

We are interested in computing $- E \left[ \frac{\partial \Phi (\eta)}{\partial \mu_t} \right]^{-1}$ where the expectation is with respect to the random indicators. First, $- E \left[ \frac{\partial \Phi (\eta)}{\partial \mu_t} \right]$ is equal to
$$
\left (
\begin{array}{c c c c}
1-FP-FN & \frac{1}{N} \sum_{j=1}^N \zeta_{t,j} (1-\pi_{j,t}) X_{j,t}^\top & 1 - \mu_t & \mu_t \\
0 & \frac{1}{N} \sum_{t^\prime=1}^T K_{t,t^\prime} \sum_{j=1}^N \pi_{j,t} (1-\pi_{j,t}) X_{j,t} X_{j,t}^\top & 0 & 0\\
0 & 0 & \left( FP (1-FP) \right)^{-1}  & 0\\
0 & 0 & 0 & \left( FN (1-FN) \right)^{-1}
\end{array}
\right )
$$
where $\zeta_{t,j} = (Y_{j,t} - FP - (1-FP-FN) \cdot \mu_t)$. This matrix can be written as a diagonal matrix equal to the diagonal of $\partial \Phi_n (\eta)/\partial \eta$ and then a rank-three matrix that is zero except for the first row on the off-diagonal. Let $A + B$ denote the sum broken into these 2 components.  Then Woodbury matrix identity gives us
$$
(A+B)^{-1} = A^{-1} - A^{-1} B A^{-1}
$$
In particular, the resulting inverse is of the form
$$
\left (
\begin{array}{c c c c}
\frac{1}{1-FP-FN} & b_1 & b_2  & b_3 \\
0 & \left[\frac{1}{N} \sum_{t^\prime=1}^T K_{t,t^\prime} \sum_{j=1}^N \pi_{j,t} (1-\pi_{j,t}) X_{j,t} X_{j,t}^\top\right]^{-1} & 0 & 0\\
0 & 0 & FP (1-FP)   & 0\\
0 & 0 & 0 & FN (1-FN)
\end{array}
\right)
$$
where $b_1, b_2,$ and $b_3$ can be calculated using the Woodbury identity.

Next, note that $\text{Var}(\Phi(\eta))$ can be decomposed into $A_{1,t} + A_{2,t} + A_{3,t} + A_{4,t}$

$$
A_{1,t} = \frac{1}{N} \left( \begin{array}{c}
\sum_{j=1}^N \frac{I_{j,t} (Y_{j,t} - FP - (1-FP-FN) \mu_t)}{\pi_{j,t}} \\
\sum_{t^\prime = 1}^T K_{t,t^\prime} \left[ \sum_{j=1}^N I_{j,t^\prime} X_{j,t^\prime} - \pi_{j,t^\prime} X_{j,t^\prime} \right] \\
0 \\
0
\end{array}
\right)
$$
and
$$
A_{2,t} = \frac{1}{N} \left( \begin{array}{c}
0 \\
\sum_{t^\prime = 1}^T K_{t,t^\prime} \left[ \sum_{j=1}^N \pi_{j,t^\prime} X_{j,t^\prime} - \sum_{j=1}^N \tilde I_{j,t^\prime} \pi_{j,t^\prime} X_{j,t} \right] \\
0 \\
0
\end{array}
\right),
$$
and
$$
A_{3,t} = \left( \begin{array}{c}
0 \\
0 \\
\frac{1}{n_{FP}} \sum_{j=1}^{n_{FP}} \frac{Z_j}{FP} - \frac{1-Z_j}{1-FP} \\
0
\end{array}
\right),
$$
and
$$
A_{4,t} = \left( \begin{array}{c}
0 \\
0 \\
0 \\
\frac{1}{n_{FN}} \sum_{j=1}^{n_{FN}} \frac{\tilde Z_j}{FN} - \frac{1-\tilde Z_j}{1-FN} \\
\end{array}
\right),
$$
By independence of these components, the variance can be calculated by summation over variance of each term individually yielding. First, $V_{1,t}$
{
\tiny
$$
 N^{-2} \sum_{i=1}^N \left(
\begin{array}{c c c c}
\frac{(1-\pi_{j,t})}{\pi_{j,t}} (Y_{j,t} - FP - (1-FP-FN) \mu_{j,t})^2 & \sum_{t^\prime} K_{t,t^\prime} (1-\pi_{j,t^\prime}) \zeta_{t^\prime, j} X_{j,t^\prime}^\top & 0 & 0 \\
\sum_{t^\prime} K_{t,t^\prime} (1-\pi_{j,t^\prime}) \zeta_{t^\prime, j} X_{j,t}&
\sum_{t^\prime} K_{t,t^\prime} \pi_{j,t^\prime} (1-\pi_{j,t^\prime}) X_{j,t} X_{j,t}^\top
& 0 & 0 \\
0 & 0 & 0 & 0 \\
0 & 0 & 0 & 0 \\
\end{array}
\right)
$$
}
and $V_{2,t}$ is a block-diagonal matrix with only one non-zero block that is equal to $D_{j,t} = N^{-2} \sum_{t^\prime} K_{t,t^\prime}^2 V_p\left( \sum_{i=1}^n \tilde W_{i,t^\prime} \tilde I_{i,t^\prime}  \pi_{j,t^\prime} X_{j,t^\prime} \right)$ which is the design-based variance-covariance matrix under the probability sampling design; and
$$
V_{3,t} = \left( \begin{array}{c}
0 \\
0 \\
\frac{1}{n_{FP}} \frac{1}{FP(1-FP)} \\
0
\end{array}
\right),
$$
and
$$
V_{4,t} = \left( \begin{array}{c}
0 \\
0 \\
0 \\
\frac{1}{n_{FN}} \frac{1}{FN(1-FN)}
\end{array}
\right),
$$

The asymptotic variance for the IPW estimator is then the first diagonal element of the matrix $E \left[ \phi_n (\eta_0) \right]^{-1} \times \left[ V_{1,t} + V_{2,t} + V_{3,t} + V_{4,t} \right] E \left[ \phi_n (\eta_0) \right]^{-1}$.
\section{IRLS}
\label{app:irls}
First derivative with respect to $\theta$
$$
\nabla E(\theta) = \sum_{t^\prime=1}^T K_h(|t^\prime - t|) \left[ \sum_{j=1}^N I_{j,t^\prime} X_{j,t^\prime} + \sum_{j=1}^N \tilde W_{j,t^\prime} \tilde I_{j,t^\prime} \pi_t (X_{j,t^\prime}; \theta) X_{j,t^\prime} \right]
$$
Second derivative with respect to $\theta$
$$
H(\theta) = \sum_{t^\prime=1}^T K_h(|t^\prime - t|) \sum_{j=1}^N \tilde W_{j,t^\prime} \tilde I_{j,t^\prime} \pi_t (X_{j,t^\prime}; \theta) \left( 1- \pi_t (X_{j,t^\prime}; \theta) \right) X_{j,t^\prime} X_{j,t^\prime}^\top
$$
Then the Fisher scoring method sets
$$
\theta^{(t+1)} = \theta^{(t)} + H \left(\theta^{(t)} \right)^{-1} \nabla E \left(\theta^{(t)} \right)
$$

\section{Indiana COVID-19 Analysis: Additional details}
\label{app:in_add_details}

Table~\ref{tab:comparison} demonstrates minimal bias in the Delphi’s COVID-19 Trends and Impact Survey (CTIS) with respect to symptom distributions.

 \begin{table}[!th]
 \centering
 \begin{tabular}{c | c c c | c}
 & \multicolumn{3}{c}{CTIS} & Random \\ \cline{2-4}
 & Lower CI & Estimate & Upper CI & Estimate \\ \hline
 Fever &  0.073 & 0.097 & 0.012 & 0.018 \\
 Cough &  0.140 & 0.150 & 0.159 & 0.149 \\
 Shortness & 0.055 & 0.062 & 0.068 & 0.062 \\ \hline
 \end{tabular}
 \caption{Comparison of CTIS and random sample restricting CTIS to surveys collected from April 25--29th, 2020.}
 \label{tab:comparison}
 \end{table}

Next, we describe the two mean-imputation methods.
\subsection{Model 1}

\begin{figure}[!th]
\centering
\begin{subfigure}{.5\textwidth}
 \centering
 \includegraphics[width=.9\linewidth]{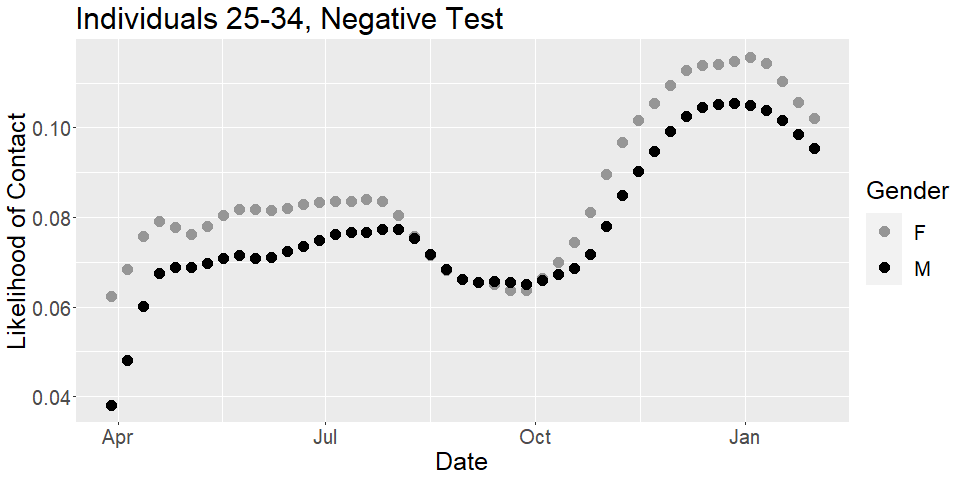}
 \caption{Contact Likelihood Given Negative Test}
 \label{fig:contactlik1}
\end{subfigure}%
\begin{subfigure}{.5\textwidth}
 \centering
\includegraphics[width=.9\linewidth]{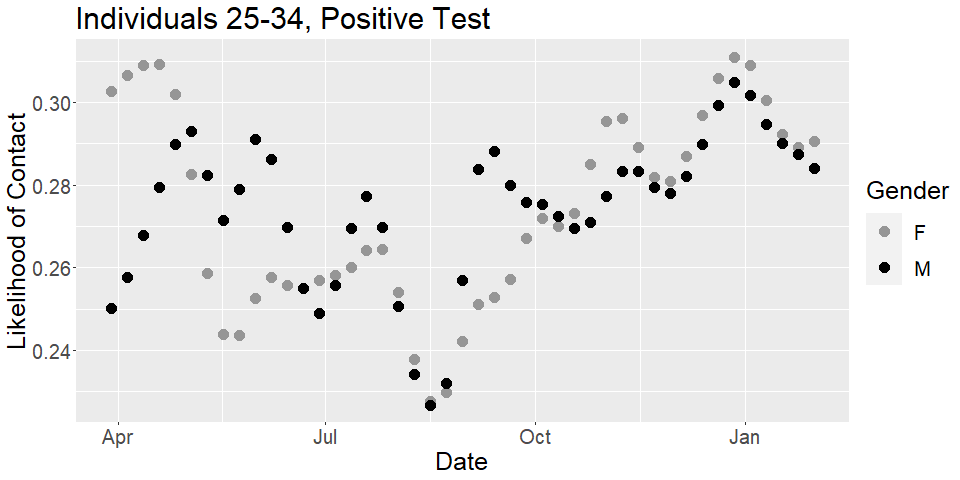}
 \caption{Contact Likelihood Given Positive Test}
 \label{fig:contactlik2}
\end{subfigure}
\caption{Likelihood of COVID-19 contact}
\label{fig:contactlik}
\end{figure}

First, we use the pseudo-likelihood to compute the likelihood of contact given age, gender, and test result.  Figure~\ref{fig:contactlik1} and~\ref{fig:contactlik2} plots the likelihood for 25-34 years old given a negative test and positive test respectfully.  We see that the likelihood of COVID-19 is much higher for those with a positive test and the likelihood is time-varying.

\begin{figure}[!th]
\centering
\begin{subfigure}{.5\textwidth}
 \centering
 \includegraphics[width=.9\linewidth]{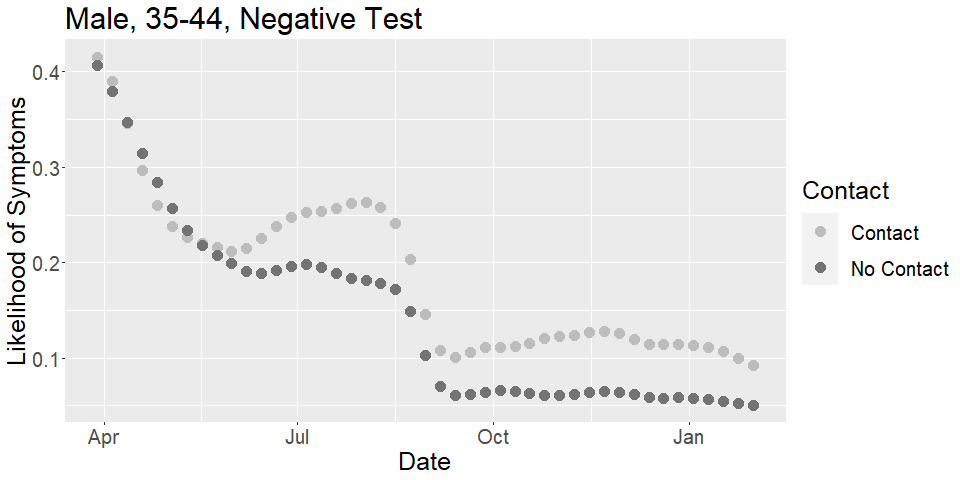}
 \caption{Symptom Likelihood Given Negative Test}
 \label{fig:symptomlik1}
\end{subfigure}%
\begin{subfigure}{.5\textwidth}
 \centering
\includegraphics[width=.9\linewidth]{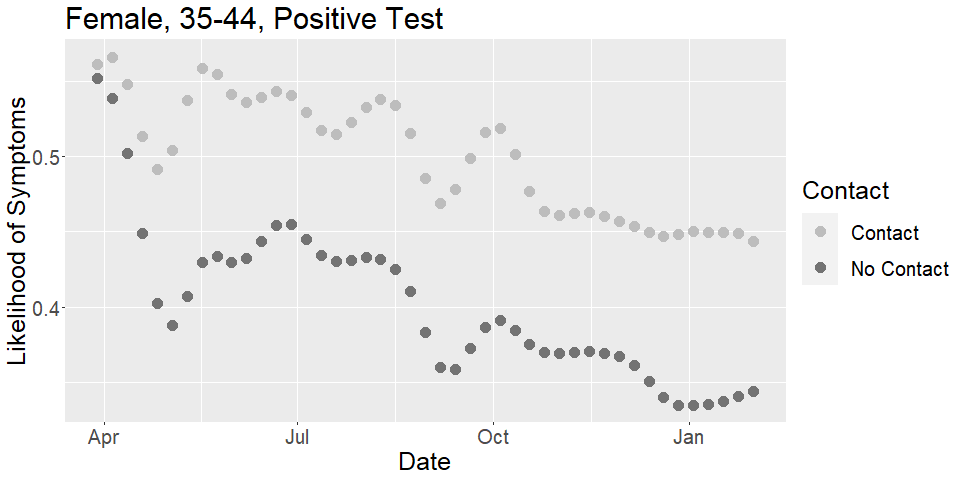}
 \caption{Symptom Likelihood Given Positive Test}
 \label{fig:symptomlik2}
\end{subfigure}
\caption{Likelihood of reported COVID-19 symptoms}
\label{fig:symptomlik}
\end{figure}

We then use the pseudo-likelihood to compute the likelihood of fever given age, gender, COVID-19 contact status and test result.  Figure~\ref{fig:symptomlik1} and~\ref{fig:symptomlik2} plots the likelihood for 35-44 year olds given a negative test and positive test respectfully.  We see that the likelihood of fever depends heavily on whether they had a COVID-19 contact and their test results. Again, the likelihood is time-varying for most configurations.

\begin{figure}[!th]
\centering
\begin{subfigure}{.5\textwidth}
 \centering
 \includegraphics[width=.9\linewidth]{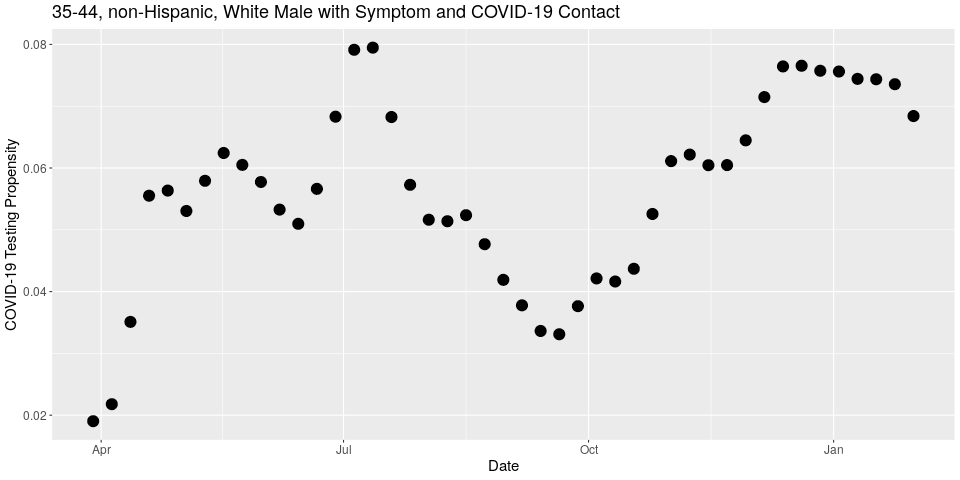}
 \caption{Testing Likelihood Given Fever and \\COVID-19 Contact}
 \label{fig:testinglik1}
\end{subfigure}%
\begin{subfigure}{.5\textwidth}
 \centering
\includegraphics[width=.9\linewidth]{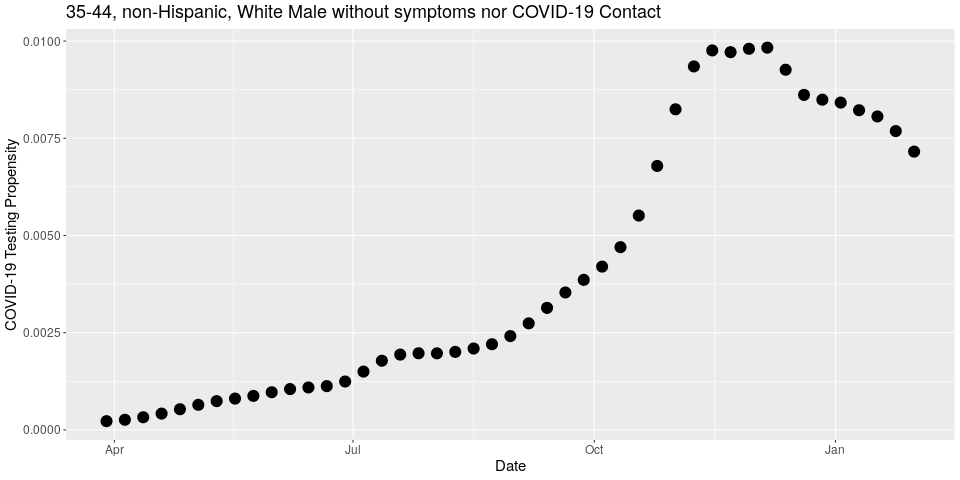}
 \caption{Testing Likelihood Given no Fever nor \\  COVID-19 Contact}
 \label{fig:testinglik2}
\end{subfigure}
\caption{Testing Likelihood Given no Fever nor COVID-19 Contact}
\label{fig:testinglik}
\end{figure}

Based on mean-imputation of COVID-19 contact and fever indicators, we can use the proposed pseudo-likelihood approach to compute the likelihood of getting tested for COVID-19 given age, ethnicity, race, gender, fever status, and COVID-19 contact indicator.  Figure~\ref{fig:testinglik1} and~\ref{fig:testinglik2} plots the likelihood of testing given fever and COVID-19 contact and no fever nor COVID-19 contact respectfully.  We see that the likelihood is time-varying and depends heavily on both fever and COVID-19 contact indicators.

\subsection{Model 2 (Hospitalization)}

\begin{figure}[!th]
\centering
\begin{subfigure}{.5\textwidth}
 \centering
 \includegraphics[width=.9\linewidth]{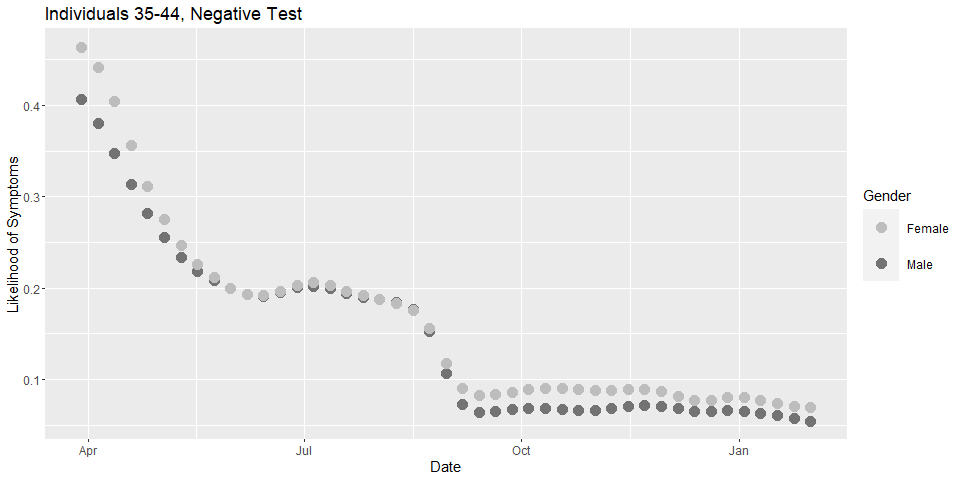}
 \caption{Symptom Likelihood Given Negative Test}
 \label{fig:symptomlik1_model2}
\end{subfigure}%
\begin{subfigure}{.5\textwidth}
 \centering
\includegraphics[width=.9\linewidth]{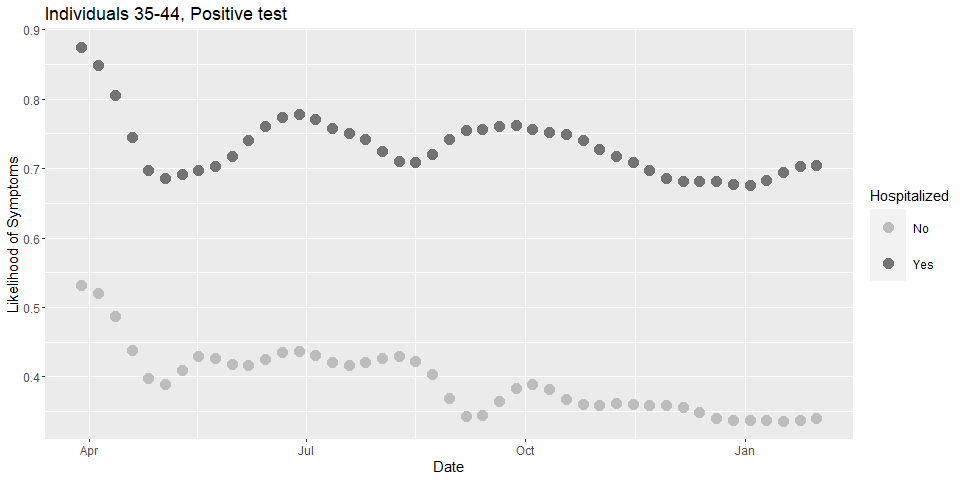}
 \caption{Symptom Likelihood Given Positive Test}
 \label{fig:symptomlik2_model2}
\end{subfigure}
\caption{Likelihood of COVID-19 contact}
\label{fig:symptomlik_model2}
\end{figure}

This model uses hospitalization records to try and improve our mean-imputation strategy.  First, we use the pseudo-likelihood to compute the likelihood of fever given age, gender, and test result.  The model only depends on hospitalization status for positive tests.  Figure~\ref{fig:symptomlik1_model2} and~\ref{fig:symptomlik2_model2} plots the likelihood for 35-44 years old given a negative test and positive test respectfully.  We see that the likelihood of fever is time-varying.  For a negative test, the likelihood of ever fever is very high in early April, which accounts for testing restrictions. We see that hospitalization significantly increases the risk of fever given a positive test.

First, we computed the probability of fever given COVID-19 positive test and hospitalization as well as COVID-19 positive test and no hospitalization.  We then computed the likelihood of symptom given a COVID-19 positive test by weighting these two likelihoods by the fraction of COVID-19 positive hospitalizations to COVID-19 total positive cases per week.  Mean imputation of fever status for COVID-19 negative tests was performed based on the above model without use of hospitalization data.

Based on mean-imputation of COVID-19 fever using hospitalization records indicators, we can use the proposed pseudo-likelihood approach to compute the likelihood of getting tested for COVID-19 given age, ethnicity, race, gender, and fever status.  Figure~\ref{fig:testinglik1_mainpaper} and~\ref{fig:testinglik2_mainpaper} in Section~\ref{section:tvipw} of the manuscript plots the likelihood of testing given fever and no fever respectfully for all age ranges.  We see that the likelihood is time-varying and depends heavily on both fever status.

\subsubsection{Propensity plots}
\label{section:propplots}

\begin{figure}[!th]
\centering
\begin{subfigure}{.5\textwidth}
 \centering
 \includegraphics[width=.9\linewidth]{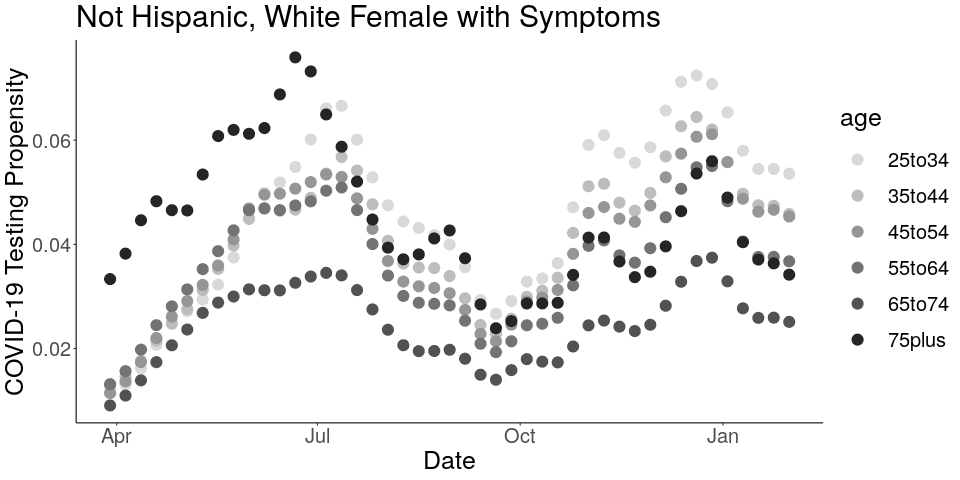}
 \caption{Symptom Likelihood Given Negative Test}
\end{subfigure}%
\begin{subfigure}{.5\textwidth}
 \centering
\includegraphics[width=.9\linewidth]{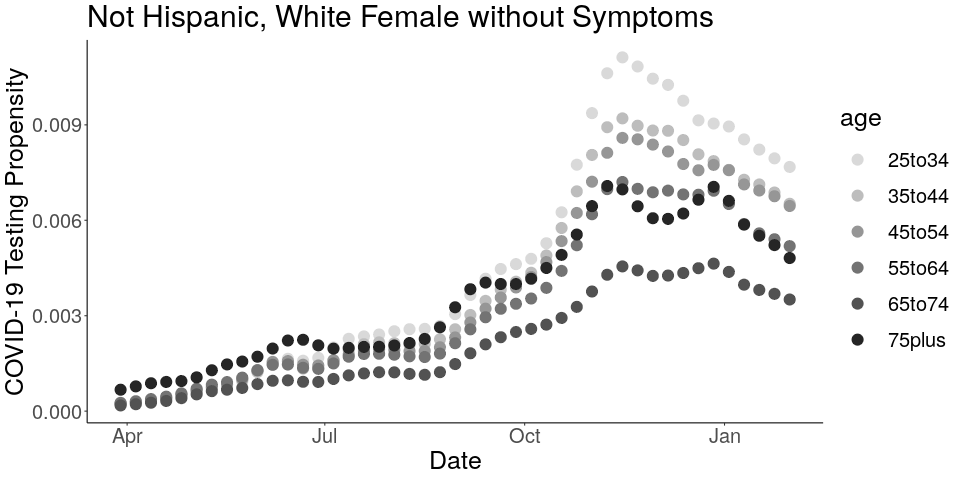}
 \caption{Symptom Likelihood Given Positive Test}
\end{subfigure}
\caption{Likelihood of COVID-19 contact for Non-Hispanic, White Females}
\label{fig:nonh-white-female}
\end{figure}

\begin{figure}[!th]
\centering
\begin{subfigure}{.5\textwidth}
 \centering
 \includegraphics[width=.9\linewidth]{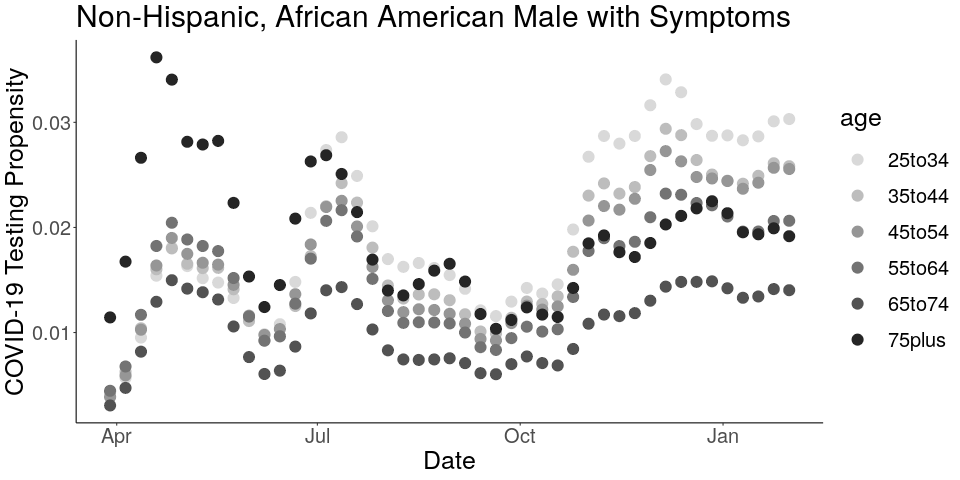}
 \caption{Symptom Likelihood Given Negative Test}
\end{subfigure}%
\begin{subfigure}{.5\textwidth}
 \centering
\includegraphics[width=.9\linewidth]{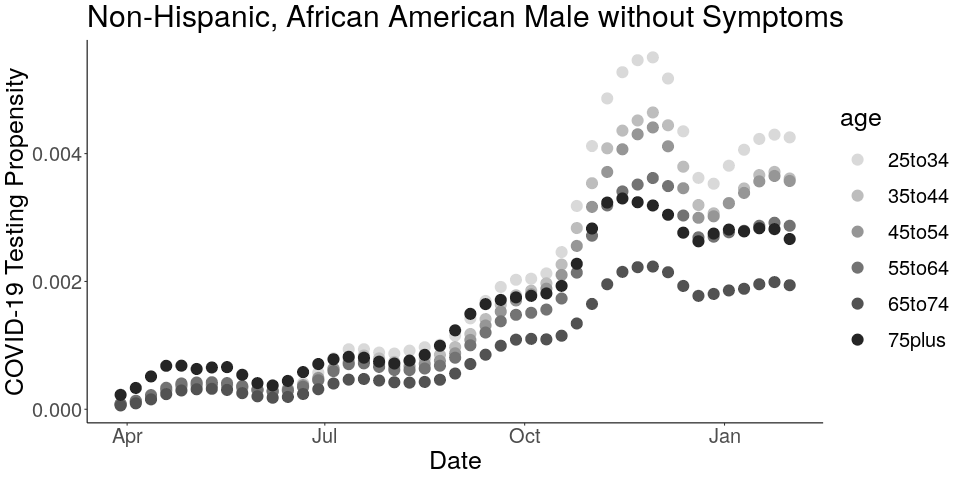}
 \caption{Symptom Likelihood Given Positive Test}
\end{subfigure}
\caption{Likelihood of COVID-19 contact for Non-Hispanic, African American Males}
\label{fig:nonh-aa-male}
\end{figure}

\begin{figure}[!th]
\centering
\begin{subfigure}{.5\textwidth}
 \centering
 \includegraphics[width=.9\linewidth]{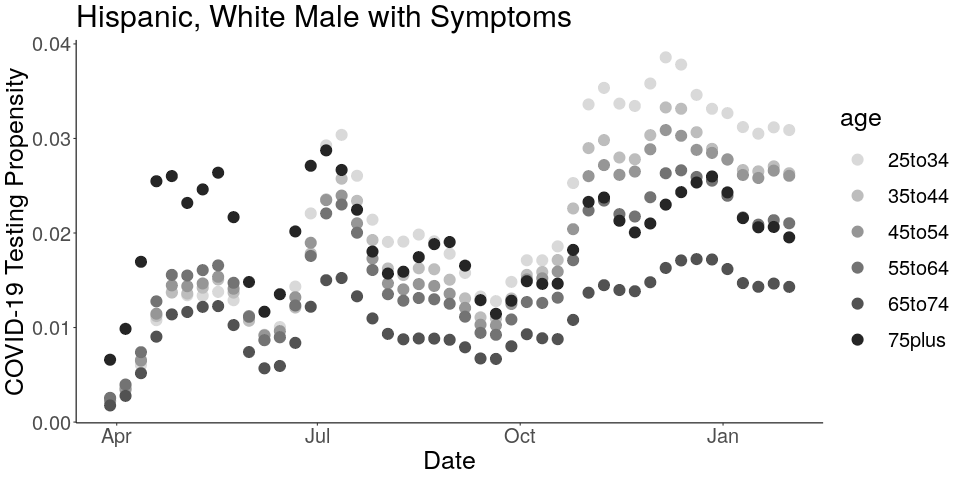}
 \caption{Symptom Likelihood Given Negative Test}
\end{subfigure}%
\begin{subfigure}{.5\textwidth}
 \centering
\includegraphics[width=.9\linewidth]{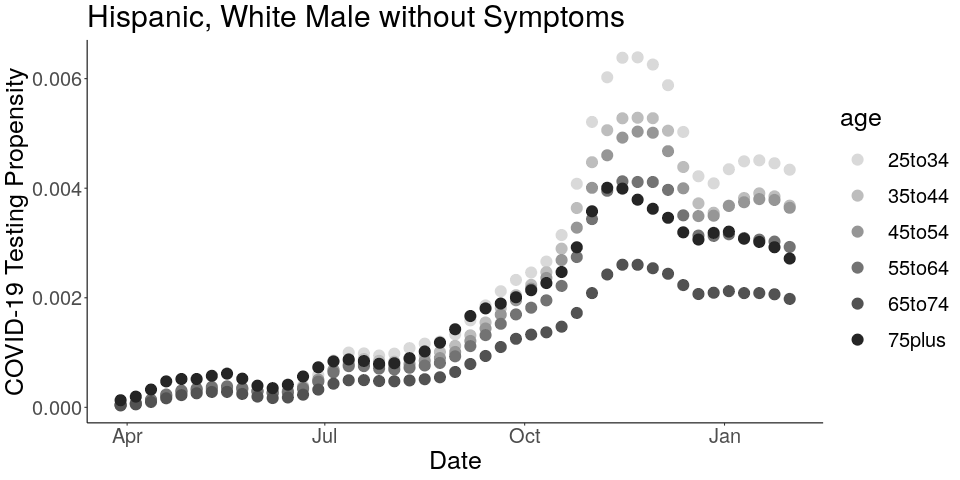}
 \caption{Symptom Likelihood Given Positive Test}
\end{subfigure}
\caption{Likelihood of COVID-19 contact for Hispanic, White Males}
\label{fig:h-white-male}
\end{figure}

\begin{figure}[!th]
\centering
\begin{subfigure}{.5\textwidth}
 \centering
 \includegraphics[width=.9\linewidth]{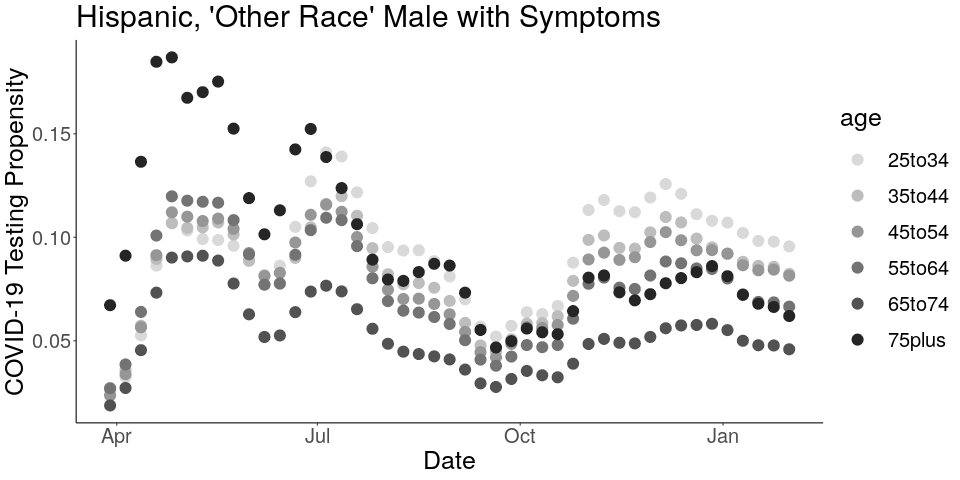}
 \caption{Symptom Likelihood Given Negative Test}
\end{subfigure}%
\begin{subfigure}{.5\textwidth}
 \centering
\includegraphics[width=.9\linewidth]{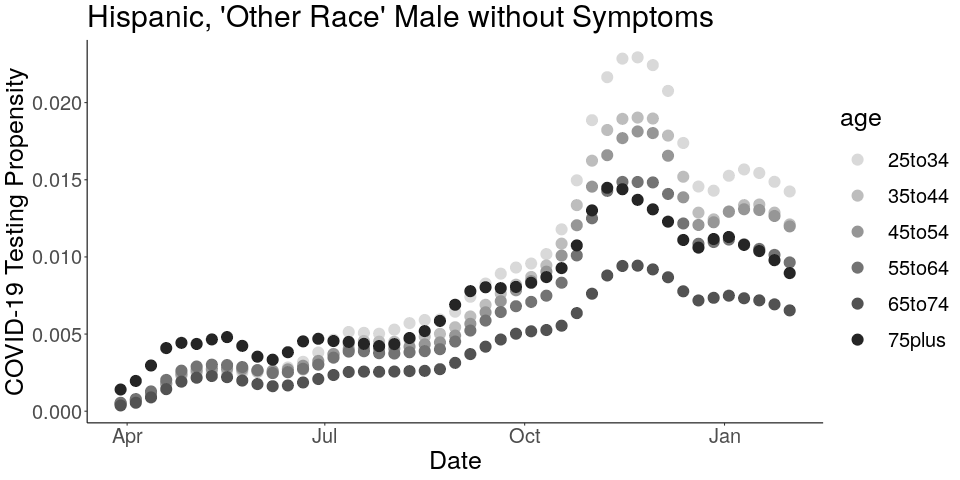}
 \caption{Symptom Likelihood Given Positive Test}
\end{subfigure}
\caption{Likelihood of COVID-19 contact for Hispanic, Males who select `Some Other Race'.}
\label{fig:h-other-male}
\end{figure}

Here we present the testing propensity for four strata: (1) non-Hispanic, white female, (2) non-Hispanic, African American male, (3) Hispanic, White male, and (4) Hispanic male who selects `Some other Race'.   Strata (1) presented in Figure~\ref{fig:nonh-white-female} demonstrates a small gender difference in testing propensities.  Strata (2) and (3) presented in Figures~\ref{fig:nonh-aa-male} and~\ref{fig:h-white-male} demonstrates a lower rate of testing among African Americans and Hispanic white men compared to non-Hispanic, white men respectively (approximately 2 times lower). Strata (4) presented in Figure~\ref{fig:h-other-male} demonstrates a much higher testing rate for Hispanic men who select ``Some Other Race''.

\newpage
\subsubsection{Confidence Intervals}

Figure~\ref{fig:air_cis} in the supplementary materials presents the confidence intervals per time point for the IPW2 estimator.  The confidence interval length decreases substantially over time, reflecting the increased testing capacity.  Due to the number of surveys per week, under correct model specification there is minimal uncertainty in the parameter estimates.  As the number of tests per week increases to over one hundred thousand, there is minimal uncertainty in the active infection rate estimates. This points to the importance of the statistical decomposition~\eqref{eq:statdecomp2} and the discussion in Section~\ref{section:IPWerrordecomp}.

\begin{figure}[!th]
 \centering
 \includegraphics[width=.6\linewidth]{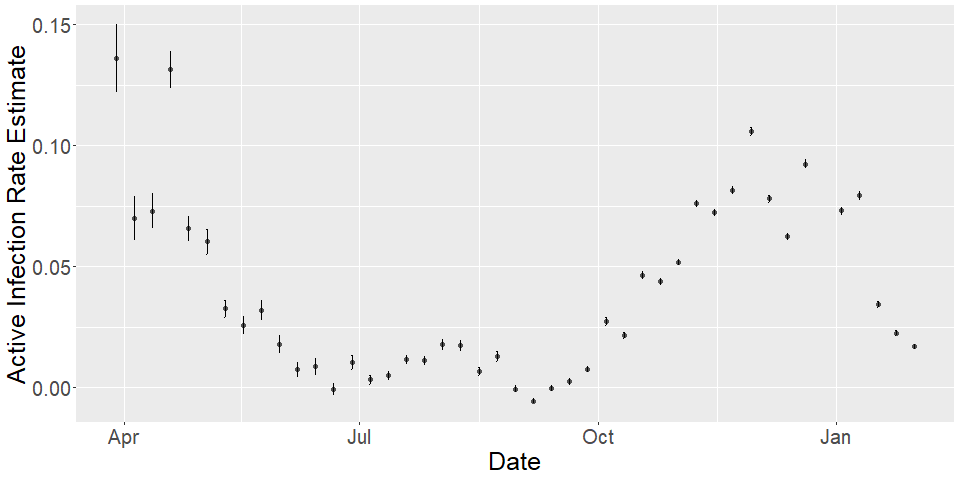}
 \caption{IPW2 estimate with confidence intervals}
 \label{fig:air_cis}
\end{figure}


\section{Model-based: Prior specification}
\label{app:prior_modelbased}

For simplicity, we list the priors used in STAN below:
\begin{itemize}
  \item \code{beta $\sim$ Normal(1.5, 1) T[0,];}
  \item \code{gamma $\sim$ normal(0.3, 0.5) T[0,];}
  \item \code{sigma $\sim$ normal(0.4, 0.5) T[0,];}
  \item \code{phi\_inv $\sim$ exponential(5);}
  \item \code{i0 $\sim$ normal(0, 10);}
  \item \code{e0 $\sim$ normal(0, 10);}
  \item \code{eta $\sim$ normal(1, 1) T[0,];}
  \item \code{eta\_two $\sim$ normal(1, 1) T[0,];}
  \item \code{eta\_three $\sim$ normal(1, 1) T[0,];}
  \item \code{nu $\sim$ exponential(1./5);}
  \item \code{nu\_two $\sim$ exponential(1./5);}
  \item \code{nu\_three $\sim$ exponential(1./5);}
  \item \code{xi\_raw $\sim$ beta(1, 1);}
  \item \code{phi = 1/phi\_inv;}
  \item \code{xi = xi\_raw + 0.5;}
\end{itemize}
where \code{beta, eta, eta\_two, eta\_three} refer to the four values that form the time-varying parameter $\beta_t$ when combined with \code{xi, eta, eta\_two, eta\_three} in the SEIR model as described in Section~\ref{section:modelbased}. Following~\cite{Song2020}, we employ Runge-Kutta (RK4) approximations for discretization.  Due to the low number of deaths per strata, i.e., based on Race, Age, Sex and Ethnicity, we employ a kernel smoothing of the new infections per age strata into these sub-strata using relative number of observed deaths.  This ensured a suitably complex model that fits the observed death data well while generating strata-specific active infection rates that can be used in the doubly-robust estimation procedure.

\begin{figure}[!th]
 \centering
 \includegraphics[width=.6\linewidth]{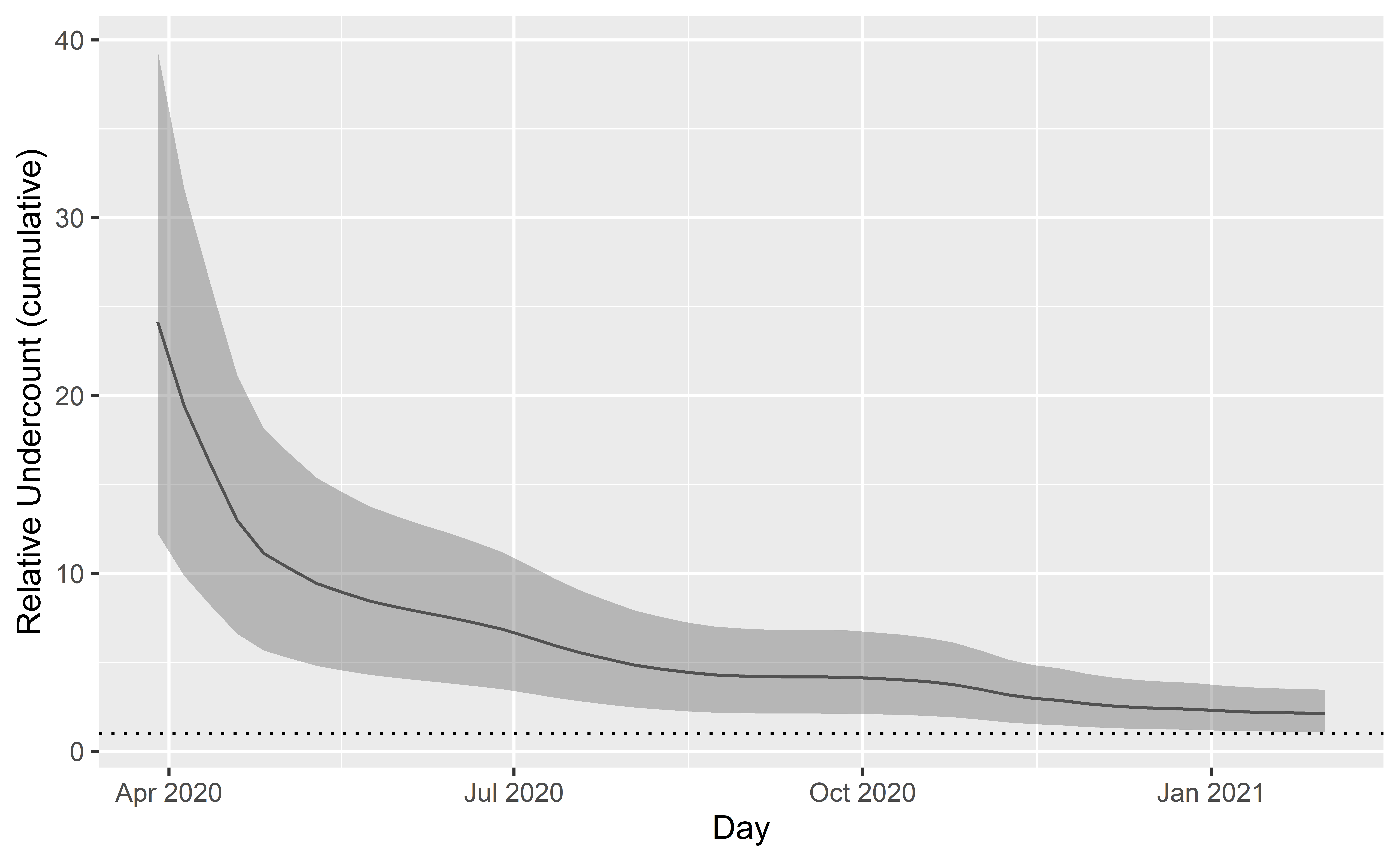}
 \caption{Cumulative undercount based on SEIR model}
 \label{fig:undercounting}
\end{figure}

\begin{figure}[!th]
 \centering
 \includegraphics[width=.6\linewidth]{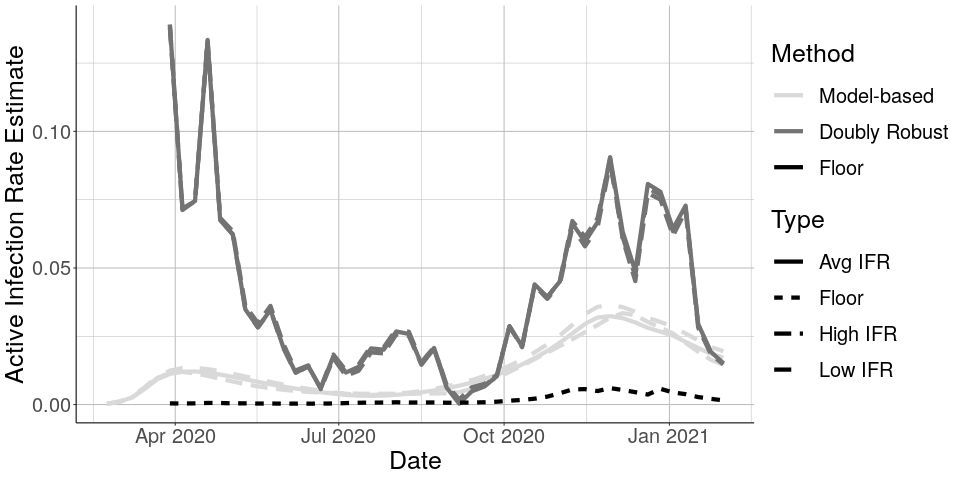}
 \caption{Model-based and Doubly Robust active infection rate estimates under~\cite{Ironse2103272118} marginal IFR estimate of $0.84$\%, under a 10\% higher marginal IFR of $0.924$\%, and under a 10\% lower marginal IFR of $0.756$\%.}
 \label{fig:tv_air_sens}
\end{figure}

Figure~\ref{fig:undercounting} is a plot of undercount based on the SEIR model.  We plot relative undercount on a cumulative basis since January due to the case counts reflecting active infections. Figure~\ref{fig:tv_air_sens} is a sensitivity analysis of the active infection rate under SEIR model with 10\% increase and decrease in the average IFR.  We see that the impact on the doubly robust estimates is minimal.  A floor based on weekly case count divided by the Indian population size is also presented.

\section{Alternative estimator of the instantaneous reproductive number}
\label{app:cori_rt}

Here we present estimates of the instantaneous reproductive number using the approach of~\cite{Cori20113} under the SEIR model from Section~\ref{section:r0-estimation}.  Let~$I_t$ denote the total infectiousness of infected individuals at time~$t$.  Then~$E[I_t] = R_t \sum_{s=1}^{t} I_{t-s} w_s$ where $w_{u}$ is the infectivity function.  Here, we follow~\cite{Cori20113} and choose a discretized shifted Gamma distribution with mean $7$ days and standard deviation $2$ days.  Then a moment-based estimator can be obtained by $\hat R_t = I_t / \sum_{s=1}^t I_{t-s} w_s$, which takes the form of the ratio estimators in Section~\ref{section:rates}.  Therefore, the bias can be readily obtained from the associated Taylor series decomposition, with the terms related to time~$t-1$ replaced by a weighted version.

\begin{figure}
 \centering
  \includegraphics[width=.75\linewidth]{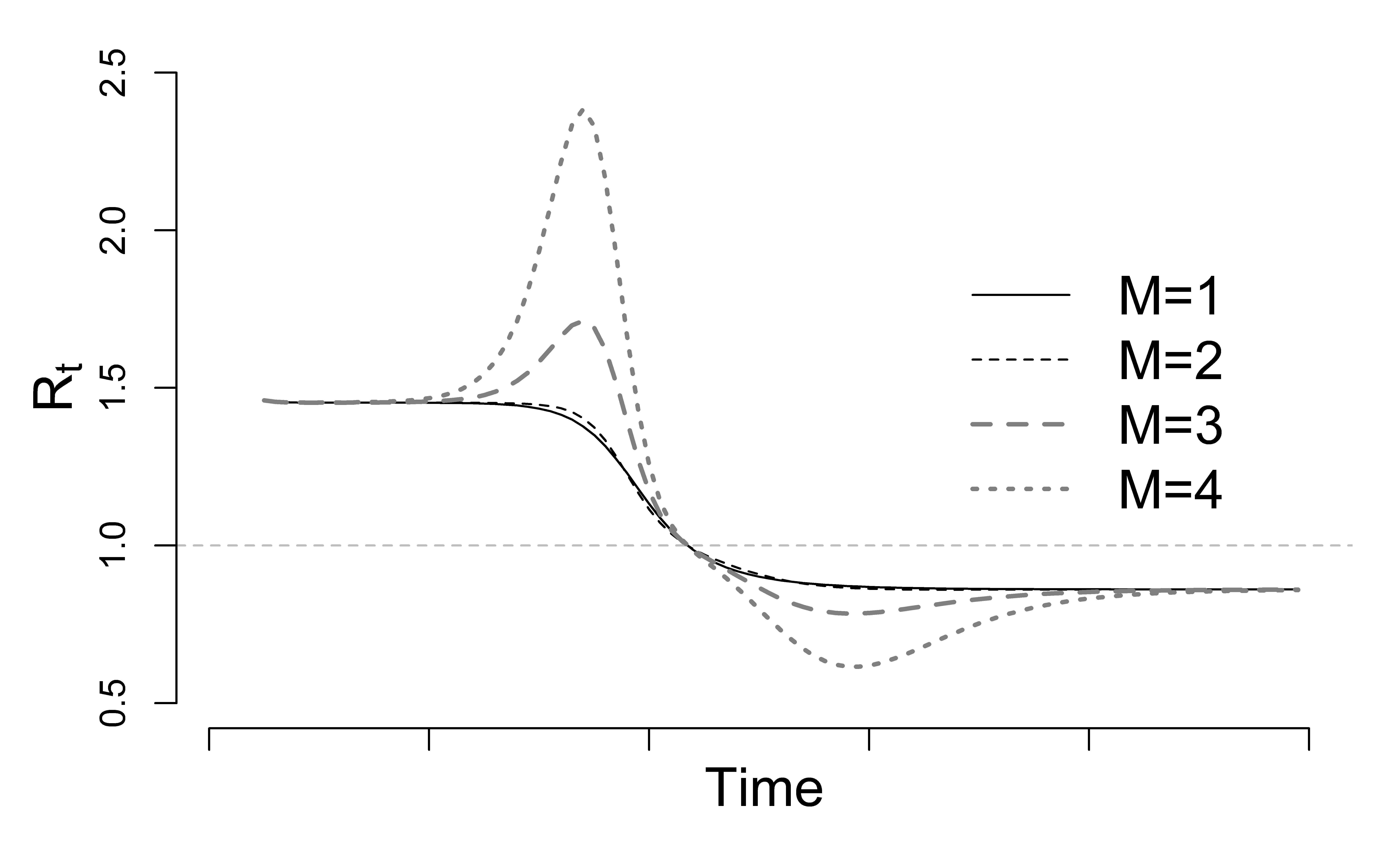}
  \caption{Effective reproductive rate estimator}
 \caption{Potential bias in instantaneous reproductive rate estimator based on~\cite{Cori20113} under an SEIR model with $\beta = 1.2$, $\gamma = 0.15$, and $\sigma = 0.3$.  Here, $f = 0.02$, $FP = 0.024$, $FN = 0.13$, and a range of relative sampling fractions $M = f_1/f_0$ are considered.}
 \label{fig:cori-bias}
 \end{figure}

Figure~\ref{fig:cori-bias} presents the potential bias in instantaneous reproductive rate estimator.  The general conclusions follow similarly as in  Figure~\ref{fig:ratio-bias}. The rate is overestimated prior to the peak and underestimated afterwards.  Estimates at the peak time appear to have minimal bias.

\section{Sensitivity Analysis to Unmeasured Confounding}
\label{sec:sensitivity_confounders}

The proposed approach relies on the assumption $I_{j,t}^{(NR)} \indep Y_{j,t} \, | \, X_{j,t}$, i.e., the sampling indicator in the non-probabilistic survey is conditionally independent of the outcome given the covariates.  However, estimates of population quantities based on inverse-weighting and/or model-based estimates will be biased in the presence of `unobserved confounding'.  Here we introduce a sensitivity analysis to address this concern.  To this end, we consider the logit-linear model by~\cite{Imbens2003}, adjusted to the time-varying, self-selection setting from the single time-point, causal inference setting.  That is, consider the model
\begin{align*}
\logit \pi \left( I_{t} = 1  | x_{t}, u_{t} \right) &= h_t(x_t) + \alpha_t u_t \\
E[ Y_t | i_t, x_t, u_t] &= l_t(x_t) + \delta_t u_t
\end{align*}
for functions~$h_t$ and $l_t$ that depend on time. Re-arranging terms we have
$$
E[ Y_t | i_t, x_t, u_t ] = l_t(x_t)  + \frac{\delta_t}{\alpha_t} \left( \logit \pi \left( I_{t} = 1  | x_{t}, u_{t} \right) - h_t (x_t) \right)
$$
The key idea in~\cite{Imbens2003} is that positing a distribution on $\pi(I_t = 1 | x_y, u_t)$ directly allows one to circumvent the need to specify a distribution for~$U_t$, a highly non-trivial task.

The logit-linear model above unfortunately does not lead to a tractable sensitivity analysis.  Instead, we consider a sensitivity model based on work by~\cite{Veitch2020}, again adjusted to deal with the current setting of self-selection, given by
\begin{equation}
\begin{aligned}
\label{eq:sensmodel}
\pi_t(X_t, U_t) &\sim \text{Beta} \left( \pi_t (X_t) (1/\alpha_t - 1), (1-\pi_t(X_t)) (1/\alpha_t - 1) \right) \\
I_t | X_t, U_t &\sim \text{Bern} (\pi (X_t,U_t)) \\
\rho_t(X_t, U_t) &= Q_t(X_t,1) + \delta_t \left( \logit \pi_t(X_t,U_t) - \EE \left[ \logit \pi_t(X_t,U_t) | X_t, I_t = 1 \right] \right) \\
Y_t | I_t, X_t, U_t &\sim \text{Bern}(\rho_t(X_t, U_t)),
\end{aligned}
\end{equation}
where~$Q_t(X_t,1)$ is the conditional expectation of the outcome given covariates~$X_t$ and self-selection into the nonprobability sample, i.e.,~$I_t = 1$. The time-varying sensitivity parameter~$\alpha_t \in (0,1)$  controls the influence of the unobserved confounder~$U_t$ on selection propensity. In particular, $\alpha_t$ measures the change in belief of how likely an individual self-selects into the non-probabilistic sample at time~$t$:
$$
\alpha_t = E[ \pi_t (X_t,U_t) | I_t = 1] - E[ \pi_t (X_t,U_t) | I_t = 0].
$$
The sensitivity model satisfies the requirement that the self-selection propensity and conditional expectation outcome model can match the observed data:
\begin{align*}
P(I_t = 1 | X_t ) &= E[ E[ I_t | X_t, U_t ] | X_t ] = E[ \pi_t(X_t, U_t) | X_t ] = \pi_t (X_t) \\
E[Y_t | X_t, I_t = 1 ] &= E[ E[ Y_t | X_t, U_t, I_t =1 ] | X_t, I_t = 1 ] =: Q_t(1,X_t).
\end{align*}
Moreover, the model satisfies that $Y_t$ does not depend on $I_t$ given $(X_t, U_t)$. This is a key adaptation to the non-probabilistic survey setting with potential time-varying confounding and a binary outcome.

\noindent {\bf IPW estimator:} By assumption, observing $X_t$ and $U_t$ together leads to consistent estimation via the estimator:
\begin{align*}
\EE\left[\pi^{-1}_t (X_{J,t}, U_{J,t}) I_{J,t} Y_{J,t} \right]
= &\EE[ E[ Y_{J,t} | X_{J,t}, U_{J,t}] ] \\
= &\EE \left[ Q(1,X_{J,t}) + \delta_t \left( \logit \pi_t (X_{J,t}, U_{J,t}) - \EE \left[ \logit \pi_t (X_{J,t}, U_{J,t}) | X_{J,t}, I_{J,t} = 1 \right] \right) \right]
\end{align*}
\noindent {\bf IPW-based estimator:} Investigating the IPW estimator using only~$X_t$, we have:
\begin{align*}
\EE\left[\pi^{-1}_t (X_{J,t}) I_{J,t} Y_{J,t} \right]
=&\EE\left[\frac{\pi_t (X_{J,t}, U_{J,t})}{\pi_t (X_{J,t})} \pi^{-1} (X_{J,t}, U_{J,t}) I_{J,t} Y_{J,t} \right] \\
= &\EE\left[ \frac{\pi_t (X_{J,t}, U_{J,t})}{\pi_t (X_{J,t})} \EE[ Y_{J,t} | X_{J,t}, U_{J,t}] \right].
\end{align*}
Using the fact that $\EE \left[ \pi_t (X_{J,t}, U_{J,t}) | X_{J,t} \right] = \pi_{t} (X_{J,t})$, the \emph{bias} is
$$
\delta_t \EE \left[ \logit \pi_t (X_{t,J}, U_{t,J}) - \frac{\pi_t (X_{J,t}, U_{J,t})}{\pi_t (X_{J,t})} \logit \pi_t (X_{t,J}, U_{t,J}) \right].
$$
Using the fact that~$Z \sim Beta(\alpha, \beta)$ then $E[ ln(Z) ] = \psi(\alpha) - \psi(\alpha + \beta)$, $E[ ln(1-Z) ] = \psi(\beta) - \psi(\alpha + \beta)$ then~$E[ \logit (Z) ] = \psi(\alpha) - \psi(\beta)$ where~$\psi$ is the digamma function.  Moreover,~$\EE[ Z ln Z ] = \frac{\alpha}{\alpha + \beta} \left[ \psi(\alpha + 1) - \psi(\alpha + \beta +1) \right]$ and
\begin{align*}
\EE \left[ (1-Z) ln(1-Z) \right] &= \frac{\beta}{\alpha + \beta} \left[ \psi(\beta + 1) - \psi(\alpha + \beta +1) \right] \\
\EE \left[ ln(1-Z) \right] &= \left[ \psi(\beta) - \psi(\alpha + \beta) \right] \\
\Rightarrow - \EE \left[ Z ln(1-Z) \right] &= \frac{\beta}{\alpha + \beta} \left[ \psi(\beta + 1) - \psi(\alpha + \beta +1) \right] - \left[ \psi(\beta) - \psi(\alpha + \beta) \right].
\end{align*}
and thus
\begin{align*}
\EE \left[ Z \logit(Z) \right] =& \frac{\alpha}{\alpha + \beta} \left[ \psi(\alpha + 1) - \psi(\alpha + \beta +1) \right] + \frac{\beta}{\alpha + \beta} \left[ \psi(\beta + 1) - \psi(\alpha + \beta +1) \right] \\
&- \left[ \psi(\beta) - \psi(\alpha + \beta) \right] \\
=& \psi(\alpha + \beta) - \psi(\alpha + \beta +1) + \frac{\alpha}{\alpha + \beta} \left[ \psi(\alpha + 1) - \psi(\beta+1) \right] + \frac{1}{\beta} \\
=& \frac{\alpha}{\alpha + \beta} \left[ \psi(\alpha + 1) - \psi(\beta+1) \right] + \left[ \frac{1}{\beta} - \frac{1}{\alpha + \beta} \right].
\end{align*}
Plugging in $\alpha = \pi_t (X_t) (1/\alpha_t -1)$ and $\beta = (1-\pi_t (X_t))(1/\alpha_t - 1)$ yields:
$$
\pi_t (X_{t,J}) \left[ \psi \left(\pi_t (X_{t,J}) \left(\frac{1-\alpha_t}{\alpha_t}\right) + 1\right) - \psi \left((1-\pi_t (X_{t,J})) \left(\frac{1-\alpha_t}{\alpha_t}\right)+1\right) \right] + \frac{\alpha_t}{1-\alpha_t}\frac{\pi_t(X_{t,J})}{1-\pi_t(X_{t,J})}.
$$
Since $\pi_t(X_t)$ cancels with the denominator term, the first term will match the form of the $E[ \logit (Z)]$ and therefore using the fact that $\psi(x+1) = \psi(x) + 1/x$, we can write the bias simply as
$$
-\delta_t \frac{\alpha_t}{1-\alpha_t} \EE \left[ \frac{1}{\pi_t(X_{t,J})} \right]
$$
Note that this is an expectation over the population.  Thus, estimation requires use of the probabilistic sample in order to estimate the potential bias.

\subsection{Reparametrization}

Following~\cite{Veitch2020}, we re-express the outcome-confounder strength in terms of the partial coefficient of determination
$$
R_t^2 (\alpha_t, \delta_t) = \frac{\EE [( Y_{J,t} - Q_t(1,X_{J,t}) )^2 | I_{J,t}=1 ] - \EE[ (Y_{J,t} - \EE[ Y_{J,t} | X_{J,t}, U_{J,t} ])^2 | I_{J,t} = 1]}{E[(Y_{J,t} - Q_t(1,X_{J,t}))^2 | I_t = 1]}
$$
in terms of $\delta_t^2$. To do so, write
\begin{align*}
&\EE [ (Y_{J,t} - \EE[ Y_{J,t} | X_{J,t}, U_{J,t}] )^2 | I_{J,t} = 1 ] \\
= &\EE [ (Y_{J,t} - Q_t(1,X_{J,t}))^2 | I_{J,t} = 1 ] \\
&- 2 \delta \EE \left[ (Y_{J,t} - Q_t(1,X_{J,t})) (\logit \pi(X_{J,t},U_{J,t}) - \EE[ \logit \pi (X_{J,t},U_{J,t}) | X_{J,t}, I_{J,t}=1 ]) | I_{J,t} = 1 \right] \\
&+ \delta^2 \EE \left[ (\logit \pi_t(X_{J,t},U_{J,t}) - \EE[ \logit \pi_t (X_{J,t},U_{J,t}) | X_{J,t}, I_{J,t}=1 ])^2 | I_{J,t} = 1 \right] \\
= &\EE [ (Y_{J,t} - Q_t(1,X_{J,t}))^2 | I_{J,t} = 1 ] - \delta^2 \EE \left[ \text{var} \left( \logit \pi_t(X_{J,t},U_{J,t}) \right) | X_{J,t}, I_{J,t} = 1 \right]
\end{align*}
By Beta-Bernoulli conjugacy, the second term is the variance of the logit-transformed Beta distribution which has an analytic expression:
$$
\text{var} \left( \logit \pi_t(X_{J,t},U_{J,t}) | X_{J,t}, I_{J,t} = 1 \right) = \psi_1 ( \pi(X_{J,t}) (\alpha_t - 1) + 1 ) + \psi_1 ( (1- \pi(X_{J,t})) (1/\alpha_t - 1))
$$
where $\psi_1$ is the trigamma function. This implies the following relationship:
\begin{equation}
\label{eq:partialcoef}
R_t^2 (\alpha_t, \delta_t) = \delta_t^2 \frac{\EE [\psi_1 ( \pi(X_{J,t}) (\alpha_t - 1) + 1 ) + \psi_1 ( (1- \pi(X_{J,t})) (1/\alpha_t - 1)) | I_{J,t} = 1 ] }{E [(Y_{J,t}- Q_t(1,X_{J,t}))^2 | I_{J,t} = 1]}
\end{equation}
Note that unlike the prior expectation, these are expressed conditional on self-selection, i.e., $I_{J,t} = 1$.  This was done since the binary outcomes are only measured in the non-probabilistic sample.

Following from~\cite[Theorem 5]{Veitch2020}, the parameter~$\alpha_t$ can be re-expressed as
\begin{equation}
\label{eq:alphaest}
\alpha_t = 1 - \frac{\EE \left[ \pi (X_{t,J},U_{t,J}) \left( 1 - \pi (X_{t,J},U_{t,J}) \right)  \right]}{\EE\left[ \pi (X_{t,J}) \left( 1 - \pi (X_{t,J}) \right) \right] },
\end{equation}
which is a more convenient form when trying to estimate the parameter from observed data.

\subsection{Calibration of sensitivity parameters}

Based on~\eqref{eq:partialcoef} and~\eqref{eq:alphaest}, one can use the probability and nonprobability samples to calibrate the sensitivity parameters to be in line with the observed covariate influence on treatment and outcome.  That is, for a given observed covariate~$Z_{J,t}$, we wish to measure the degree of influence it has on self-selection and outcome given the other observed covariates~$X_{J,t} \backslash Z_{J,t}$.  Note that this dependence is likely time-varying, which is why it is important that~$\alpha_t$ and~$R_t^2$ are functions of time~$t$.

For the outcome, we can measure the partial coefficient of determination as:
$$
R_{t, Z}^2 := \frac{\frac{1}{n} \sum_{i=1}^n (y_{i,t} - \hat Q_{t,Z} (1, x_{i,t} \backslash z_{i,t}) )^2 - \frac{1}{n} \sum_{i=1}^n (y_{i,t} - \hat Q_{t} (1, x_{i,t}) )^2}{\frac{1}{n} \sum_{i=1}^n (y_{i,t} - \hat Q_{t,Z} (1, x_{i,t} \backslash z_{i,t}) )^2},
$$
where~$\hat Q_t$ is the conditional expectation given all covariates and~$\hat Q_{t,Z}$ is the conditional expectation given all covariates except~$Z$.  Note these are sums averages the non-probabilistic sample.  We can measure influence of $Z$ on self-selection propensity given $X_{J,t} \backslash Z_{J,t}$ by
$$
\hat \alpha_{t, Z} := 1 - \frac{\sum_{j=1}^N I_{j,t}^R W_{j,t}^R \cdot \hat \pi(X_{J,t}) (1- \hat \pi(X_{J,t}))}{\sum_{j=1}^N I_{j,t}^R W_{j,t}^R \cdot \hat \pi(X_{J,t} \backslash Z_{J,t}) (1- \hat \pi(X_{J,t} \backslash Z_{J,t}))}
$$
where~$\hat \pi_t (\cdot)$ is the fitted self-selection probability based on~\eqref{eq:auxinfoprob} in Section~\ref{subsec:auxprob} using the appropriately chosen set of observed covariates.

\subsection{IPW Sensitivity Analysis of COVID-19 Active Infection Prevalence in Indiana}

\begin{figure}[!th]
 \centering
 \begin{subfigure}{.5\textwidth}
  \centering
  \includegraphics[width=.9\linewidth]{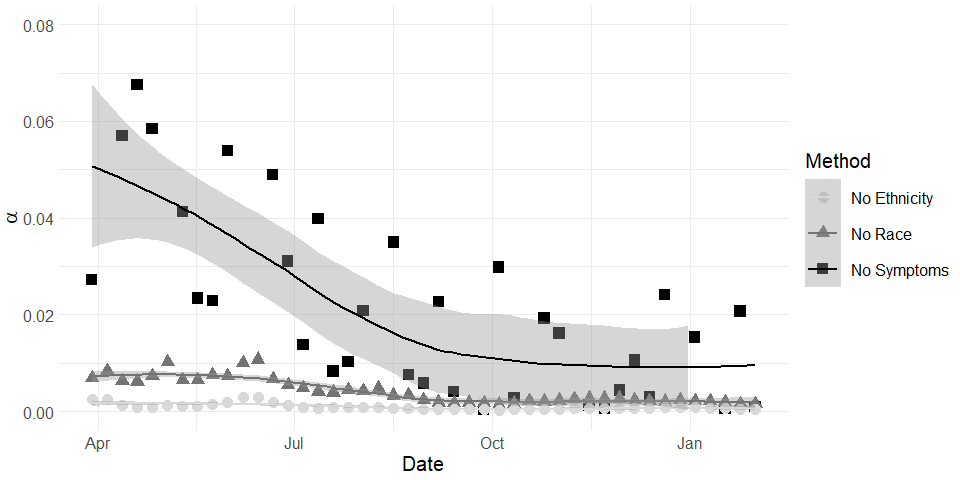}
  \caption{Calibration curves for~$\alpha$ per week}
  \label{fig:alpha_sensitivity}
 \end{subfigure}
 \begin{subfigure}{.45\textwidth}
  \centering
  \includegraphics[width=.9\linewidth]{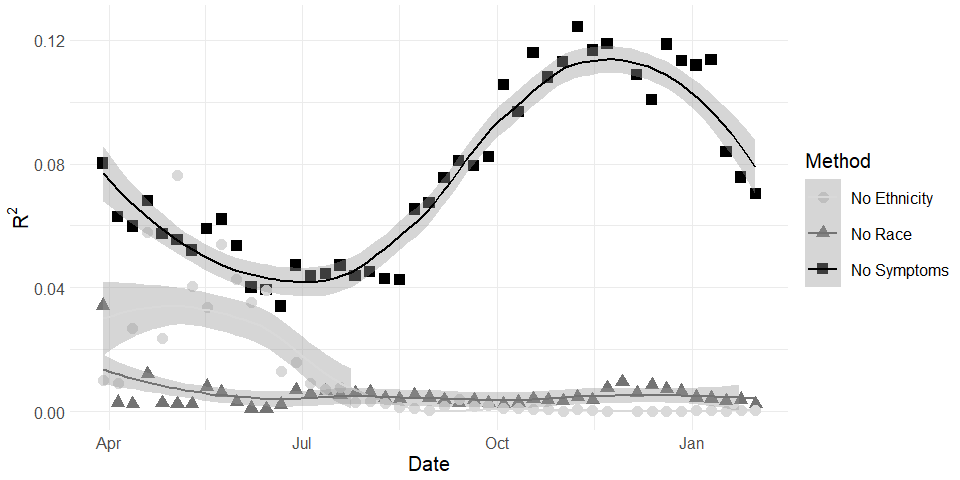}
 \caption{Calibration curves for~$R^2$ per week}
 \label{fig:rsq_sensitivity}
 \end{subfigure} \\ [1ex]
 \begin{subfigure}{\linewidth}
 \centering
 \includegraphics[width=.5\linewidth]{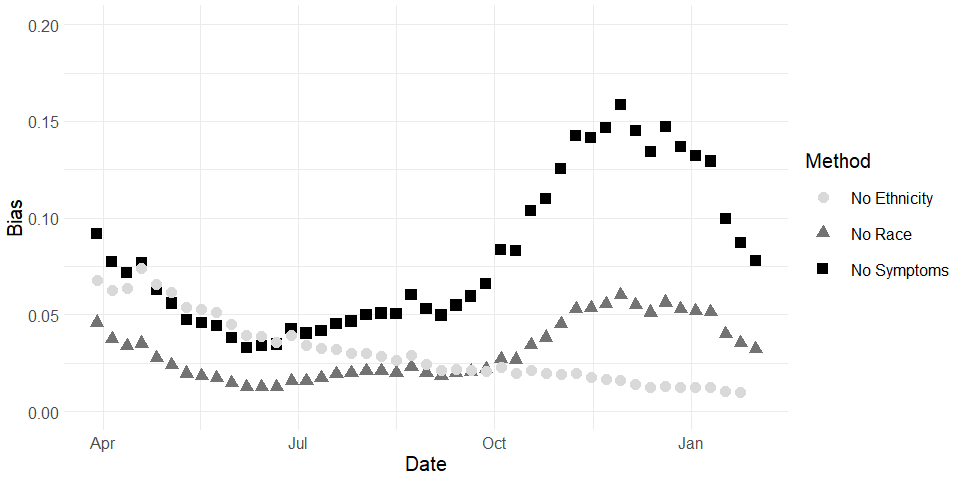}
  \caption{Bias as a function of~$\alpha$ and~$R^2$ calibration curves}
  \label{fig:bias_sensitivity}
 \end{subfigure}
 \caption{Sensitivity analysis for time-varying active infection rate estimates}
 \label{fig:sensitivity}
 \end{figure}

Here we perform a sensitivity analysis of the IPW estimator of COVID-19 active infection prevalence in Indiana.  While~\cite{Veitch2020} use Austen plots, we do not have such a simple visualization due to time-dependence.  Here, we plot  $\alpha_{t,Z}$ and $R^2_{t,Z}$ as functions of time for three covariates that demonstrate different patterns: (1) Symptoms, (2) Race, and (3) Ethnicity.  Figures~\ref{fig:alpha_sensitivity} and~\ref{fig:rsq_sensitivity} present these ``calibration curves.'' Figure~\ref{fig:bias_sensitivity} presents the bias under these three curves.  Here, we see that if an unobserved confounder has similar strength in selection-confounder and outcome-confounder relationships as ``Symptom Status'', then bias would be high even into early 2021.  However, bias under an unobserved confounder with similar strength to ``Ethnicity'' dissipates significantly by late 2020.

\end{document}